\documentclass[fleqn,usenatbib]{mnras}

\usepackage{fontawesome5}
\usepackage{hyperref}

\newcommand{\github}[1]{
   \href{#1}{\faGithubSquare}
}

\usepackage{newtxtext,newtxmath}

\usepackage[T1]{fontenc}

\usepackage{pdflscape}

\DeclareRobustCommand{\VAN}[3]{#2}
\let\VANthebibliography\thebibliography
\def\thebibliography{\DeclareRobustCommand{\VAN}[3]{##3}\VANthebibliography}

\usepackage{amsmath}
\usepackage{wasysym}

\usepackage{rotating}
\usepackage{xcolor}
\usepackage{sidecap}

\usepackage{hyperref}

\usepackage{graphics}
\usepackage{relsize}
\usepackage{natbib}

\usepackage{graphicx}	
\usepackage{amsmath}

\title[Mergers in SDSS]{A declining major merger fraction with redshift in the local Universe from the largest-yet catalog of major and minor mergers in SDSS}

\author[R. Nevin et al.]{R. Nevin,$^{1}$\thanks{E-mail: rnevin@fnal.gov}
L. Blecha,$^{2}$
J. Comerford,$^{3}$
J. Simon,$^{3}$\thanks{NSF Astronomy and Astrophysics Postdoctoral Fellow}
B. A. Terrazas,$^{4}$
R. S. Barrows,$^{3}$\newauthor
J. A. V\'{a}zquez-Mata$^{5}$
\\
$^{1}$Fermi National Accelerator Laboratory, P.O. Box 500, Batavia, IL 60510, USA\\
$^{2}$Department of Physics, University of Florida, Gainesville, FL 32611, USA\\
$^{3}$Department of Astrophysical and Planetary Sciences, University of Colorado, Boulder, CO 80309, USA \\
$^{4}$Columbia Astrophysics Laboratory, Columbia University, 550 West 120th Street, New York, NY 10027, USA\\
$^{5}$Departamento de F\'isica, Facultad de Ciencias, Universidad Nacional Aut\'onoma de M\'exico, Ciudad Universitaria, CDMX, 04510, M\'exico\\
}

\date{Accepted Feb 16, 2023. Received YYY; in original form ZZZ}

\pubyear{2022}

\begin{document}
\label{firstpage}
\pagerange{\pageref{firstpage}--\pageref{lastpage}}
\maketitle

\begin{abstract}
It is difficult to accurately identify galaxy mergers and it is an even larger challenge to classify them by their mass ratio or merger stage. In previous work we used a suite of simulated mergers to create a classification technique that uses linear discriminant analysis (LDA) to identify major and minor mergers. Here, we apply this technique to 1.3 million galaxies from the SDSS DR16 photometric catalog and present the probability that each galaxy is a major or minor merger, splitting the classifications by merger stages (early, late, post-coalescence). We present publicly-available imaging predictor values and all of the above classifications for one of the largest-yet samples of galaxies. We measure the major and minor merger fraction ($f_{\mathrm{merg}}$) and build a mass-complete sample of galaxies, which we bin as a function of stellar mass and redshift. For the major mergers, we find a positive slope of $f_{\mathrm{merg}}$ with stellar mass and negative slope of $f_{\mathrm{merg}}$ with redshift between stellar masses of $10.5 < M_* (log\ M_{\odot}) < 11.6$ and redshifts of $0.03 < z < 0.19$. We are able to reproduce an artificial positive slope of the major merger fraction with redshift when we do not bin for mass or craft a complete sample, demonstrating the importance of mass completeness and mass binning. We determine that the positive trend of the major merger fraction with stellar mass is consistent with a hierarchical assembly scenario. The negative trend with redshift requires that an additional assembly mechanism, such as baryonic feedback, dominates in the local Universe.

\end{abstract}


\section{Introduction}

The $\Lambda$CDM model of structure growth predicts that galaxies grow hierarchically through mergers, but uncertainty still surrounds the impact of mergers on physical processes in galaxies. For instance, while theory predicts that mergers contribute to the growth of stellar bulges and elliptical galaxies \citealt{Springel2000,cox08}, trigger star formation (\citealt{DiMatteo2008A&A...492...31D}) and active galactic nuclei (AGN, \citealt{Hopkins2006ApJ...652..864H}), and even quench star formation (\citealt{DiMatteo2005,Hopkins2008ApJS..175..390H}), observational work often disagrees about the importance of mergers for driving these evolutionary processes (e.g. for whether mergers trigger AGN and/or star formation, see \citealt{cister11,Knapen2015MNRAS.454.1742K,Ellison2019MNRAS.487.2491E,Pearson2019A&A...631A..51P}). This is a critical tension: the implication is that our models and/or our current methods for identifying mergers are incorrect.

In order to determine the role of mergers in driving galaxy evolution, reconcile simulations with observations, and test the $\Lambda$CDM cosmological model, the galaxy-galaxy merger rate and merger fraction are key diagnostic tools. The merger rate, which will be the focus of future work (Simon et al. 2023, in prep), is measured using the merger fraction and the merger observability timescale (\citealt{Lotz2011}), both of which vary as a function of redshift, mass, mass ratio, and critically, the technique used to identify mergers.

Characterizing the merger fraction as a function of mass, redshift, and mass ratio is critical for understanding the relative contributions of both major and minor mergers to the growth of different types of galaxies over cosmic time. For instance, we can use the mass- and redshift-dependent merger fraction to constrain the relative contribution of major and minor mergers to the growth of the most massive galaxies, which are predicted to assemble at late times by $\Lambda$CDM. It is therefore an important test of $\Lambda$CDM cosmology. We can also use the merger fraction to test the predictions of other structure formation channels (see \S \ref{discuss:theory} for a review).

Many different techniques exist to measure the evolution of the major merger fraction with redshift, including close-pair (e.g. \citealt{Patton1997ApJ...475...29P,Lin2004ApJ...617L...9L,Kartaltepe2007ApJS..172..320K,Bundy2009ApJ...697.1369B}), clustering (e.g. \citealt{Bell2006bApJ...652..270B,Robaina2010ApJ...719..844R}) and morphological techniques (e.g. \citealt{Lotz2008,Conselice2009}). The majority of these studies find that the major merger fraction peaks at earlier times, in agreement with the above theoretical measurements. Other work focuses on the evolution of the major merger fraction with stellar mass (e.g. \citealt{Xu2012ApJ...747...85X,Casteels2014}), finding either an increasing or decreasing merger fraction with stellar mass. For a thorough review of past results, see \S \ref{discuss:empirical}.

Most of the literature has focused either on the mass- or redshift-dependence of the merger fraction separately. Also, most of the redshift-dependent studies only cover higher redshifts. In this work we focus on constraining the mass- and redshift-dependent merger fraction for galaxies in the Sloan Digital Sky Survey (SDSS). Our focus is on the local Universe, which will allow us to avoid the uncertainties that plague many of the above studies due to small sample sizes. We additionally use a carefully calibrated morphologically-based technique that avoids incompleteness issues due to fiber overlap.

While most past work has focused on the more easily measured major merger fraction, the minor merger fraction is also an important quantity. Past work finds that the minor merger fraction is several times higher than the major merger fraction (e.g. \citealt{Lotz2011,Lopez-Sanjuan2011A&A...530A..20L,Bluck2012ApJ...747...34B,Kaviraj2014aMNRAS.440.2944K,Kaviraj2014bMNRAS.437L..41K,Rodriguez-Gomez2015}), indicating that minor mergers have a critical role to play in building mass in disk galaxies, the envelopes of massive ellipticals, and the bulges of lower mass galaxies without destroying the merger remnant (\citealt{Hopkins2010}). In this work we set out to constrain not only the major merger fraction but also the minor merger fraction and how they both vary as a function of stellar mass and redshift.

In addition to providing constraints on the importance of galaxy mergers for galaxy evolution, the galaxy-galaxy merger fraction and rate are crucial for constraining the predicted supermassive black hole (SMBH) merger rate. The SMBH merger rate will be measured by upcoming gravitational wave observatories such as the (evolved) Laser Interferometer Space Antenna (eLISA, LISA), which is anticipated to detect SMBH mergers out to $z \sim 10$ (\citealt{Amaro-Seoane2017arXiv170200786A,GravitationalWaveAdvisory2016APS..APRJ12002M,LISA2022LRR....25....4A}), and indirectly measured by pulsar timing arrays through the gravitational wave background (e.g. \citealt{Hobbs2010CQGra..27h4013H,NanoGrav2015ApJ...813...65N,Nanograv2020ApJ...905L..34A}), which is dominated by the signal from binary SMBHs, which form following major galaxy mergers, with masses $M_{SMBHB} > 10^7\ M_{\odot}$ out to $z\sim2$ (e.g. \citealt{Sesana2013CQGra..30x4009S,Simon2016ApJ...826...11S}).

The galaxy-galaxy merger rate is also important for breaking degeneracies in the gravitational wave signal. For instance, \citet{Siwek2020MNRAS.498..537S} find that the chirp mass of SMBH binaries is degenerate with the merger rate, so separately constraining the galaxy-galaxy merger rate can complement gravitational wave background measurements, break these degeneracies, and constrain SMBH accretion models. A strength of the LDA technique used in this work to identify mergers is that it is created from detailed temporal simulations of mergers, hence we have a solid understanding of the merger observability timescale. In future work (Simon et al. 2023, in prep), we plan to combine the observability timescales from this work with the merger fractions also measured in this work to derive the galaxy-galaxy merger rate and make predictions for the expected gravitational wave background signal from merging binary SMBHs in the local universe.

In this paper, we address the above challenges using a statistical learning tool calibrated on well-understood hydrodynamical models of merging galaxies from \citet{Nevin2019} (henceforth N19). We apply this automated merger classification technique to the 1.3 million galaxies in the Sloan Digital Sky Survey (SDSS) DR16 photometric sample (\S \ref{data}). The strength of this approach lies in the massive statistical sample of mergers identified using a morphological-based technique that exceeds previous morphological techniques in accuracy and completeness to classify different types of mergers (\S \ref{methods}). The focus of this paper is twofold: 1) We present publicly-available catalogs of different types of mergers identified by both stage and mass ratio (major/minor, early, late, and post-coalescence) and 2) We estimate the galaxy merger fraction as a function of mass ratio, mass, and redshift (\S \ref{results}). We end by discussing our results in the context of cosmological models, past empirical studies of the merger fraction, and future directions (\S \ref{discuss}). A cosmology with $\Omega_m = 0.3$, $\Omega_\Lambda = 0.7$, and $h = 0.7$ is assumed throughout.

\section{Data}
\label{data}

Here we present an overview of the data set. We describe how we create image cutouts and the properties of the photometric sample in \S \ref{cutouts}. We present our process for measuring imaging predictor values from these image cutouts in \S \ref{measure_predictors}.

\subsection{Creating image cutouts of galaxies in SDSS}
\label{cutouts}
\begin{sloppypar}
The Sloan digital sky survey (SDSS, \citealt{Gunn2006}) is an all-sky spectroscopic and imaging survey. To construct our sample of galaxies, we use the $r-$band imaging data from data release 16 (DR16, \citealt{DR16}), which is the fourth data release of SDSS-IV (\citealt{Blanton2017}). Using CasJobs, we select all galaxies from the DR16 photometric catalog that have an $r-$band magnitude less than or equal to 17.77, the completeness limit of SDSS. We do not restrict the selection to objects that also have a spectroscopic object ID, maximizing the number of objects in the sample. We also do not restrict the sample by redshift. The redshift range of the mass complete sample (described in \S \ref{methods:mass}) is $0.03 < z < 0.19$.

The exact SQL search is as follows:
\end{sloppypar}

\noindent\rule{84mm}{0.5mm}
\\
\noindent \textbf{select} po.objID, po.ra, po.dec,\\
\indent (po.petroMag\_r - po.extinction\_r) as dered\_petro\_r \\
\noindent \textbf{into} MyDB.five\_sigma\_detection\_saturated\_mode1 \\
\noindent \textbf{from} PhotoObj as po \\
\noindent \textbf{where} (po.petroMag\_r)<=17.77 and po.type = 3 \\
\indent and ((flags\_r \& 0x10000000) != 0) \\
\indent and (flags\_r \& 0x40000) = 0 and mode=1

\noindent\rule{84mm}{0.5mm}

This query restricts the search to galaxies (po.type=3), eliminates galaxies that are detected at less than 5$\sigma$ (flags\_r 0x10000000!=0) and galaxies for which no petrosian radius could be determined in the $r-$band (flag\_r 0x40000 = 0), and removes duplicates using mode=1. This search returns 1393923 galaxies.

We use the Skycoords utility from \texttt{astropy} to create $80\farcs0$ by $80\farcs0$ square cutout $r-$band images for each galaxy from the frame images. After eliminating a small fraction ($\sim$0.4\%) of the cutouts that are blank, corrupted, or at the edge of the frame, we have a total of 1388533 galaxy cutout images.

\subsection{Measuring predictor values from the SDSS cutout images}
\label{measure_predictors}

For each galaxy image, we measure seven imaging predictor values: $Gini$, M$_{20}$, Concentration ($C$), Asymmetry ($A$), Clumpiness ($S$), Sersic index ($n$), and shape asymmetry ($A_s$). We use the same procedure as N19 to measure the imaging predictors, which incorporates \texttt{SourceExtractor} (\citealt{Bertin1996}), \texttt{GALFIT} (\citealt{Peng2002,Peng2010}), and \texttt{statmorph} (\citealt{Rodriguez-Gomez2019}). We also use \texttt{statmorph} to measure the average S/N value (<S/N>) within the segmentation maps. After extracting the imaging predictors, the sample size is 1344677 galaxies; we lose about 3\% of the sample due to either \texttt{GALFIT} or \texttt{statmorph} failing to converge on a good fit.

We next flag galaxies for unreliable predictor values; these galaxies are included in both the predictor and the classification tables but are excluded from our analysis of the merger fraction. Excluding the galaxies with one or more flags, there are 938892 galaxies with clean photometry. We employ three separate flags:

\begin{enumerate}

\item The `low S/N' flag is thrown when the average S/N value is below 2.5, which is the cutoff value quoted in N19 below which the classification is significantly different. 

\item The `outlier predictor' flag is thrown when one or more imaging predictors are outside the range of predictor values from the simulated galaxies. The range of simulated values is: $0.44 < Gini < 0.72$, $-2.70 < M_{20} < -0.50$, $1.32 < C < 5.57$, $-0.24 < A < 0.76$, $-0.24 < S < 0.16$, $0.47 < n < 5.14$, and $0.0 < A_s < 1.21$.
\item The `segmap' flag is thrown when the segmentation map does not include the central pixel or for when the segmentation map extends beyond the edge of a clipped image. This identifies images for which the predictor values are actually measuring a brighter foreground galaxy or star.
\end{enumerate}

\begin{figure*}
    \centering
    \includegraphics[scale=0.65]{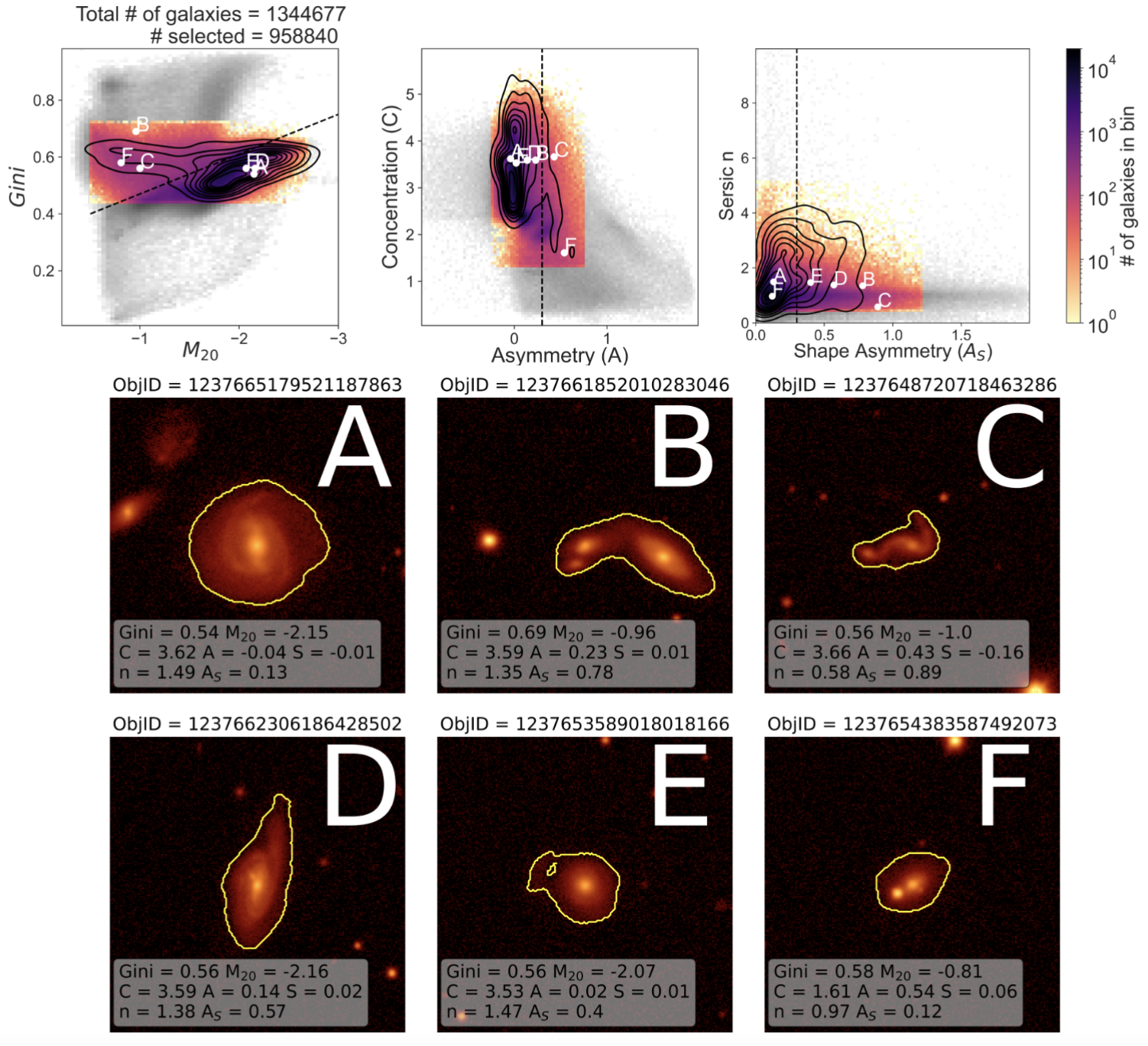}
    \caption{Distributions of predictor values for the full SDSS DR16 sample of galaxies (top, grey distribution), the simulated galaxies (black contours), and the selected non-flagged sample of galaxies (color distribution). We show six example galaxies with predictor values and segmentation maps (bottom) and overplot the locations of these galaxies on the top panels. All galaxy image panels are $80\farcs0\times80\farcs0$.}
    \label{fig:distabc}
\end{figure*}

We present the predictor values for six galaxies in Table \ref{tab:preds}. We plot the distributions of predictor values for the full sample in Figure \ref{fig:distabc} alongside the six example galaxies from Table \ref{tab:preds} identified with capital letters A-F.

\begin{table*}
    \centering
    \begin{tabular}{c|c c c c c c c | c | c c c}
        & \multicolumn{7}{c}{Predictor Values$^b$}\ &  & \multicolumn{3}{c}{Flags$^d$}\\
         SDSS ObjID$^a$&$Gini$ & $M_{20}$ & $C$ & $A$ & $S$& $n$ & $A_s$ & S/N$^c$ & low S/N & outlier predictor & segmap\\
         \hline
         1237665179521187863 (A) &	0.54 & 	-2.15 & 	3.62 &	-0.04 &-0.01 &	1.49 &	0.13&9.98 &	0	&0	&0\\
         1237661852010283046 (B) & 0.69 & 	-0.96 &	3.59 &	0.22 & 0.01 &	1.32 & 0.78 &	12.49 &	0	& 0	& 0\\
         1237648720718463286 (C) & 0.56 & -1.0 & 3.66 & 0.43 & -0.16 & 0.58 & 0.89 & 6.4 & 0 & 0 & 0\\
         
         1237662306186428502 (D) & 0.56 & 	-2.16 & 3.59 & 0.14 & 	0.02 & 	1.38 & 	0.57 &	16.35 & 	0&	0	&0\\
         1237653589018018166 (E) & 0.56 & 	-2.07 & 	3.53 & 0.02 & 0.01 & 	1.47 &0.40 & 14.31 & 	0 & 	0 & 	0\\
         1237654383587492073 (F) & 0.58 & -0.81 & 1.61 & 0.54 & 0.06 & 0.97 & 0.12 & 54.27 & 0 & 0 & 0 \\

    \end{tabular}
    \caption{Six galaxies from the table of predictor values alongside their identification letters (A-F) that will be used throughout this paper. \\\hspace{\textwidth}$^a$The SDSS photometric object ID from DR16 \\\hspace{\textwidth}$^b$The pre-standardized predictor values \\\hspace{\textwidth}$^c$Average S/N for the area of the galaxy enclosed by the segmentation mask \\\hspace{\textwidth}$^d$Flags have a value of 1 when activated}
    \label{tab:preds}
\end{table*}

\section{Methods}
\label{methods}

With predictor values in hand for 1.344 million galaxies, we are ready to classify the galaxies using the LDA imaging classification technique (\citealt{Nevin2019}). We review the classification technique and discuss some relevant changes in \S \ref{methods:review}. We describe how we further split the classification by merger stage in \S \ref{methods:sepbystage}. We apply the different classifications to the measured predictor values in \S \ref{methods:apply} and describe how we account for all possible merger priors in \S \ref{methods:marginalizing}, which is critical for the direct comparison of $p_{\mathrm{merg}}$ values across the different classifications as well as the calculation of the merger fraction. We present the \texttt{MergerMonger} suite in \S \ref{methods:MergerMonger}. Finally, we describe how we create a mass-complete sample in \S \ref{methods:mass}.

\subsection{Review of the LDA merger identification technique}
\label{methods:review}

The merger classification technique is built on a Linear Discriminant Analysis (LDA) framework that is trained to separate mock images of simulated nonmerging from merging galaxies using their imaging predictors. The full details of the technique are presented in \citet{Nevin2019} and \citet{Nevin2021} (henceforth, N19 and N21). N19 presents the imaging side of the approach, and N21 presents the kinematic side of the approach and some relevant changes to the N19 method. Here we will briefly review the results of these earlier papers.

The classification was trained using a suite of five \texttt{SUNRISE/GADGET-3} simulations of merging galaxies. The galaxies in this suite are best described as initially disk-dominated intermediate mass galaxies ($3.9-4.7 \times 10^{10} \ \mathrm{M}_{\odot}$). They span a range of stellar mass ratios ($\mu_* = {0.1,0.2,0.333,0.333,0.5}$), have gas fractions of 0.1 and 0.3, and have initial bulge-to-total-mass ratios of 0 and 0.2. While the simulated training set is limited in morphological parameter space, this does not significantly affect our main results (see \S \ref{discuss:robust}).

\begin{figure*}
    \centering
    \includegraphics[scale=0.15]{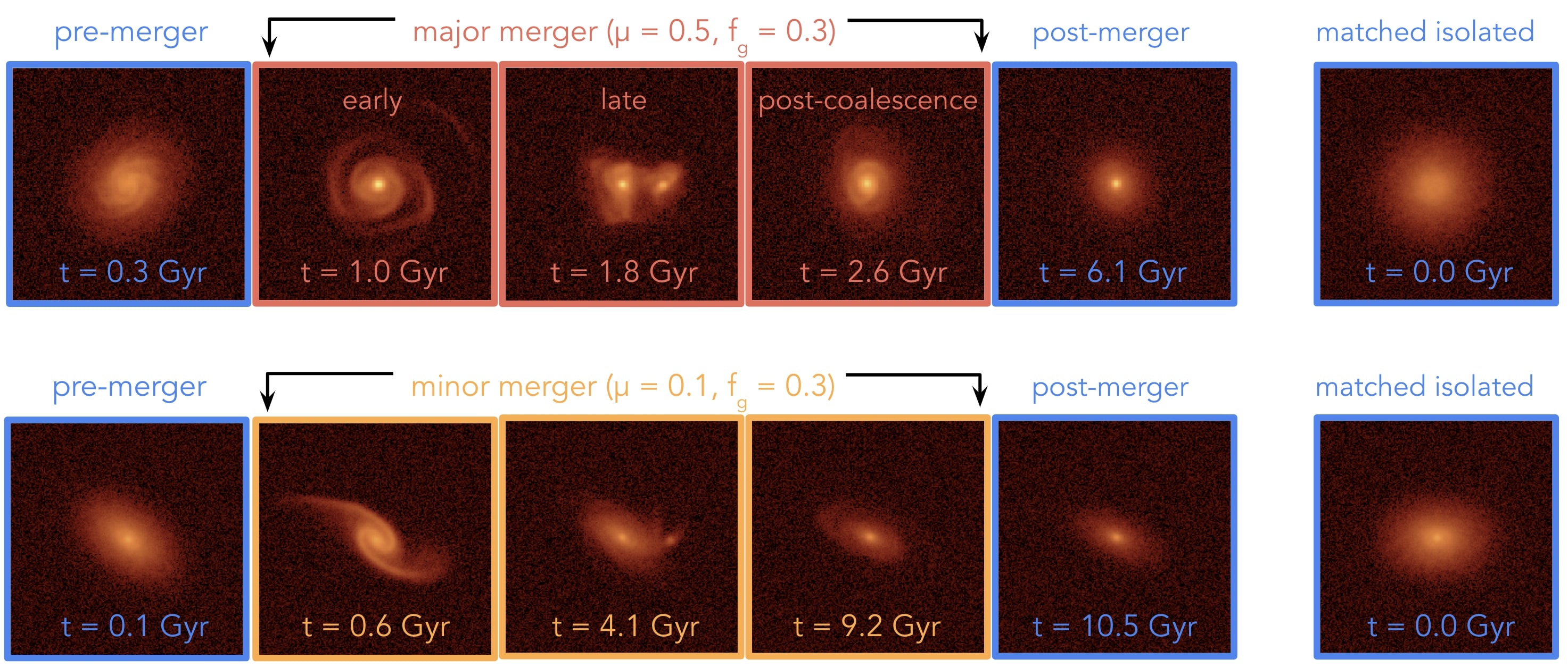}
    \caption{Snapshots from the $\mu = 0.5$ major merger (top row) and $\mu = 0.1$ minor merger (bottom row) with merging snapshots in pink and orange, respectively, and non-merging snapshots in blue. The non-merging snapshots include the pre-merger snapshots (before first pericentric passage), the post-merger snapshots ($>0.5$Gyr after coalescence), and the matched isolated galaxies (right column), which are matched to the initial conditions of each merger simulation in mass and gas fraction.}
    \label{fig:sim}
\end{figure*}

Each simulation spans 3-10 Gyr and contains a total of 100-200 snapshots in time, with a spacing of $\sim$10 Gyr. For each snapshot in time, we sample the merger at seven isotropically spaced viewpoints. We show example snapshots from the $\mu_*=0.5$ major merger and the $\mu_* = 0.1$ minor merger in Figure \ref{fig:sim}, where $\mu_*$ is the stellar mass ratio of the two merging galaxies.

In order to build the classification, we also required a set of simulated nonmerging galaxies which consist of isolated galaxies that were matched in gas fraction and stellar mass to each simulated merger as well as merging snapshots before first pericentric passage and 0.5 Gyr after final coalescence (pre- and post-merger snapshots).

We created mock images from the simulated galaxies that match the specifications of SDSS $r-$band images and measured the seven imaging predictors from the mock images. We trained seven separate LDA classifiers to identify mergers (one each for the five simulations and one each for a combined major and combined minor merger simulation).
    
Relevant details of the LDA classification include:
\begin{itemize}
    \item The LDA relies on a prior to correct for the larger fraction of merging relative to nonmerging galaxies in the simulations. In N19, we use fiducial merger fraction priors of $f_{\mathrm{merg}} = 0.1$ and 0.3 for the major and minor merger classifications, respectively. We explore how changing the merger fraction prior affects our measured posterior merger fraction in \S \ref{results:fraction}.
    \item We include interaction terms to explore correlations between predictors. 
    \item We use $k$-fold cross-validation to obtain 1$\sigma$ errors on the predictor coefficients and to measure the performance statistics of the classifications.
    \item In order to select which coefficients are necessary for the classification, we use a forward step-wise selection technique, which orders and includes only the relevant terms and interaction terms.
    
\end{itemize}

For complete details, including the full mathematical formulation for the LDA, see N19 and N21.

There are two key differences between the imaging LDA presented in N19 and the classification we use in this work that result in slightly different merger classifications and performance metrics. First, updates to the \texttt{scikit-learn} software (we are now using version 0.24.2, \citealt{scikit-learn}) including bug fixes and enhancements to the modeling logic result in classifications with different coefficients, terms, and slightly different performance metrics. Second, the training sets are slightly different from those used in N19; in N21 and here, we use the predictor values from all of the simulated snapshots that have measured values of imaging and kinematic predictors.

After rerunning the analysis from N19 with all of the above updates, the major merger classification is:

    \begin{equation}
    \begin{split}
        \mathrm{LD1}_{\mathrm{major}} =& \textcolor{blue}{+13.9\ A_s} \textcolor{red}{-8.0\  C*A_s - 5.4\ A*A_s} \textcolor{blue}{+ 5.1\ A} \\
        & \textcolor{blue}{+ 4.8\ C} \textcolor{red}{- 2.9\ Gini*A_s} \textcolor{blue}{+ 0.6\ M_{20}*A}\\ 
        & \textcolor{blue}{+ 0.4\ M_{20}*n + 0.4\ Gini} - 0.6 
    \end{split}
    \label{eq:major}
    \end{equation}
    
Terms with positive/negative contributions to the LD1 value are blue/red.

    The minor merger classification is:
       \begin{equation}
       \begin{split}
        \mathrm{LD1}_{\mathrm{minor}} =&  \textcolor{red}{-10.4\  C*A_s}  \textcolor{blue}{+ 8.8\ C*A} 
    \textcolor{red}{ -7.8\ Gini*S  -7.8\  A}\\
    & \textcolor{blue}{+ 6.6\ A_s + 6.5\ Gini*M_{20}} \textcolor{red}{- 6.0\ M_{20}*S} \\
    &\textcolor{red}{-5.7\ M_{20}*A_s} \textcolor{blue}{+4.9\ S} \textcolor{red}{-4.4\ M_{20}} \textcolor{blue}{+3.7\ Gini*C}\\ &\textcolor{red}{-2.9\ S*n -1.0\ n*A_s -0.2\ A*S} -0.7
        \end{split}
        \label{eq:minor}
        \end{equation}

We present the four leading coefficients for each LDA run alongside their uncertainties in Table \ref{tab:leading}.
    
We quantify the observability timescales and performance metrics for the LDA classifications using the cross-validation set of simulated mergers. We measure the observability timescale by applying each classification to the corresponding simulation and determining the length of time where the average LD1 value for consecutive snapshots is greater than zero. The observability timescale of the major/minor merger classifications is 2.31/5.36 Gyr. It is important to emphasize that the observability timescale is a performance metric that is measured by applying the derived LDA classifications applied to the simulated images. This is why the observability timescales from the early and late stage classifications do not sum to the observability timescale of the pre-coalescence classification.

Accuracy ($A$) is the fraction of true positive (TP) and true negative (TN) classifications relative to all classifications: 
$$A = \frac{TP + TN}{TP + TN + FP + FN}$$ 
where FP are false positive and FN are false negative classifications. 

Precision ($P$) quantifies the fraction of true positive classifications to all positive classifications:
$$P = \frac{TP}{TP + FP}$$
Recall is also known as the completeness and quantifies the ability of the classifier to retrieve mergers:
$$R = \frac{TP}{TP + FN}$$
The F1 score is the harmonic mean of precision and recall:
$$F1 = \frac{2TP}{2TP + FN + FP}$$ 

The major merger combined simulation has an accuracy of 0.86, a precision of 0.96, and a recall of 0.83. The minor merger combined simulation has an accuracy of 0.77, a precision of 0.93, and a recall of 0.63. We present these performance metrics and the observability timescales for all classifications in Table \ref{tab:accuracy}.

\begin{table*}
    \centering
    \begin{tabular}{c|cccc}
    Classification & Term 1 & Term 2 & Term 3 & Term 4\\
    \hline
        All Major Mergers & 13.9 $\pm$1.0 $A_s$ & -8.0 $\pm$ 0.7 $A_s*C$  &-5.4 $\pm$ 0.4   $A_s*A$  &5.1 $\pm$ 0.4 $A$     \\
        
          Major, pre-coalescence & 10.0 $\pm$ 0.6 $A_s$    &
7.5 $\pm$ 0.2 $A$    &
-6.3 $\pm$ 0.2 $A_s*A$  &
-6.1 $\pm$ 0.5 $A_s*C$ \\

        Major, early stage & 9.1 $\pm$ 0.4 $A_s$   &
-5.8 $\pm$ 0.4  $A_s*C$   &
5.3 $\pm$ 0.6 $C$   &
4.9 $\pm$ 0.5 $A$  \\
        
        Major, late stage & -8.9 $\pm$ 0.8 $A_s*A$  &
7.9 $\pm$ 0.4 $A_s$  &
7.2 $\pm$ 0.7   $Gini*A$  &
1.2 $\pm$ 0.2   $A*S$ \\

        Major, post-coalescence (0.5)& -10.8 $\pm$ 0.9   $A_s*M_{20}$  &
10.1 $\pm$ 1.1   $C*Gini$  &
-10.0 $\pm$ 1.1   $A_s*C$  &
5.0 $\pm$ 0.9   $Gini*M_{20}$  
 \\

Major, post-coalescence (1.0)& -14.3 $\pm$ 0.9   $C*n$  &
11.7 $\pm$ 1.4   $C$  &
5.9 $\pm$ 0.9   $Gini*n$  &
-1.3 $\pm$ 0.2   $A_s*M_{20}$   \\
        
        \hline
        All Minor Mergers&-10.4 $\pm$ 1.9 $A_S*C$  &
8.8 $\pm$ 0.7   $A*C$  &
-7.8 $\pm$ 3.3   $Gini*S$  &
-7.8 $\pm$ 0.6   $A$
 \\
 
  Minor, pre-coalescence & -31.3 $\pm$ 7.7   $Gini*S$  &
-28.6 $\pm$ 6.0   $Gini*n$ &
27.4 $\pm$ 5.7   $n$ &
21.0 $\pm$ 2.8   $C$\\

        Minor, early stage & 20.8 $\pm$ 3.6   $C$ &
-20.5 $\pm$ 5.4   $Gini*C$ &
-18.0 $\pm$ 2.2   $n*M_{20}$ &
-16.7 $\pm$ 2.2   $n*C$ \\

        Minor, late stage &10.1 $\pm$ 1.4 $A_s*C$    &
-5.3 $\pm$ 1.0   $A_s*Gini$ &1.9 $\pm$ 0.1 $A_s*A$ & --\\

        Minor, post-coalescence (0.5) & 2.3 $\pm$ 0.2 $A_s$ & -- & -- & --\\
        
        Minor, post-coalescence (1.0)  &2.0 $\pm$ 0.1   $Gini$  &
-1.1 $\pm$ 0.1   $A*S$  & 0.6 $\pm$ 0.1   $n$ & --\\
    \end{tabular}
    \caption{The four leading coefficients and terms and for each classification. The LD1 value for each classification is constructed by multiplying the standardized predictor value by each coefficient and summing all terms. We distinguish between the post-coalescence classifications with a 0.5 Gyr and 1.0 Gyr cutoffs after coalescence. }
    \label{tab:leading}
\end{table*}

\subsection{Classifying by Merger Stage}
\label{methods:sepbystage}

\begin{table*}
    \centering
    \begin{tabular}{c|cccc|c}
    Classification & Accuracy & Precision & Recall & F1 & $t_{\mathrm{obs}}$\\
    \hline
        All Major Mergers & 0.86 &0.96 &0.83 &0.89 & 2.31\\
        Major, pre-coalescence & 0.87 & 0.96 & 0.83 & 0.89& 2.16\\
        Major, early stage & 0.86 & 0.95 & 0.78 & 0.86 & 1.72\\
        Major, late stage & 0.94 & 0.97 & 0.84 & 0.90 &0.83\\

    Major, post-coalescence (0.5)&0.84 & 0.89 & 0.65 & 0.75& 0.40 \\
        Major, post-coalescence (1.0)& 0.90 & 0.94 & 0.85 & 0.89 & 1.26\\
        \hline
        All Minor Mergers & 0.77 & 0.93 & 0.63 & 0.75 & 5.36\\
        Minor, pre-coalescence & 0.80 & 0.89 & 0.71 & 0.79 & 5.75 \\
        
        Minor, early stage & 0.83 & 0.89 & 0.73 & 0.80 & 3.11\\
        Minor, late stage &  0.93 & 0.79 & 0.79 & 0.79 & 5.85\\
        Minor, post-coalescence (0.5)& 0.85 & 0.53 & 0.60 & 0.56 & 0.19\\
        Minor, post-coalescence (1.0)&  0.85& 0.84& 0.71 & 0.77 & 0.96\\
    \end{tabular}
    \caption{Accuracy, precision, recall, F1 score, and observability timescale for each classification measured from the cross-validation sample of simulated mergers. }
    \label{tab:accuracy}
\end{table*}

In N19, the classification is applied to the entire duration of the merger (from early to post-coalescence stages). In this work, we further split the classification into multiple different stages (pre-coalescence, further subdivided into early and late, and post-coalescence). Splitting the classification by merger stage will enable other work to address if and how galaxy mergers drive time-dependent evolutionary processes. 

Our definitions of merger stage are based on previous theoretical and observational work that define merger stages using both morphological and evolutionary (i.e. star formation) properties. \citet{Moreno2015} establish a sequence of merger stages for the pre-coalescence stages of the merger based on triggered star formation: a) Incoming, b) First pericentric passage, c) Apocenter, and d) Second approach. Other theoretical work to identify mergers in cosmological simulations such as IllustrisTNG is limited in temporal sampling and tends to distinguish more coarsely between pre-coalescence and post-coalescence mergers, where the time since merger varies based on the study (\citealt{Hani2020,Bickley2021}).

Observational work most often defines merger stage based on projected separation. \citet{Ellison2013} distinguish between pre- and post-coalescence mergers in a sample of 10,800 spectroscopic close pairs in SDSS, where pre-coalescence mergers have projected separations less than 80 kpc. \citet{Pan2019} define a merger sequence based on morphological disturbance and separation; 1) well-separated pairs without disturbance, 2) close pairs with strong interaction signs, 3) well-separated pairs with weak distortion (apocenter), and 4) strong distortion (final coalescence) and single galaxies with morphological remnants from merging (post-mergers). 

We divide our classification into pre- and post-coalescence stages to match the methodology of cosmological merger identification schemes. The early and late stages roughly correspond to the stages from \citet{Moreno2015} and \citet{Pan2019} of first pericentric passage and apocenter (early) and final approach (late). We also implement a sliding timescale for the definition of the post-coalescence stage; we use the time cutoff of 0.5 Gyr after coalescence and then additionally implement a time cutoff of 1 Gyr. The 1 Gyr cutoff is motivated by the work of \citet{Bickley2021}, who find that the morphology of IllustrisTNG galaxies is disturbed for up to 2.5 Gyr following a merger.

To reconstruct the separate classifications, we eliminate all merger snapshots that are not from the stage in question. For example, for the major merger combined early stage classification, we eliminate all of the merger snapshots belonging to the late and post-coalescence stages, but we retain the pre- and post-merger snapshots as examples of nonmergers. In this way, we are training the classification to recognize traits of a specific stage while discouraging it from learning a strict cutoff between stages. 

It is important to mention that since the merger stage classifications are all trained separately, there may be overlap between stages, i.e. certain galaxies will have high probabilities of belonging to multiple merger stages. We discuss how to directly compare $p_{\mathrm{merg}}$ values from different classifications in \S \ref{results:interpret_stage} and quantify this overlap in \S \ref{results:fraction}.

We present the accuracy, precision, recall and F1 score for the new classifications in Table \ref{tab:accuracy} and the four leading coefficients for each new classification in Table \ref{tab:leading}.

\subsection{Classifying SDSS image cutouts}
\label{methods:apply}
\begin{figure*}
    \centering
  
    \includegraphics[scale=0.4]{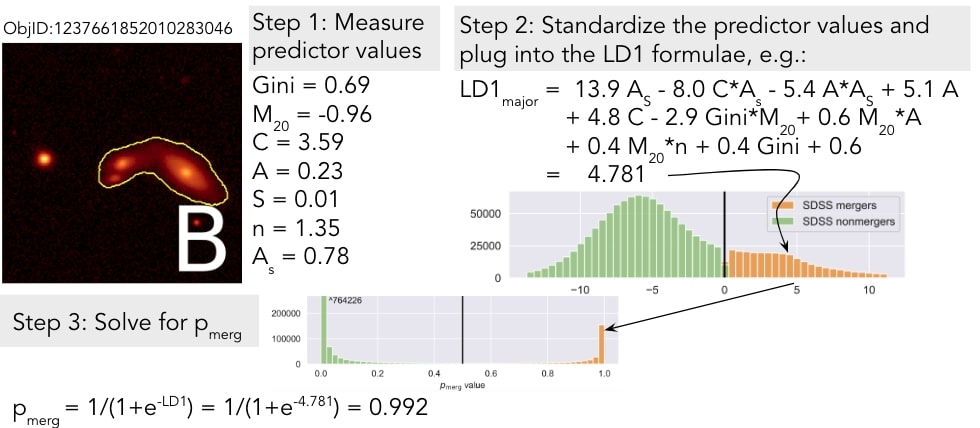}
    \caption{Schematic showing the classification steps for an example galaxy (left). Our first step is to measure the imaging predictor values (top middle). We then standardize these values and plug them into each LD1 formula. We show this (top right) for the major merger classification. The LD1 value for this galaxy is 4.781, which places it to the right of the decision value in the histogram of LD1 values (right). Our final step is to assign each galaxy a probability value (bottom left).}
    \label{fig:schematic}
\end{figure*}

The next step is to measure the LD1 values for each SDSS galaxy and to assign each galaxy a probability of merging for each merger classification. To calculate LD1 for each galaxy, we standardize the measured predictor values using the mean and standard deviation for each classification. We then determine the value of LD1 for each galaxy by summing the coefficients and standardized predictor values for each classification. We present a schematic of this process in Figure \ref{fig:schematic}, which demonstrates how this process works for one example image for the major merger combined classification.

We assign a probability of merging to each galaxy. From N19, the probability of a galaxy belonging to the merging class is:

\begin{equation}
    p_{\mathrm{merg}} = \frac{e^{\hat{\delta}_{\mathrm{merg}}}}{e^{\hat{\delta}_{\mathrm{merg}}}+e^{\hat{\delta}_{\mathrm{nonmerg}}}}
    \label{eq:prob}
\end{equation}
where $\hat{\delta}_{\mathrm{merg}}$/$\hat{\delta}_{\mathrm{nonmerg}}$ is the score of a galaxy for the merging/nonmerging class. 

Linear discriminant axis 1, or LD1, can be written in terms of $\hat{\delta}_{\mathrm{merg}}$ and $\hat{\delta}_{\mathrm{nonmerg}}$:

\begin{equation}
    \mathrm{LD1} = \hat{\delta}_{\mathrm{merg}} - \hat{\delta}_{\mathrm{nonmerg}}
    \label{eq:LD1}
\end{equation}

where the decision boundary is at LD1 = 0 and if $\hat{\delta}_{\mathrm{merg}} > \hat{\delta}_{\mathrm{nonmerg}}$, then the galaxy will be classified as merging.

Using equation \ref{eq:LD1}, equation \ref{eq:prob} can be re-written in terms of LD1:
\begin{equation}
    p_{\mathrm{merg}} = \frac{1}{1+e^{-\mathrm{LD1}}}
\end{equation}

For the 1344677 galaxies in SDSS DR16, we calculate the value of LD1 and the merger probability for the major and minor merger classifications and for all of the stage-specific classifications (early/late/pre-coalescence/post-coalescence). We present these results in \S \ref{results:class}.

\subsection{Marginalizing the calculation of the merger fraction over all merger priors}
\label{methods:marginalizing}

Critical to this paper is a discussion of the merger fraction priors ($\pi$) that are incorporated into the calculation of the $p_{\mathrm{merg}}$ values. In N19, we adopt a fiducial merger fraction prior of $\pi = 0.1$ for the major merger classifications and $\pi = 0.3$ for the minor merger classifications, meaning that we expect 10\% and 30\% of galaxies in the local Universe to be experiencing major and minor mergers, respectively. These priors are based on observations and simulations (e.g. \citealt{Rodriguez-Gomez2015,Lotz2011,Conselice2009,Lopez-Sanjuan2009,Shi2009,Bertone2009}). 

The fiducial priors are used to measure the LD1 and the $p_{\mathrm{merg}}$ values in the previous section. The choice of this input prior affects the distribution of LD1 and $p_{\mathrm{merg}}$ values for the full sample and therefore also affects the individual values. It is therefore particularly important to consider which $\pi$ value is used when comparing $p_{\mathrm{merg}}$ values between classifications and when calculating the merger fraction $f_{\mathrm{merg}}$, which is the focus of this paper.

To approach the comparison of $p_{\mathrm{merg}}$ values and the calculation of $f_{\mathrm{merg}}$ in the cleanest and most agnostic (to input prior) way possible, we perform a Bayesian marginalization where we re-calculate the $p_{\mathrm{merg}}$ values for all possible input priors in a range $0.05 < \pi < 0.5$ (we fully justify this range of priors in \S \ref{results:fraction}). The implication is that we redo the previous $p_{\mathrm{merg}}$ calculation for 46 different input priors, returning 46 different $p_{\mathrm{merg}}$ values for each galaxy in SDSS. From these, we calculate the 16th, 50th (median), and 84th percentile of the posterior distribution for each galaxy, which we present in \S \ref{results:class}. We present the results for the overall merger fraction calculation based on these measurements in \S \ref{results:fraction}.

\subsection{The \texttt{MergerMonger} Suite}
\label{methods:MergerMonger}

We prepare a suite of tools \textcolor{gray}{\github{https://github.com/beckynevin/MergerMonger-Public}  }(\texttt{MergerMonger})\footnote{\href{https://github.com/beckynevin/MergerMonger}{https://github.com/beckynevin/MergerMonger}} that applies the LDA method to classify major and minor merging galaxies from optical images. \texttt{MergerMonger} includes four main utilities: 

\begin{enumerate}
    \item \texttt{GalaxySmelter}: A tool for measuring imaging predictors from simulated or observed galaxy images.
    \item \texttt{Classify}: A tool that creates the LDA classification using the predictor values from the simulated training set. 

    \item \texttt{MergerMonger}: A tool that applies the LDA classification to observed galaxies, measuring merger probabilities. 
    \item Utilities that help with the interpretation of the predictor and probability values for each galaxy.
\end{enumerate}

In this work we apply the \texttt{MergerMonger} suite to SDSS $r-$band imaging. However, the classification is designed with broader use in mind. The classification can be re-created using new sets of simulated images (i.e. simulated images created to match the specifications of LSST or DESI imaging) or new imaging filters. For example, to apply the classification to LSST images, one could design their own set of mock LSST mergers and extract the training data using \texttt{GalaxySmelter}. Then, they could use \texttt{Classify} to train their own LDA classification and finally classify LSST galaxies using \texttt{MergerMonger}.

\subsection{Galaxy stellar masses}
\label{methods:mass}
To measure the stellar masses for the SDSS galaxies, we use the empirical relation from \citet{Bell2003} that relates the SDSS $u$, $g$, $r$, $i$, and $z$ band luminosities and colors to the stellar mass-to-light (M/L) ratio using the k-correction: $log_{10}(M/L) = a_{\lambda} + (b_{\lambda} \times color)$, where the color in units of AB magnitudes and the luminosity is in solar units.  We use the values for SDSS $g-r$ color because \citet{Du2019RAA....19..171D} find that the $g-r$ color provides an almost unbiased $M/L$ value for many different galaxy types and regions. We use the $a_r = -0.840$ and $b_r = 1.654$ from \citet{Zibetti2009MNRAS.400.1181Z}, which incorporates an TP-AGB star correction and revised SFHs for bursty galaxies, improving upon the prescription from \citet{Bell2003}.  

To conduct this calculation, we rely on photometric-based redshifts, which are available for the full SDSS sample (1035607 available photometric redshifts versus 437094 spectroscopic redshifts). In Appendix \ref{ap:mass} we further explore the differences between using photometric and spectroscopic redshifts to determine the stellar mass. Although there are biases inherent to using the photometric-based redshifts (especially at low redshift), we find that our results remain unchanged when we measure the merger fraction as a function of redshift (\S \ref{results:properties}).

\begin{figure}
    \centering
  \includegraphics[scale=0.04, trim = 6cm 0cm 12cm 7cm, clip]{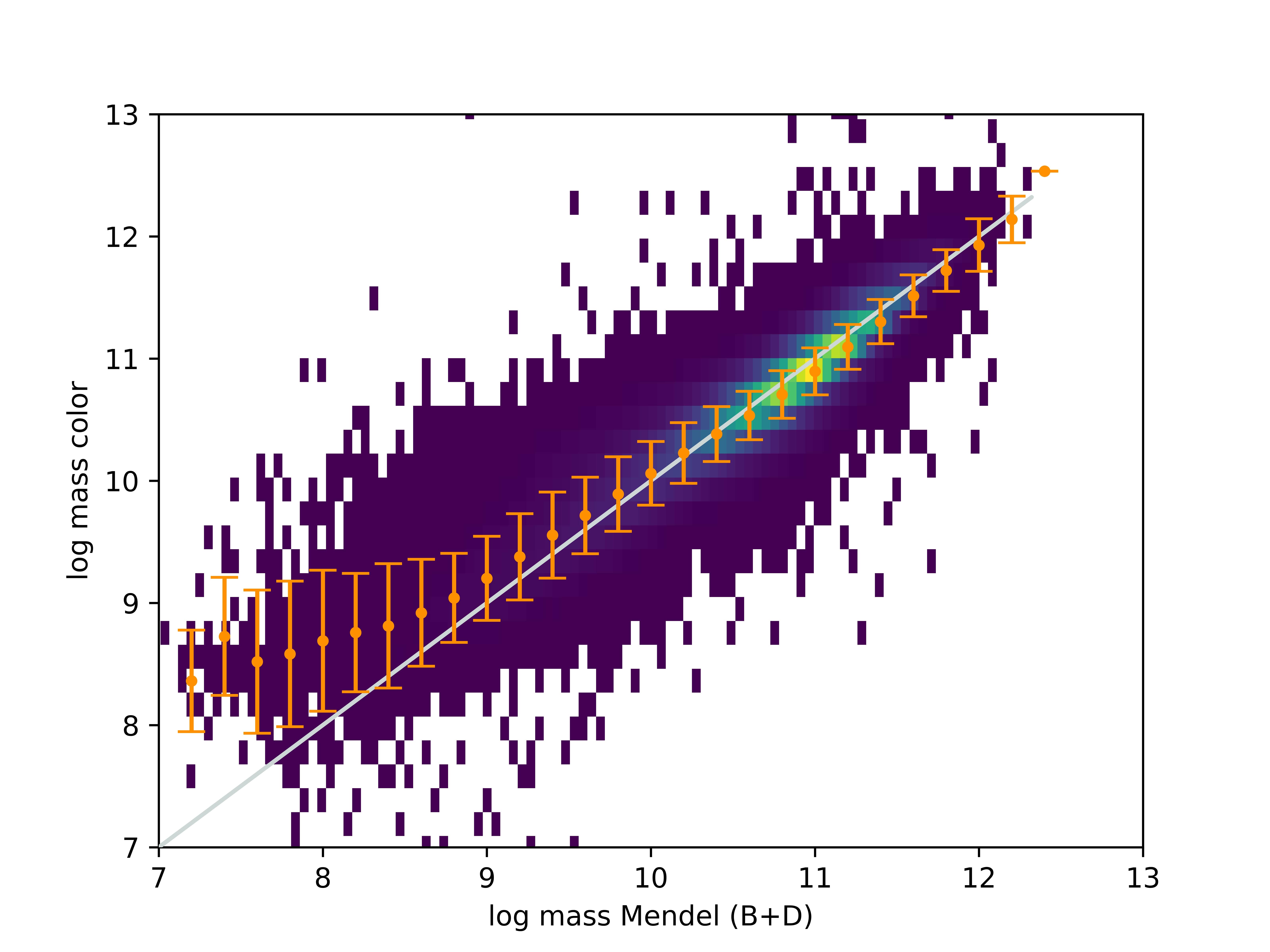}
    \caption{Comparing the stellar masses derived from the \citet{Mendel2014} method (x-axis) to those derived from using the empirical color method from \citet{Zibetti2009MNRAS.400.1181Z}. }
    \label{fig:mass_compare}
\end{figure}

Our method for measuring stellar mass shows good agreement with the SED-based approach of \citet{Mendel2014}, which uses a stellar population synthesis approach to measure the stellar mass using SDSS SEDs and S\'ersic models of the bulge and disk components. We present this comparison in Figure \ref{fig:mass_compare}, where the mean stellar masses agree above a stellar mass of $\sim10^9$.

Next, we determine the mass completeness limit as a function of redshift using the technique from \citet{Darvish2015ApJ...805..121D}. For each redshift bin\footnote{We use the redshift bins presented in \S \ref{results:properties}.}, we compute the lowest stellar mass ($M_{\mathrm{lim}}$) that could be detected for each galaxy given the magnitude limit of SDSS ($r = 17.77$): $log (M_{\mathrm{lim}}) = log (M) + 0.4\times(r - 17.77)$, where $r$ is the apparent (rest-frame) $r-$band magnitude of each galaxy and $M$ is the stellar mass. The mass completeness limit at each redshift bin is the mass at which 95\% of the limiting masses are below the mass completeness limit, meaning that only 5\% of galaxies would be missed in the lowest mass end of the mass function. 

\begin{figure}
\includegraphics[scale=0.2]{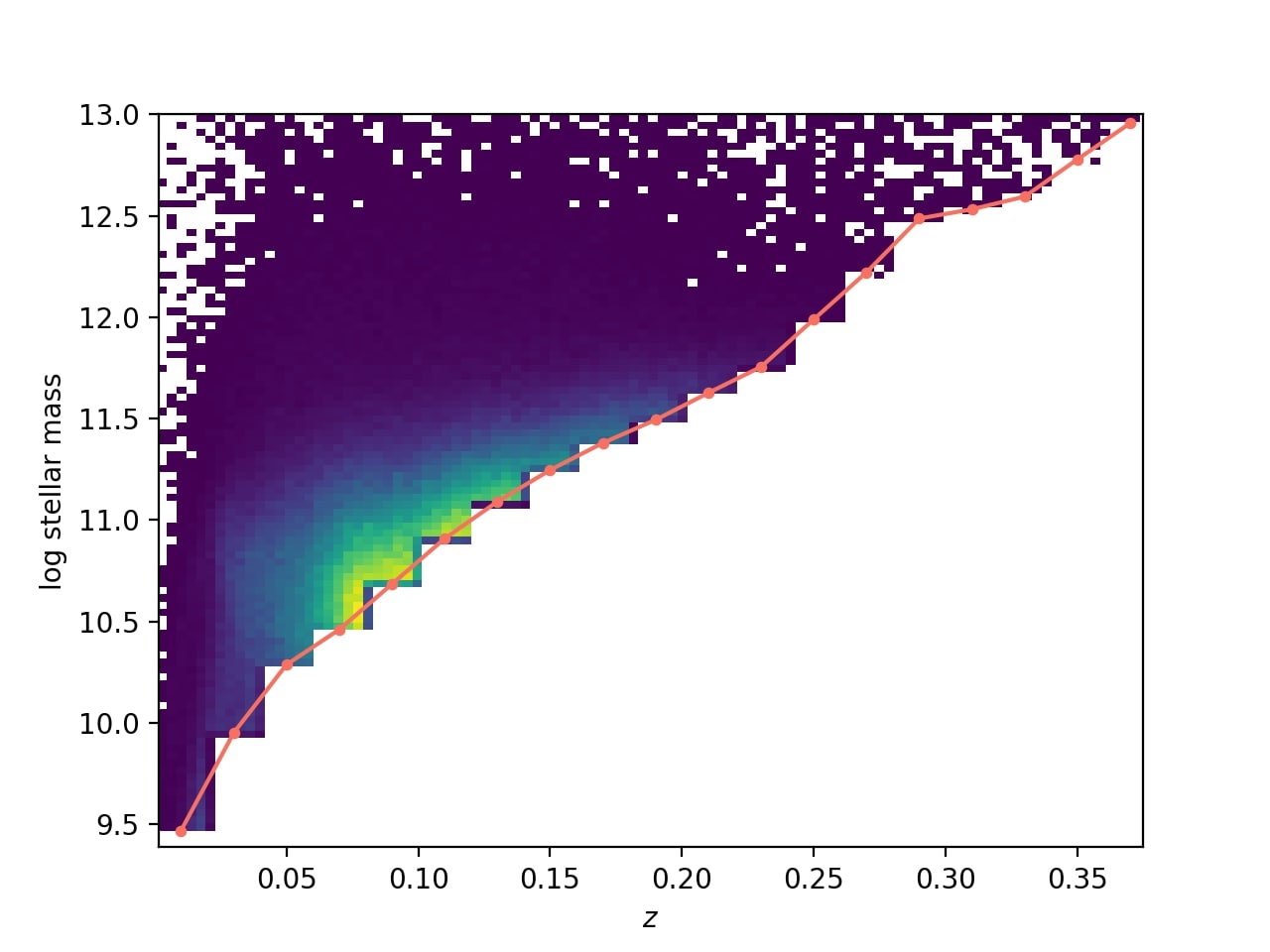}
\caption{Mass completeness as a function of redshift for redshift bins with spacing $\Delta z = 0.02$. For each redshift bin, we determine the 95\% completeness limit (pink line) and eliminate all galaxies below this point. For the distribution of masses at each redshift bin, see Appendix \ref{ap:mass}.}
\label{fig:mass_complete}
\end{figure}

Our final step is to eliminate all galaxies below the mass completeness limit at each redshift bin. We show this process in Figure \ref{fig:mass_complete}. This reduces our sample by roughly a factor of three from 958840 photometrically clean galaxies with measured masses to 362216 galaxies in a mass-complete sample. The factor of $\sim$3 reduction in sample size induced by the mass completeness correction is similar to the sample reduction in \citealt{Cebrian2014MNRAS.444..682C}, which applies a similar mass completeness correction to the NYU-VAGC catalog of SDSS DR7 galaxies.

\section{Results}
\label{results}

We present the classification results in \S \ref{results:class}, and provide a guide for interpreting the predictors that influence the classification in \S \ref{results:interpret}. We also provide a guide for deciding between merger stages and types in \S \ref{results:interpret_stage} and a guide for dealing with cases where by-eye classification and the LDA classification are in conflict in \S \ref{results:intuition}. 

We then analyze the properties of the merger sample in \S \ref{results:properties_of_sample} and compare our results to previous SDSS merger selections in \S \ref{results:compare}. 

We constrain the observed merger fraction using all of the different merger classifications in \S \ref{results:fraction} and explore how the major merger fraction varies as a function of galaxy mass and redshift in \S \ref{results:properties}. We explore if S/N or galaxy morphology (bulge-to-total mass ratio and color) are confounding the redshift-dependent major merger fraction in \S \ref{results:S_N} and \ref{results:morphology}, respectively. We explore how the minor merger fraction varies as a function of stellar mass and redshift in \S \ref{results:properties_minor}. We discuss the influence of contamination of the major and minor merger fraction calculations by mergers of the opposite type (minor and major, respectively) in \S \ref{results:double_count}. We run numerous sanity checks in \S \ref{results:sanity} (more details can be found in Appendix \ref{ap:sanity}) to confirm the main result of how the major merger fraction trends with mass and redshift. Finally, we end with a discussion of the importance of mass binning to our result in \S \ref{sanity:no_mass_bins}, where we find a different result in the absense of mass binning.

\begin{landscape}
    \begin{table}
    \centering
    \begin{tabular}{c|c c c c c c c c c c c }
         & & & & && & & & & low S/N  \\
         & & & & && & & & & / outlier predictor / \\
         ID	& LD1	&$p_{merg}$ & 	CDF & 	Leading term 1 & 	Leading coef 1 & Leading term 2 & Leading coef 2 & Leading term 3 & Leading coef 3 &  segmap\\	
         
         \hline
         
                  1237665179521187863 (A) & 	-4.137	&0.016	&0.510 &$A_s$ &	-11.3 & 	$A$ &	-3.8 & $C$ & 		-0.5 & 	0/0/0\\
                  
                  1237661852010283046 (B) &	4.781&	0.992& 0.919& $A_s$ &	31.8	& $A$ & 5.2	& $Gini$ &	0.9	& 0/0/0\\
                  
                  1237648720718463286 (C) & 2.081	& 0.889	&0.839& $A_s$ & 39.4 & $A$ & 	12.4 & $n*M_{20}$ & 0.5 & 0/0/0\\	

                  1237662306186428502 (D) & 4.235 &	0.986 &0.907 &	$A_s$ &	17.9	& $A$ &	2.2 & $n*M_{20}$ &		0.2	 & 0/0/0 \\
                  
                  1237653589018018166 (E)	&	1.784 &	0.856 & 0.830&	$A_s$ &	6.9	 & $A_s*A$ & 2.2	& $A*M_{20}$ &	0.2 &	0/0/0\\
                  
                  1237654383587492073 (F) & 0.678 &	0.663 &	0.792 & $A$ & 	16.0 & $A_s*C$ & 	9.1	& $A_s*Gini$ & 	2.4	 & 0/0/0 \\

    \end{tabular}
    \caption{Classification results for the six galaxies presented in Figure \ref{fig:distabc}. Here we provide the LD1 value and corresponding $p_{\mathrm{merg}}$ value for the major merger classification. We also list the leading (most influential) term in each classification and the contribution from this term, which is the product of the standardized predictor value and the LD1 coefficient for that term. We bold the classifications where a galaxy is classified as a merger ($p_{\mathrm{merg}} > 0.5$). The online-available tables provide these values for all six merger classifications (major, major pre-coalescence, major post-coalescence, minor, minor pre-coalescence, and minor post-coalescence).}
    \label{tab:classifications}
    \end{table}

\end{landscape}

\subsection{LDA classification results}
\label{results:class}

Here we present three data products: 
\begin{enumerate}
\item For each galaxy in the 1,344,577 DR16 sample, we provide all of the predictor values and the flag values. This table was previously described in \S \ref{data} and presented in Table \ref{tab:preds}.

\item For each merger classification, we provide the fiducial LD1, $p_{\mathrm{merg}}$, and CDF (described below) values for each galaxy in the 1,344,577 SDSS DR16 galaxy sample accompanied by explanatory information such as the most important (leading) terms in the classification and the coefficients associated with these leading terms. Our intent is that these tables can be used to ascertain why a galaxy is classified as a merging or non-merging galaxy according to the different fiducial classifications. We describe how this explanatory analysis might work in \S \ref{results:interpret}. Table \ref{tab:classifications} presents the major merger classification results for the six galaxies from Figure \ref{fig:distabc}.

\item We also provide a table (Table \ref{tab:marginalized_p}) that presents the 16th, 50th, and 84th percentile of the posterior $p_{\mathrm{merg}}$ distribution (and accompanying CDF value) for all photometrically clean galaxies (958,840) from the marginalization analysis described in \S \ref{methods:marginalizing}. This single table includes these results for all of the merger classifications. Using this table, the user can directly compare $p_{\mathrm{merg},50}$ values across different classifications. 
\end{enumerate}

\begin{figure*}
    \centering
    \includegraphics[scale=0.6, trim = 0.5cm 0cm 1.7cm 0cm, clip]{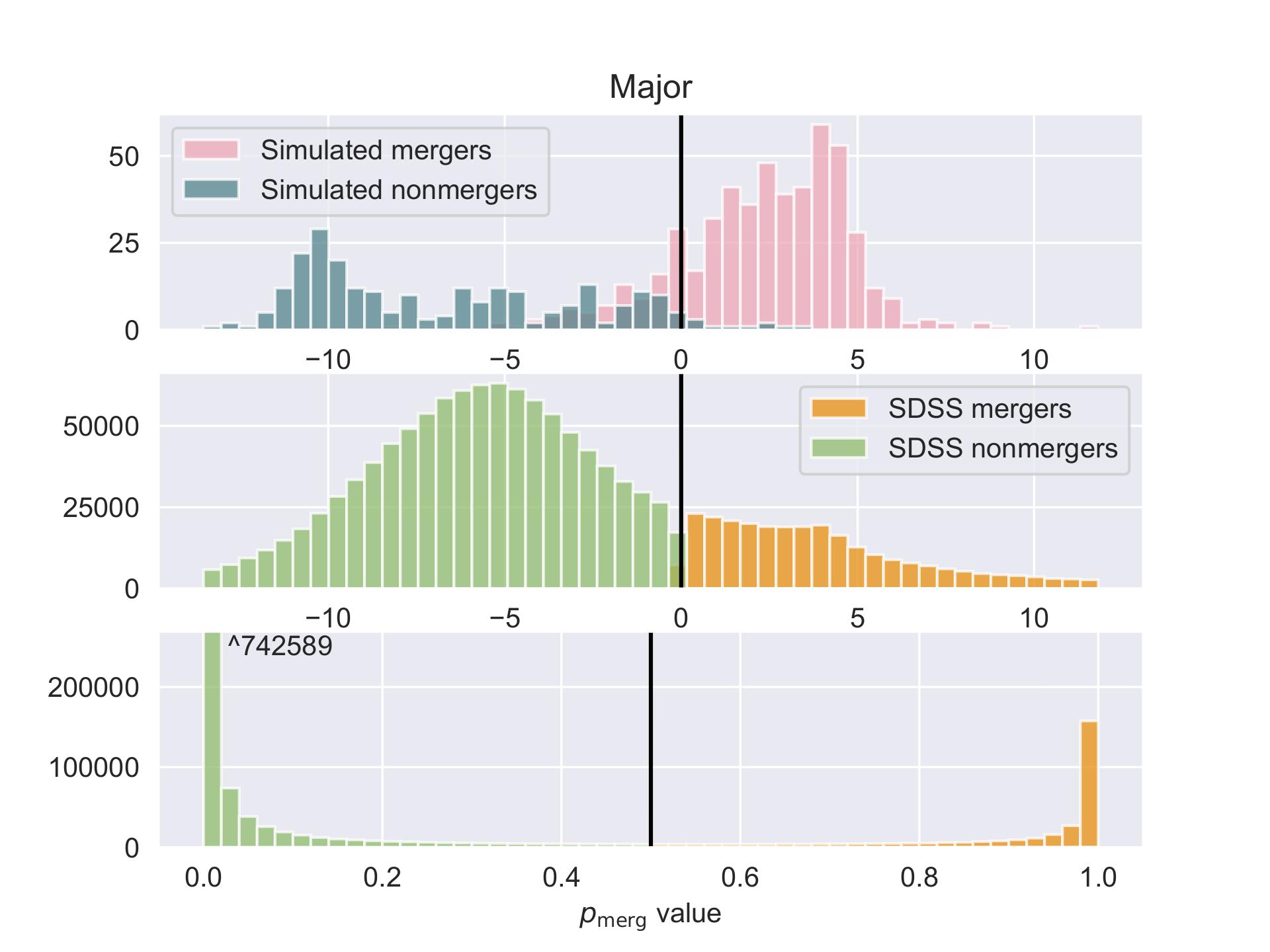}
    \includegraphics[scale=0.6, trim = 0.5cm 0cm 1.7cm 0cm, clip]{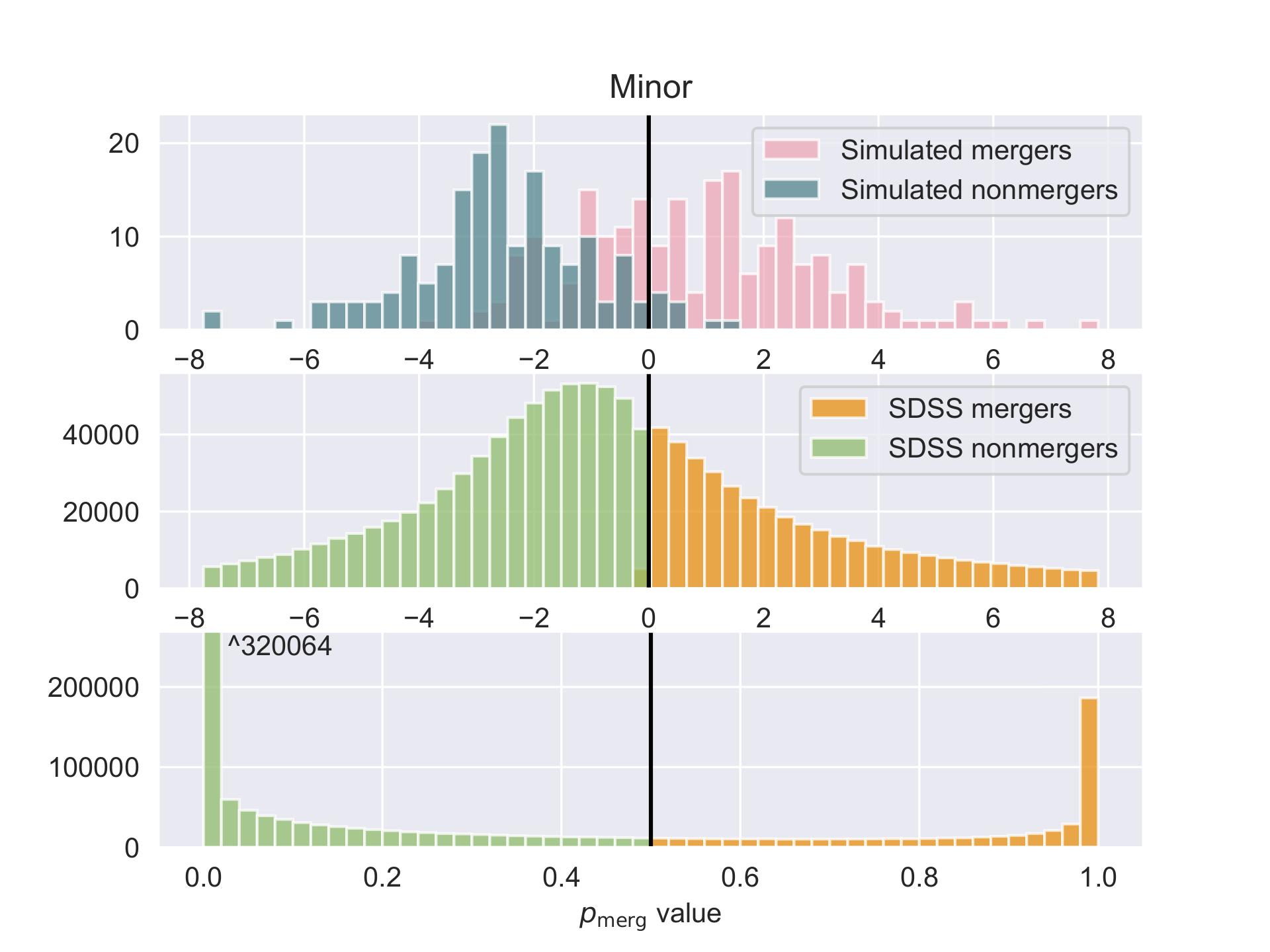}
    \caption{Distribution of LD1 values for the simulated suite (top panel) and the SDSS sample (middle panel) and the corresponding distribution of $p_{\mathrm{merg}}$ values for the SDSS sample (bottom panel) for the major (left) and minor (right) merger classifications. In all cases, the y-axis is the number of galaxies. In the bottom panels, we zoom in on the distributions and the inset numbers give the number of galaxies in the largest bin. }
    \label{fig:LDA_major}
\end{figure*}

In Figure \ref{fig:LDA_major}, we present histograms of the fiducial LD1 values and the corresponding $p_{\mathrm{merg}}$ values for the training set of simulated galaxies and the SDSS galaxies classified by the major and minor merger classifications. Since the LDA technique is designed to find the hyperplane of maximal separation between two populations, the distribution of probability values in the bottom panels of Figure \ref{fig:LDA_major} peak very near to 0 and 1. This makes direct interpretation of these probability values very difficult. We therefore provide a complementary cumulative distribution function (CDF) analysis (which is part of data products 2 and 3) to compare individual $p_{\mathrm{merg}}$ values to the the $p_{\mathrm{merg}}$ values of all SDSS galaxies for a given classification. For instance, if we examine the major merger classifications in Table \ref{tab:classifications}, galaxy A has a $p_{\mathrm{merg}}$ value of 0.016, which corresponds to an CDF value of 0.510, meaning that 51\% of galaxies in SDSS have a lower $p_{\mathrm{merg}}$ value. In Table \ref{tab:thresholds}, we list the $p_{\mathrm{merg}}$ values that correspond to the 5\%, 10\%, 90\%, and 95\% values of the CDF for the fiducial merger classifications.

\begin{table*}
    \centering
    \begin{tabular}{c|c c c c }
        & & \multicolumn{3}{c}{$p_{\mathrm{merg,16}}$/$p_{\mathrm{merg,50}}$/$p_{\mathrm{merg,84}}$ (CDF) }\\
        ID& Type& All & Pre-coalescence & Post-coalescence (1.0) \\

         \hline
         
                1237648720718463286 & Major & 0.84/0.99/1.0 (0.96) &0.67/0.88/0.99 (0.85) &0.0/1.0/1.0 (0.99) \\ 
 & Minor & 0.0/0.12/1.0 (0.51) &0.0/0.04/0.84 (0.44) &0.46/1.0/1.0 (0.98) \\ 
1237653589018018166 & Major & 0.79/0.88/0.92 (0.89) &0.81/0.89/0.94 (0.85) &0.88/0.97/0.99 (0.83) \\ 
 & Minor & 0.88/0.95/0.98 (0.88) &0.88/0.96/0.99 (0.87) &0.74/0.93/1.0 (0.74) \\ 
1237654383587492073 & Major & 0.52/1.0/1.0 (0.98) &0.96/1.0/1.0 (0.97) &0.0/0.0/0.0 (0.0) \\ 
 & Minor & 0.0/0.0/0.91 (0.18) &0.0/0.0/1.0 (0.17) &0.1/0.89/1.0 (0.7) \\ 
1237661852010283046 & Major & 0.93/0.98/0.99 (0.94) &0.99/1.0/1.0 (0.94) &0.02/1.0/1.0 (0.92) \\ 
 & Minor & 0.04/0.89/1.0 (0.85) &0.33/0.98/1.0 (0.89) &0.19/1.0/1.0 (1.0) \\ 
1237662306186428502 & Major & 0.99/1.0/1.0 (0.98) &0.99/1.0/1.0 (0.95) &0.98/1.0/1.0 (0.93) \\ 
 & Minor & 0.78/0.98/1.0 (0.91) &0.71/0.99/1.0 (0.91) &0.63/0.98/1.0 (0.84) \\ 
1237665179521187863 & Major & 0.03/0.09/0.17 (0.56) &0.01/0.02/0.07 (0.51) &0.29/0.63/0.76 (0.68) \\ 
 & Minor & 0.13/0.36/0.56 (0.67) &0.18/0.37/0.57 (0.68) &0.17/0.46/0.62 (0.55) \\

    \end{tabular}
    \caption{Marginalized $p_{\mathrm{merg}}$ values and accompanying CDF values for the six galaxies from Figure \ref{fig:distabc}. We list the $p_{\mathrm{merg}}$ corresponding to the 16th, 50th (median), and 84th percentiles of the marginalized posterior $p_{\mathrm{merg}}$ distributions for each galaxy using each classification. We also list the CDF value that corresponds to the 50th percentile in parenthesis. For each galaxy, we list only the combined minor/major merger classifications and the pre- and post-coalescence (1.0) results. In the online-available table, we also include the early, late, and post-coalescence (0.5) results. In the online-available table, each of the 16/50/84 percentile values is its own column. }
    \label{tab:marginalized_p}
\end{table*}

\begin{table*}
    \centering
    \begin{tabular}{c|ccccccccc}
         & \multicolumn{9}{c}{CDF threshold value}\\
      Classification  & 0.01 & 0.05& 0.1&0.25 & 0.5 & 0.75 &  0.9 & 0.95 & 0.99\\
         \hline

    Major merger & 3.2e-8 & 1.6e-7 & 3.2e-7 &7.4e-7 & 0.01 & 0.39&  0.9891260 & 0.999999353 & 0.9999998720\\
    Minor merger & 5.6e-8 & 2.8e-7 & 5.7e-7 & 0.02 & 0.24 & 0.79 & 0.996 & 0.99999950 & 0.99999989\\

    \end{tabular}
    \caption{Probability of merging  ($p_{\mathrm{merg}}$) that correspond to different thresholds of the CDF for all of the merger classifications. This table is provided to enable user interpretation of individual $p_{\mathrm{merg}}$ values, which evolve exponentially and their interpretation can be assisted with careful consideration of the CDF values. For instance, if a galaxy has a $p_{\mathrm{merg}}$ value of 0.01 for the major merger classification, this corresponds to a CDF value of 0.5, meaning that about half of our SDSS sample is more likely to be a non-merger. 
    }
    \label{tab:thresholds}
\end{table*}

\begin{figure}
    \centering
    \includegraphics[scale=0.5]{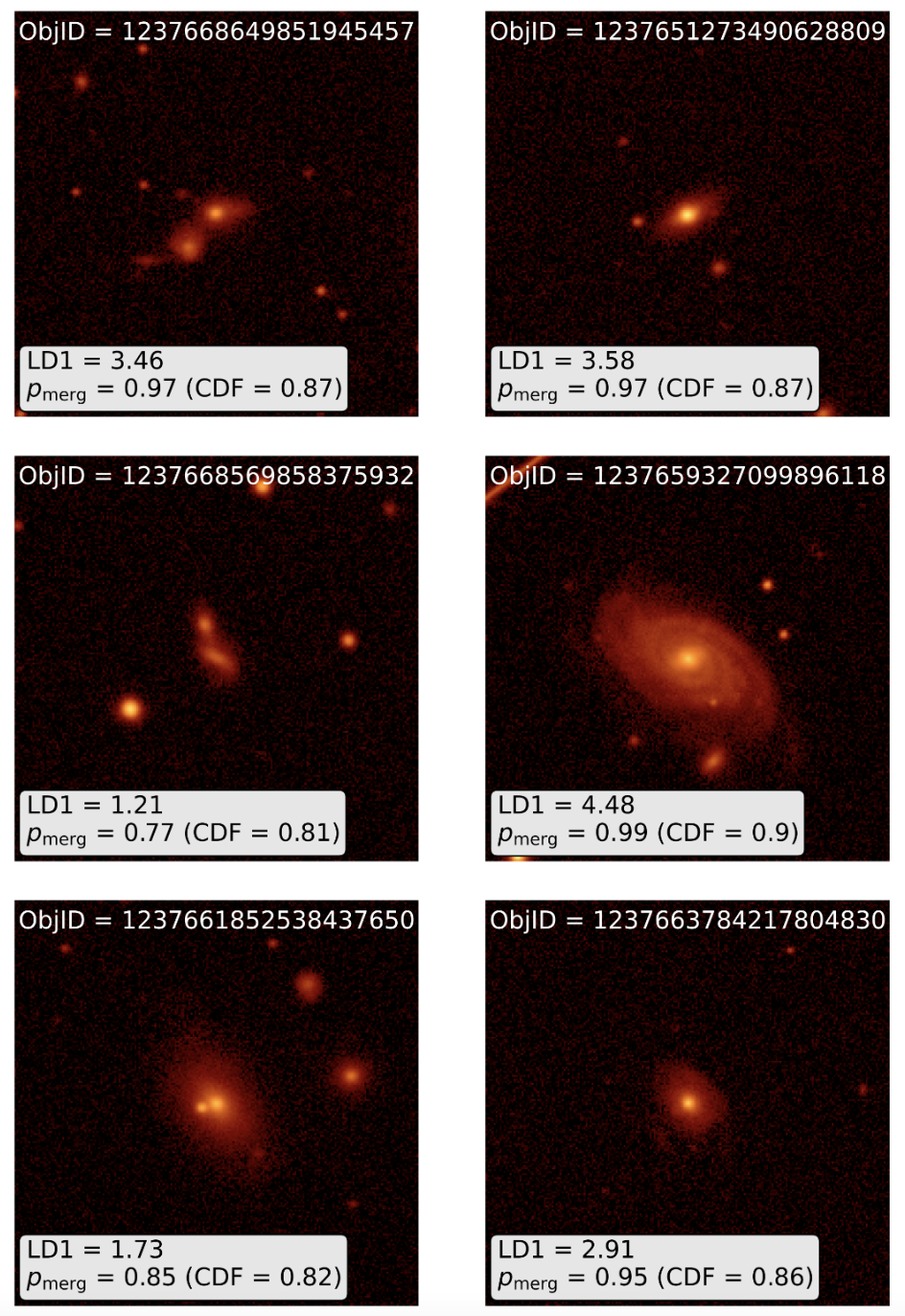}
    \caption{Merging galaxies ($p_{\mathrm{merg}} > 0.5$) according to the major merger LDA technique. The inset panels list the LD1 value and its accompanying $p_{\mathrm{merg}}$ value and CDF value. All panels are $80\farcs0\times80\farcs0$.}
    \label{fig:mergers}
\end{figure}

\begin{figure}
    \centering
    \includegraphics[scale=0.5]{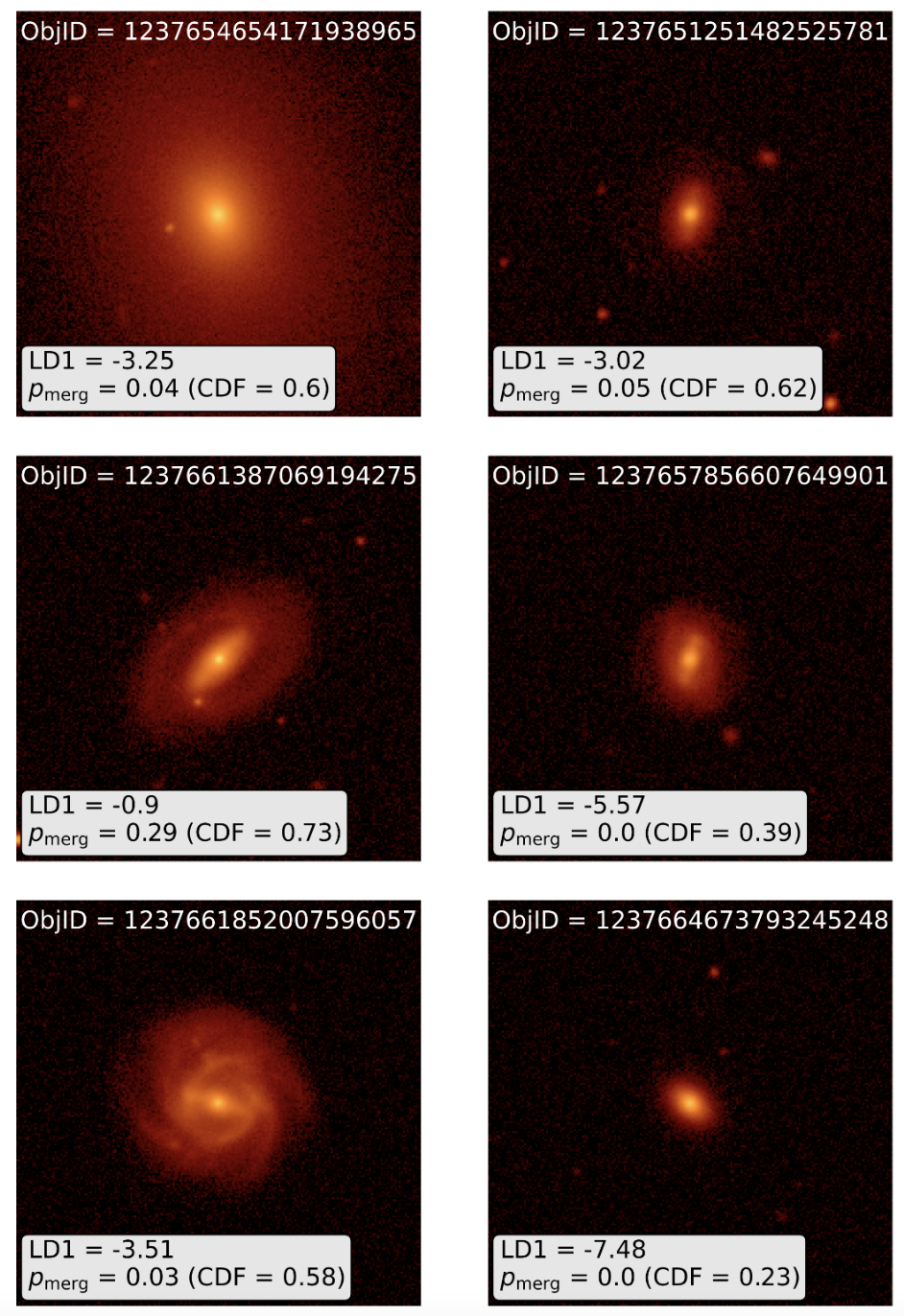}
    \caption{Non-merging galaxies ($p_{\mathrm{merg}} < 0.5$) according to the major merger LDA technique. The inset panels are described in Figure \ref{fig:mergers}.}
    \label{fig:nonmergers}
\end{figure}

Finally, we provide visual examples of a randomly selected sample of merging galaxies (Figure \ref{fig:mergers}) and non-merging galaxies (Figure \ref{fig:nonmergers}) according to the fiducial major merger LDA classification.

\subsection{A guide for interpreting classification results}
\label{results:interpret}

The LDA classification method was designed with the interpretability of individual results as one of its central goals. In this section, we discuss how to use the additive linear terms that compose LD1 to understand why a galaxy is classified as merging or non-merging. To assist users with this interpretation, we provide Table \ref{tab:classifications}, which lists the $p_{\mathrm{merg}}$ and CDF values for the major merger classification for individual galaxies alongside the most influential predictors and coefficients.

We include an utility within \texttt{MergerMonger} that calculates CDF values for $p_{\mathrm{merg}}$ values and vice versa. This is useful if the user wants to create a `superclean' merger sample that has minimal non-merger contamination. They could either do this by defining an CDF threshold or by deciding on a $p_{\mathrm{merg}}$ threshold (i.e. $p_{\mathrm{merg}} > 0.95$) and using Table \ref{tab:thresholds} or the \texttt{MergerMonger} utility to determine the corresponding $p_{\mathrm{merg}}$ or CDF value. It is then possible to re-run the LDA classifications using \texttt{MergerMonger} and a different $p_{\mathrm{merg}}$ value as the threshold to identify mergers.

\begin{figure}
    \centering
    \includegraphics[scale = 0.4]{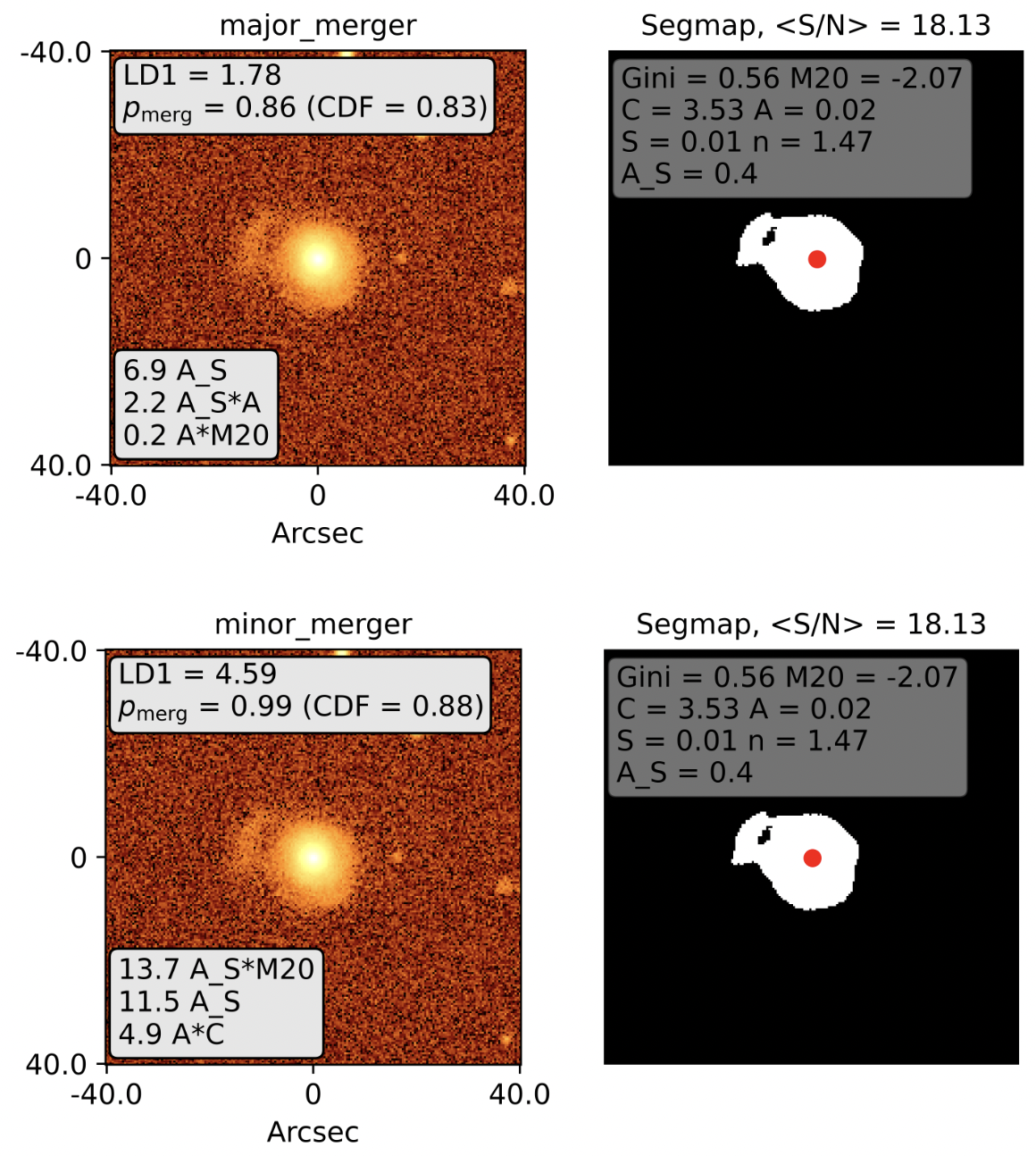}
    
    \caption{Diagnostic classification diagrams for the major (top) and minor (bottom) merger classification results for galaxy E. The three most influential terms and their contribution to LD1 are given in the bottom left panels. We describe how to interpret these leading terms in the text.}
    \label{fig:1237653589018018166}
\end{figure}

We also provide a diagnostic tool within \texttt{MergerMonger} (\texttt{find\_galaxy.py}) that accepts single or multiple galaxy SDSS Object ID(s) as input. This utility then presents the predictor values, the most influential predictors in the classification, and the classification results in a diagnostic diagram that includes the individual galaxy image and segmentation map. We show an example of two diagnostic diagrams in Figure \ref{fig:1237653589018018166} for the major (top) and minor (bottom) merger classifications for galaxy F from Figure \ref{fig:distabc}.

This galaxy is classified as a merger by both major and minor merger fiducial classifications, with high LD1 and corresponding $p_{\mathrm{merg}}$ values in the upper left informational panel. The lower panel on the left hand image lists the three leading terms and their corresponding contribution to the value of LD1; here, shape asymmetry and asymmetry are important predictors for both classifications. The inset informational panel for the right hand segmentation maps lists all of the pre-standardized predictor values.

These diagnostic diagrams can help the user interpret why the classifications have determined that this galaxy is likely to be a merger. Looking first at the major merger panels, shape asymmetry followed by the $A_s*A$ cross term are the most influential terms. In the right panel, the asymmetry for this galaxy is low while the shape asymmetry is high. This is due to the low surface brightness of the shell feature. Since the coefficient of the $A_s$ term is positive in Equation \ref{eq:major}, this boosts the LD1 score. The coefficient of the $A_s*A$ term is negative in Equation \ref{eq:major}. This coefficient will be multiplied by the standardized $A_s$ and $A$ values, which will be positive and negative respectively (recall, the $A$ value is relatively low). The net result will be a positive contribution to LD1, meaning that this galaxy is even more likely to be detected as a merger. In this case, the $A_s*A$ term allows the LDA to better distinguish between asymmetric bright features such as spiral arms and low surface brightness asymmetric features that are more likely to be caused by a merger.

For the minor merger classification, the $M_{20}*A_s$ cross term is influential; this term has a negative coefficient in Equation \ref{eq:minor}, so for this term to have a large positive influence, either the standardized value of $M_{20}$ or $A_s$ must be very negative (meaning relatively low for SDSS galaxies). Here, this is because $M_{20}$ is quite negative, meaning that the light is concentrated. By eye, this galaxy looks like a post-coalescence merger with a shell from the merger event; the minor merger technique is relying both upon the high concentration (also measured by $M_{20}$) and the shell feature to identify it as a merger. This galaxy and others like it demonstrate that the LDA classification succeeds in the case of concentrated early-type galaxies.

\subsection{A guide for distinguishing between merger types and stages}
\label{results:interpret_stage}

Here we discuss the overlap between different merger stages and types and how to directly compare $p_{\mathrm{merg}}$ values across different classifications. Directly comparing $p_{\mathrm{merg}}$ values between the fiducial runs is not encouraged, especially between minor and major classifications. These different classifications were prepared assuming different priors, meaning that the distribution of $p_{\mathrm{merg}}$ values will be affected by this choice. We also do not recommend directly comparing the $p_{\mathrm{merg}}$ values from Table \ref{tab:classifications} between different stages of the same merger type (i.e. early versus late stage major mergers) because these tables assume the same fiducial merger prior. As we will show in \S \ref{sanity:classifications}, this is not a safe assumption. 

\textit{Best practice is therefore to use the marginalized $p_{\mathrm{merg}}$ values from Table \ref{tab:marginalized_p} to decide which stage or which merger type is most likely for a given galaxy.} This table includes the $p_{\mathrm{merg}}$ values that corresponds to the 16th, 50th , and 84th percentile of the posterior distribution of $p_{\mathrm{merg}}$ for each galaxy for the major, minor, and pre- and post-coalescence (1.0 Gyr) stages. The online-available table also includes the early, late and post-coalescence (0.5) stage results.

Here we walk the user through the process of distinguishing between merger types and stages using Table \ref{tab:marginalized_p} and the (\texttt{compare\_classifications.py}) utility within \texttt{MergerMonger}, which plots an image of a galaxy and compares the $p_{\mathrm{merg}}$ values between different classifications.

\begin{figure*}
\includegraphics[scale=0.07]{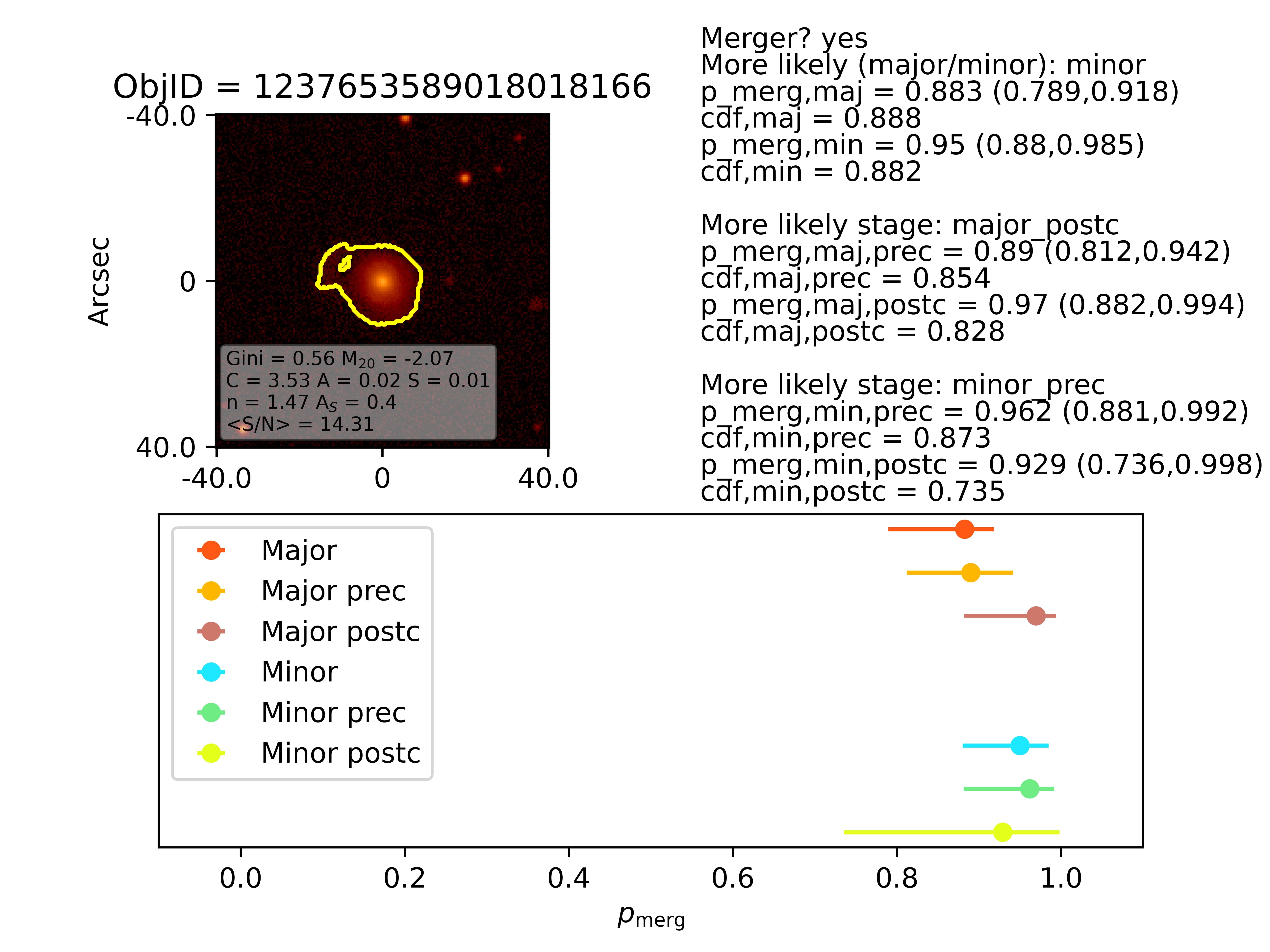}
\caption{Diagnostic diagram for determining the most likely merger type (major or minor) and the most likely merger stages for galaxy E from Figure \ref{fig:distabc}. This diagram is produced by the \texttt{compare\_classifications.py} utility. The top left panel shows the galaxy, segmentation map (yellow), and imaging predictor values. The top right panel runs through a diagnosis of merger type, beginning with diagnosing whether $p_{\mathrm{merg,maj,50}}$ or $p_{\mathrm{merg,min,50}}$ are greater than 0.5. If so, then the galaxy is identified as a merger. The next step is to identify which is more likely (a major or minor merger), which is determined using the $p_{\mathrm{merg,50}}$ values from each classification. We provide the $p_{\mathrm{merg,50}}$ values for all classifications along with the $p_{\mathrm{merg,16}}$ and $p_{\mathrm{merg,84}}$  values in the following format: $p_{\mathrm{merg,50}}$ ($p_{\mathrm{merg,16}}$,$p_{\mathrm{merg,84}}$). Finally, this diagnostic diagram decides which stage is more likely for the major followed by the minor merger classifications. Here, the post-coalescence stage is more likely for the major merger and the pre-coalescence stage is more likely for the minor merger classification. In the bottom panel, the y-axis is used to order the classification results, where different colors correspond to the median, $p_{\mathrm{merg,50}}$ values for each classification and the error bars give the the range between the 16th and 84th percentile of the $p_{\mathrm{merg}}$ value for each classification.}
\label{fig:diagnostic}
\end{figure*}

\begin{sloppypar}
Using galaxy E from Figure \ref{fig:distabc} and Table \ref{tab:marginalized_p}, we show a diagnostic diagram in Figure \ref{fig:diagnostic} created using \texttt{compare\_classifications.py} as an informative example for how to decide between merger type and stage for an individual galaxy. The \texttt{compare\_classifications.py} utility decides the most likely classifications in a hierarchical manner; first, it determines if the galaxy is more likely to be a major or minor merger by directly comparing the $p_{\mathrm{merg,50}}$ values from each classification. The utility then decides whether the galaxy is more likely to be a pre-coalescence merger or a post-coalescence (1.0 Gyr) merger. It does this for both the major and minor classifications. All of these rankings occur regardless of if the $p_{\mathrm{merg,50}}$ values are greater than 0.5.

\end{sloppypar}

For galaxy E using the major and minor merger $p_{\mathrm{merg,50}}$ values, we are able to conclude that it is more likely a minor merger. We can then further distinguish between the minor merger stages, finding that it is more likely a pre-coalescence minor merger. 

In general, we recommend following the hierarchical framework of \texttt{compare\_classifications.py}; first decide between the all-inclusive major and minor merger classifications and then decide between the sub-stages of each. We also recommend using the post-coalescence (1.0) classification as opposed to the post-coalescence (0.5) classification, which has lower performance statistics due to its short observability timescale. If the use case is to identify all early-stage major and minor mergers, then we recommend creating a new process using the code framework of \texttt{compare\_classifications.py} that requires that $p_{\mathrm{merg,50}}$ from the early stage classifications is greater than the $p_{\mathrm{merg,50}}$ values corresponding to the late and post-coalescence (1.0) stage classifications. In this case, we recommend comparing the stages of the major/minor merger classification directly to one another (i.e. major merger early is compared to major merger late and post-coalescence). 

We also provide the 16th and 84th percentile values if the user wants to develop a more conservative sample, i.e. requiring that  $p_{\mathrm{merg,major,16}} > p_{\mathrm{merg,minor,84}}$ would be a more conservative way to compare the classifications. However, there is significant overlap between different classification samples when using the full range (16th and 84th percentiles), so we recommend using the 50th percentile (median) values for simplicity. For instance, in Figure \ref{fig:diagnostic}, if we were to use the more conservative technique, all of the classifications and stage-specific classifications would overlap.

Note that there is overlap between stages and/or merger types, i.e. there will be many galaxies that have $p_{\mathrm{merg,50}}$ values that are greater than 0.5 for multiple different classifications. We discuss this overlap in more detail in \S \ref{results:fraction}, where we measure the merger fraction.

\subsection{Interpreting cases where the LDA classification disagrees with by-eye classification}
\label{results:intuition}

We acknowledge that as with any merger identification approach that relies on imaging predictors, the individual classifications may disagree with by-eye decisions. We therefore recommend that if the user is working with a relatively small sample they also examine the classifications by eye to identify potential misclassifications. 

A failure mode of the LDA major merger combined classification, for instance, is classifying equal mass major mergers that happen to be in a symmetric configuration as non-merging. This happens when the merging galaxies also have a low overall concentration. This is relatively rare and can be understood by running the interpretive \texttt{MergerMonger} utilities which reveals which predictors are responsible for the non-merger classification. We also recommend running galaxies with surprising classifications through the \texttt{compare\_classification.py} utility in order to examine the results of the different merger stage classifications. Obvious early or late stage equal mass major mergers might be classified as having a low probability of being a major merger by the overall classification (due to the unlikely combination of imaging predictor values) often have a high probability of being a major merger in the pre-coalescence stage. 

The combined LDA classifications (major and minor) are trained from an ensemble of images, meaning that they are optimized for high accuracy for all stages of the merger. The implication is that the combined classifications are best for determining bulk sample properties such as the overall merger fraction, while the individual stage classifications may be better suited for understanding smaller samples of mergers or for cases where determining the merger stage is important.

\subsection{Properties of the LDA mergers}
\label{results:properties_of_sample}

The challenge we face in validating our sample of merging galaxies is that there is no gold standard to rely upon for which galaxies are truly mergers. We therefore take the approach of checking for large systematic issues by investigating the global properties of the merger samples. We carry out this analysis in two parts: first, here we compare the properties of the (mass-complete) parent sample to those of the merger samples. Second, in \S \ref{results:compare}, we will compare the properties of the merger samples to those of other merger selection techniques.

\begin{figure*}
    \centering
    \includegraphics[scale=0.2, trim = 5cm 2cm 5cm 3cm, clip]{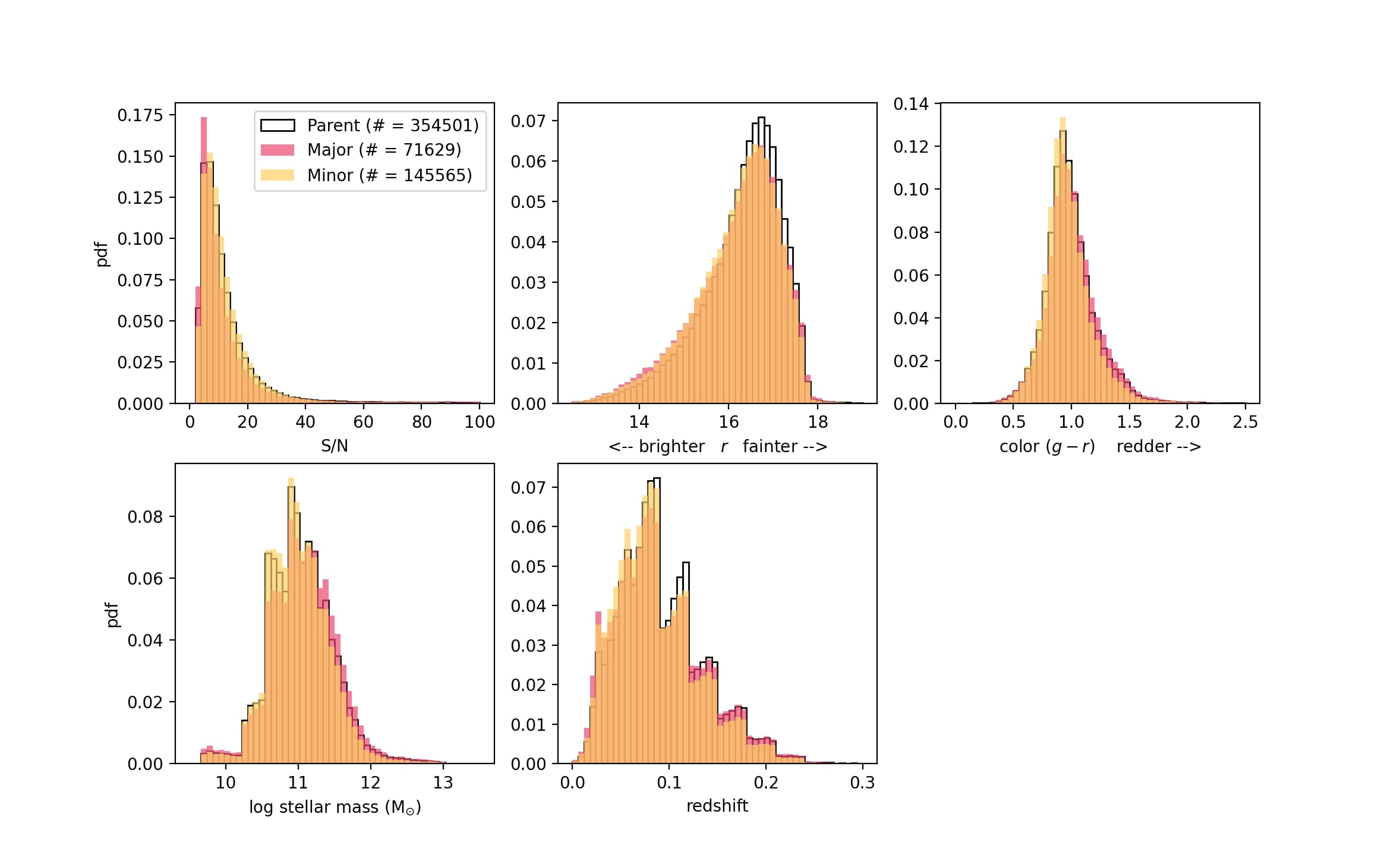}
    \caption{Probability density functions (pdfs) of the properties of the parent sample of SDSS gal
    axies (white) compared to properties of the $p_{\mathrm{merg}} > 0.5$ major merger sample (pink) and the minor merger sample (yellow). All histograms are normalized so that all bins sum to a value of one. Left to right: the distributions for average S/N ratio, $r-$band magnitude, color ($g-r$), log stellar mass, and redshift. Using the two-sample KS test, we confirm that the all distributions are statistically indistinguishable. }
    \label{fig:properties_merger_parent}
\end{figure*}

In Figure \ref{fig:properties_merger_parent} we compare the probability density functions (pdfs) for the major (pink) and minor (yellow) merger samples to that of the parent SDSS sample (white) using average S/N, $r-$band magnitude, color ($g-r$), stellar mass, and redshift. The pdfs are normalized so that all bins from a given distribution sum to a value of one. 

The mergers have properties that span the full range of properties of the parent distribution. This is a major success when we consider that our training sample of galaxies was limited in these spaces. For instance, the training set of galaxies spanned $3.9 - 4.7\times10^{10} M_{\odot}$ in stellar mass and all galaxies in the training set had their surface brightnesses and apparent sizes adjusted to a redshift of $z = 0.03$. The fact that the LDA techniques identify mergers over a large range in surface brightness, stellar mass, and redshift indicates that the LDA method is successfully able to adjust to a wider span of galaxy properties. 

Furthermore, we run two-sample Kolmogorov-Smirnov (KS) tests to compare the cumulative distribution functions (constructed from the pdfs) for each property and find that the distributions are statistically indistinguishable. Specifically, we are unable to reject the KS null hypothesis (that the distributions are identical) when we compare the parent distribution to the major and minor merger selection and when we compare the major and minor merger distributions. The implication is that while the major and minor merger classifications are using different imaging properties to identify mergers, they are not significantly biased in any of these properties.

This is a massive success of the method; previous studies have uncovered significant biases, especially related to S/N. For instance, \citet{Bickley2021} train a Convolutional Neural Network (CNN) to identify post-merger galaxies in Illustris TNG100. When they test the performance on galaxies in the Canada-France Imaging Survey, they find a deficit of very faint galaxies in the post-merger sample. Their merger technique is slightly biased towards identify more massive, brighter, and higher redshift galaxies (due to the volume-limited nature of the survey, more massive galaxies are more likely to appear at higher redshift).

Despite the KS test revealing that the distributions are statistically indistinguishable, we do notice some slight by-eye differences. The major and minor classifications have slight excesses at low (brighter) $r-$band magnitudes compared to the parent distribution. To quantify this, we measure the offset in the median value of each major/minor distribution compared to the parent distribution and find values of $\Delta r = 0.11/0.12$, where the major and minor merger distributions are slightly brighter than the parent distribution. The distributions also differ at low redshift, where the major and minor merger distributions tend towards lower redshift values ($\Delta z = 0.002/0.005$). In terms of mass, the major mergers tend to have higher masses ($\Delta log M_* (M_{\odot}) = 0.06$). 

The brighter major mergers constitute two populations; one is more massive and at higher redshift, while one is less massive and at lower redshift. Both of these populations have slightly lower S/N ratios than the parent sample. These properties could reflect a slight bias for the merger classifications to identify galaxies with lower S/N ratios as mergers, which is the opposite bias as that identified in work such as \citet{Bickley2021}. We investigate this potential bias in more depth in \S \ref{results:S_N}, where we show that the merger fraction does increase with decreasing S/N when we control for mass and redshift. However, we also show in this section that this trend does not change our finding of a decreasing merger fraction with increasing redshift.

\subsection{Properties of LDA mergers compared to previous merger samples in SDSS}
\label{results:compare}

In order to better understand the biases of our technique, we compare the mergers selected using the LDA major merger classification with those from two large SDSS merger samples: GalaxyZoo and the \citet{Ackermann2018MNRAS.479..415A} technique (from here on, A18).

First, we compare our SDSS merger catalog to the GalaxyZoo selection of mergers in SDSS imaging, which is a large publicly-available catalogs of mergers (\citealt{Lintott2008,Lintott2011}). We cross-match the GalaxyZoo catalog from DR8 (893,163 galaxies) with our clean DR16 sample and find 570,455 matches. The GalaxyZoo catalog provides $p$, or probability values, for four morphological categories (mergers, ellipticals, combined spirals, and `don't know'), corresponding to the percentage of users that selected each morphological category. We identify the morphological category with the highest $p$ value for each galaxy. We then identify the number of galaxies in each category that have a fiducial major merger probability greater than 0.5 from our classification. We use the major merger classifications from our technique for comparison because the GalaxyZoo classifications are based on visual inspection, which is more likely to identify the more obvious major mergers.

The results are as follows: for the GalaxyZoo merger category, 6626/10433 (64\%) are LDA major mergers, for the combined spirals, 25467/176213 (14\%) are LDA major mergers, for the ellipticals, 54431/378993 (14\%), and for the ambiguous category, 1413/4816 (29\%). We also build a `clean' sample of GalaxyZoo mergers, where the fraction of users that classify galaxies as mergers is greater than 95\%. Of these, 30/34 (88\%) are classified as mergers by our classification.

These results are reassuring in two ways: first, the LDA classification returns $\sim$2/3 of mergers identified in GalaxyZoo, and second, the fraction of spirals and ellipticals that are identified as mergers by the LDA method are not significantly different. This tells us that the LDA method is not significantly biased as a function of galaxy morphology.

We visually inspect mergers according to GalaxyZoo that we classify as nonmergers and find that many of them can be described as double nuclei galaxies without noticeable tidal tails. Some of these galaxies may be nonmergers that are superimposed along the line of sight and some of them may be very early stage mergers (approaching for a first encounter) or gas poor mergers. For these galaxies, the most important major merger predictors, such as shape asymmetry, have low values, resulting in a non-merger identification from the LDA technique. We discuss this particular failure mode of the LDA classification in \S \ref{results:intuition}.

We next compare the properties of mergers from our classification technique to those identified in the SDSS sample using the A18 technique, which uses transfer learning to retrain a convolutional neural network (CNN) on the \citet{Darg2010} sample of merging galaxies (from GalaxyZoo). A18 use the 3003 merger objects from \citet{Darg2010} as merger examples ($0.005 < z < 0.1$) and 10,000 GalaxyZoo galaxies with $f_m < 0.2$ as examples of nonmergers, where $f_m$ is the fraction of users who identify a galaxy as a merger. We cross-match the results from the A18 catalog, which is mass complete down to $10^{10} M_{\odot}$, with those of our mass-complete LDA classifier (we calculate completeness as a function of redshift, Figure \ref{fig:mass_complete}), and find an overlap of 98,645 galaxies. From these, we use the same method as A18 to identify galaxies with an average $p_m$ value above 0.95 as merging, where $p_m$ is the output of the CNN classifier.

\begin{figure*}
    \centering
   \includegraphics[scale=0.04, trim = 9cm 8cm 9cm 11cm]{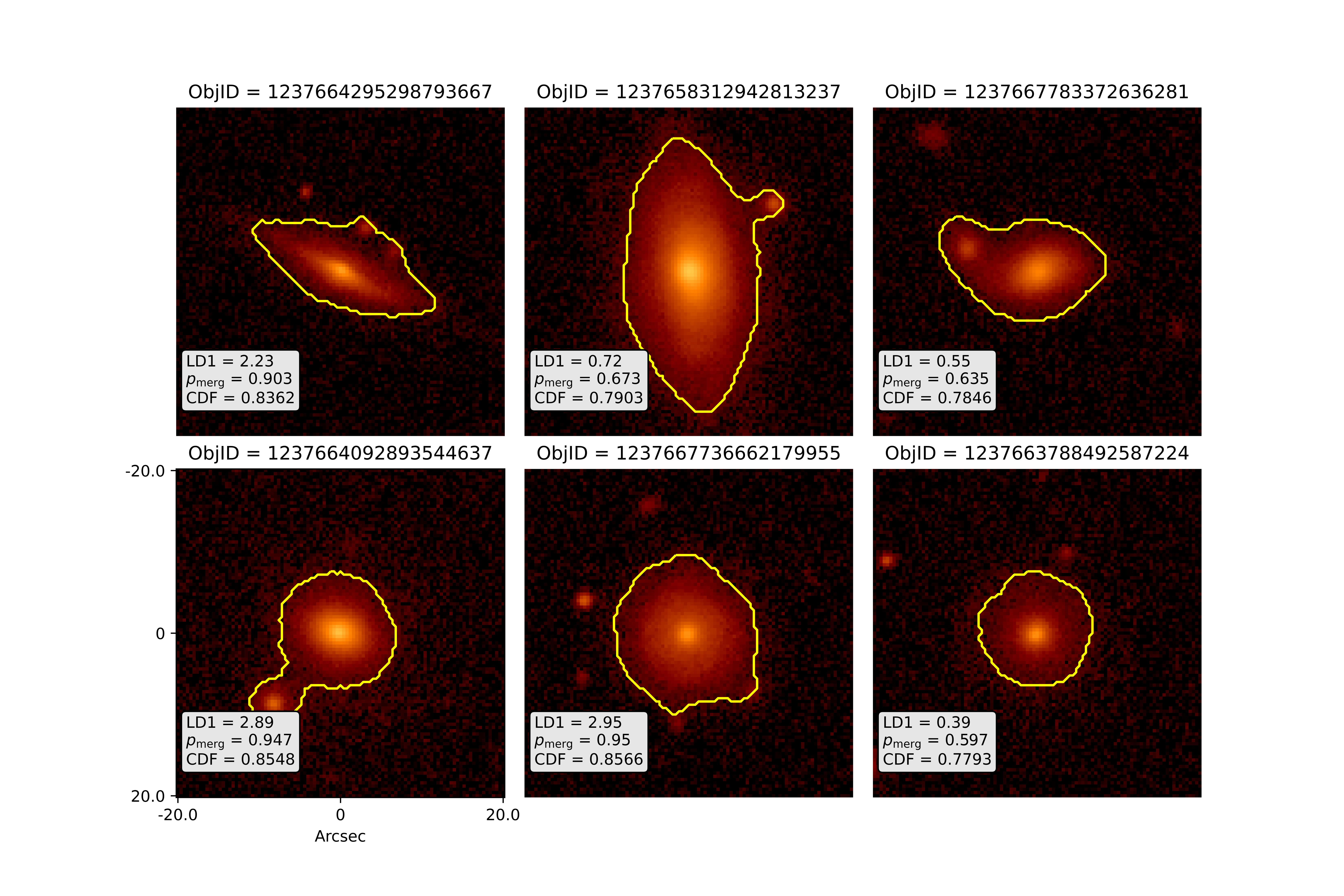}
    \caption{Galaxies classified by the LDA classification as major mergers that are classified by the A18 major merger classification as nonmerging. The yellow line marks the edge of the segmentation mask and the inset panels provide the LD1, $p_{\mathrm{merg}}$ and CDF values for each galaxy. The LDA technique identifies a large fraction of SDSS galaxies as mergers that the A18 technique does not; this is the case even when the A18 threshold is adjusted to a lower value. }
    \label{fig:ack_fps}
\end{figure*}

We first compare the overlap of the merger samples. When we measure the performance statistics of our merger sample relative to the A18 classifications (assuming the A18 classifications to be correct), we find an accuracy of 0.85, a precision of 0.11, a recall of 0.78, and an F1 score of 0.20. The precision is low because there are a large number of galaxies that we identify as mergers that A18 does not.

We present a few examples of galaxies that we classify as major mergers that A18 does not in Figure \ref{fig:ack_fps}. Using visual inspection, one of the galaxies in this example (top right) looks like a faint major merger, three appear to be minor mergers (top left\footnote{This merger and others like it could be chance projections along the line of sight. We discuss this caveat of the method in more detail in \S \ref{discuss:caveats}.}, top middle, and bottom left), and two appear to be post-merger remnants (bottom middle and bottom right). 

\textit{Figure \ref{fig:ack_fps} demonstrates something fundamental about the differences between techniques that are trained using visually-identified samples and the LDA technique presented here; techniques trained using mergers identified by eye are biased towards identifying major mergers in the early or late stages. The LDA technique on the other hand will identify a greater variety of merger stages (including the post-coalescence stage, see the result of longer merger observability timescale from N19). }

\begin{figure}
    \centering
    \includegraphics[scale=0.2, trim = 1cm 0cm 2cm 2.5cm, clip]{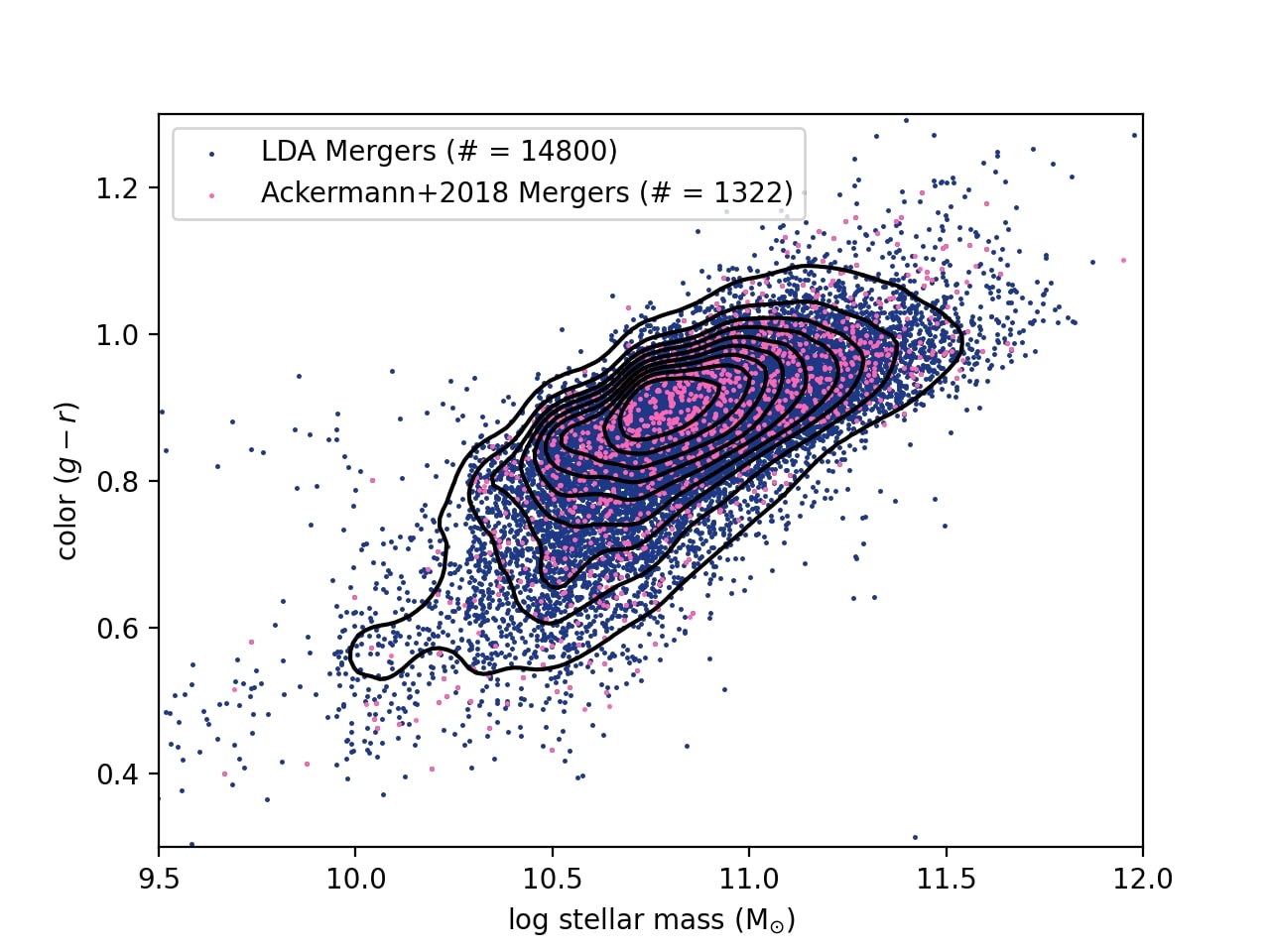}
    \caption{Color-mass diagram. We cross-match the merger catalog from A18 with the LDA catalog; the parent distribution is shown in black contours. We compare mergers selected using the LDA technique (green) to those selected using the A18 technique (orange). The mergers identified using the LDA technique span the same color and mass ranges as those identified using the transfer learning technique, indicating that the LDA technique does not introduce significant morphological biases in its merger identification.}
    \label{fig:cmd_ack}
\end{figure}

We next use the color-mass diagram (Figure \ref{fig:cmd_ack}) to compare the properties of galaxies selected as mergers by the LDA technique ($p_{\mathrm{merg}} > 0.5$) to those of the galaxies selected as mergers by the A18 technique ($p_{\mathrm{merg}} > 0.95$). The cross-matched sample is incomplete at low galaxy stellar mass ($M_* < 10^{10} M_{\odot}$) due to the A18 sample, meaning that the parent sample is almost entirely composed of red sequence galaxies. However, over the extent of the cross-matched sample, it is clear that the mergers identified using the LDA method span the same regions of color-mass space as those identified using the A18 method, further verifying that the LDA technique does not introduce significant morphological biases relative to the A18 method.

We next bin the color-mass diagram by both stellar mass and color to compare the colors and stellar masses, respectively, of our sample of mergers to the A18 mergers. Using the KS test to compare the merger distributions, we find mergers identified using the LDA technique have similar stellar masses (for a fixed color) and are slightly bluer (for a fixed stellar mass) relative to mergers identified using the A18 method. \citet{Ackermann2018MNRAS.479..415A} compare their sample of mergers to those of their training set (\citealt{Darg2010}) and find that their sample tends towards redder colors relative to the GalaxyZoo-identified mergers. We also find that the A18 sample is redder relative to our galaxies.

\subsection{Merger fraction}
\label{results:fraction}

We measure the merger fraction ($f_{\mathrm{merg}}$), which is the fraction of galaxies that have a $p_{\mathrm{merg}}$ value greater than 0.5. We do this for both the major and minor merger classifications, focusing mostly on the major merger fraction in our analysis. For the remainder of the paper, $f_{\mathrm{merg}}$ or `merger fraction' refers to the major merger fraction. We will specify if we are referring to the minor merger fraction.

More specifically, a given output merger fraction $f_{\mathrm{merg}}$, is computed from an individual LDA classification that is calibrated using an input prior $\pi$ and then applied to all of the galaxies in SDSS. Our fiducial values of $\pi$ for the major/minor merger classifications are 0.1/0.3, respectively. Therefore, the measured (output) merger fraction for the fiducial major merger classification is:

$$f_{\mathrm{merg,\pi=fiducial}} = \frac{N_{p_{\mathrm{merg} > 0.5}}}{N_{\mathrm{all}}}$$

where $p_{\mathrm{merg}}$ is the merger probability for each SDSS galaxy calculated using the major merger classification created using the fiducial prior of $\pi$ = 0.1, $N_{p_{\mathrm{merg} > 0.5}}$ is the number of SDSS galaxies with probability values greater than the threshold of 0.5, and $N_{\mathrm{all}}$ is the number of SDSS galaxies in the sample. We perform this calculation for the 363,644 galaxy subset that are photometrically clean and mass-complete. 

\begin{figure}
    \centering
    \includegraphics[scale=0.2]{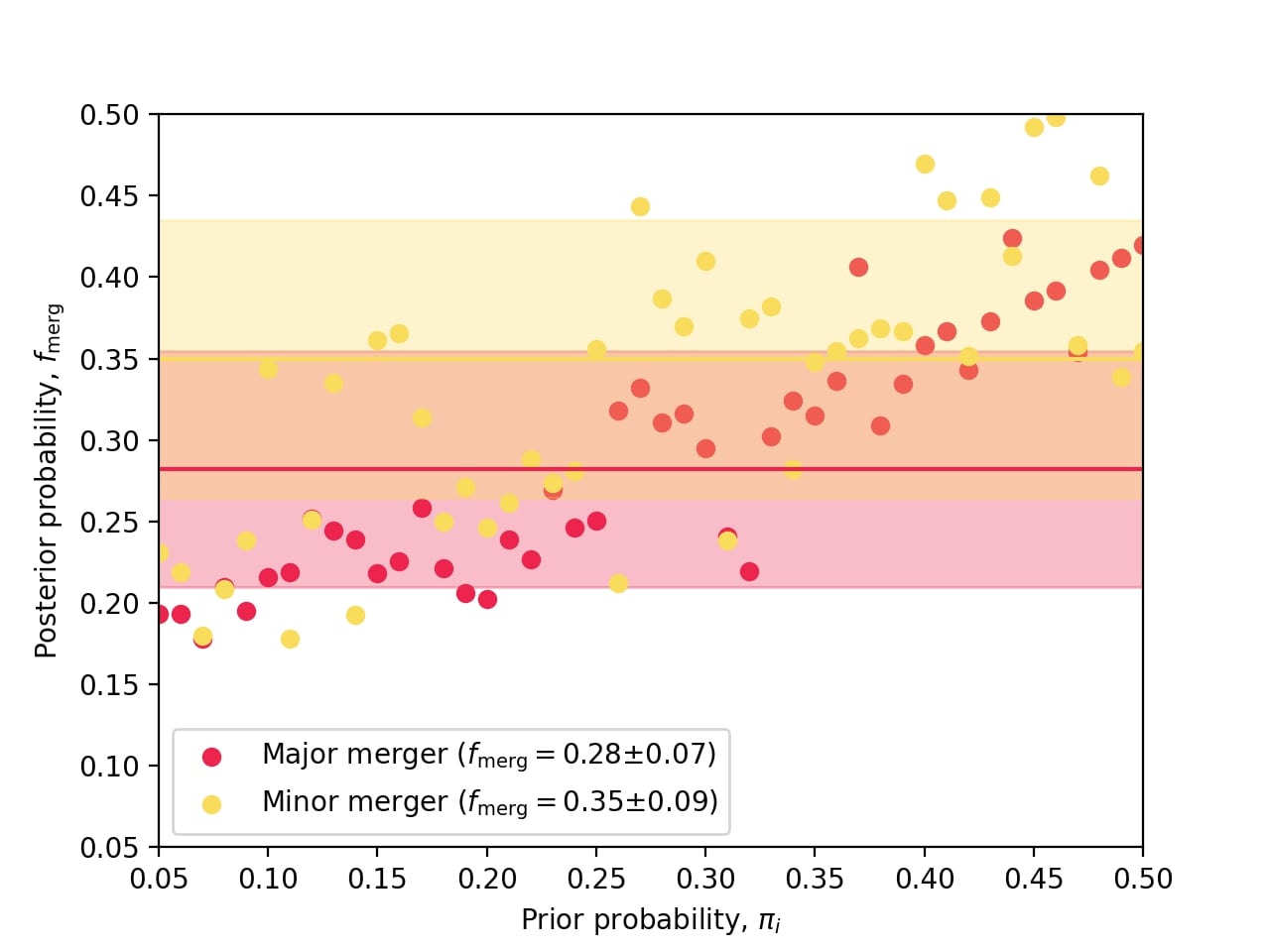}
    \caption{Measured merger fraction as a function of the input prior for the major (pink) and minor (yellow) merger classifications for the mass complete sample. We marginalize over the posterior probability (y-axis) to account for the effects of multiple possible input priors (prior probability, x-axis). The horizontal lines and shaded regions show the median and standard deviation of the merger fraction when marginalized over all input priors. The slope of this relationship is flat between a prior range of $0.05 < \pi < 0.25$ and that the slope flattens out beyond $\pi > 0.5$. This justifies the chosen prior range of $0.05 < \pi < 0.5$ and assures us that this range most likely covers the true merger fraction. }
    \label{fig:prior_adjust}
\end{figure}

As we discuss in \S \ref{methods:marginalizing}, adjusting the input prior affects the LDA classification and distribution of LD1 and $p_{\mathrm{merg}}$ values. We demonstrate this in Figure \ref{fig:prior_adjust}, where adjusting the prior ($\pi$, x-axis) affects our measurement of the posterior (merger fraction, $f_{\mathrm{merg}}$). In order to calculate the overall posterior probability, we employ the Bayesian approach described in \S \ref{methods:marginalizing}, marginalizing over the prior probability. The marginalized output merger fraction is the median of the individual merger fractions from each of the 46 priors shown in Figure \ref{fig:prior_adjust}. The error on $f_{\mathrm{merg}}$ is calculated from the standard deviation of the $f_{\mathrm{merg}}$ values for each input prior.

Figure \ref{fig:prior_adjust} demonstrates a flattening of the relationship between the input prior and the output posterior between the range $0.05 < \pi < 0.25$ for both the major and minor merger fraction. On the upper end, we rerun this calculation for major merger priors $\pi > 0.5$ and find a similar flattening in the relationship between the prior and posterior. Furthermore, we find that the median major merger fraction is unchanged when we widen the prior to $0.05 < \pi < 0.85$. This further justifies the 0.5 cutoff of the uniform prior that we introduced in \S \ref{methods:marginalizing} and assures us that we have used the appropriate prior range to recover the true merger fraction.

For each merger classification, we calculate the fiducial values of $f_{\mathrm{merg}}$ (which do not have associated errors) and the marginalized value of $f_{\mathrm{merg}}$ (uses the full posterior distribution of $p_{\mathrm{merg}}$) for both the clean and the clean and mass-complete samples of SDSS galaxies. We present these results in Table \ref{tab:nmerg} for the major and minor combined classification and the pre- and post-coalescence (1.0 Gyr) classifications for each.

Finally, it is important to note that some galaxies will be counted multiple times in this approach to calculating merger fraction. For instance, many galaxies that are classified as major mergers are also classified as minor mergers. The opposite is slightly less common and may be due to an increased minor merger fraction. Quantifying the overall major merger fraction (requiring that $p_{\mathrm{merg,50,maj}} > 0.5$), we find a major merger fraction of 0.21. When we remove all galaxies that are more likely to be minor mergers ($p_{\mathrm{merg,50,min}} > p_{\mathrm{merg,50,min}}$), we find a clean major merger fraction of 0.12. We repeat this calculation for the minor merger fraction and clean minor merger fraction and find values of 0.28 and 0.24, respectively. We find that these contamination fractions remains the same when we adjust the $p_{\mathrm{merg}}$ threshold value we use to define mergers.

We also investigate this overlap as it pertains to the calculation of the merger fraction trends with stellar mass and redshift in more depth in \S \ref{results:double_count}. We ultimately find that the contamination of minor mergers in the major merger sample does not affect our results about the merger fraction trends.

We also find that many galaxies are classified as multiple different stages of mergers. For instance, users should be aware that if they select for major mergers in the early stage ($p_{\mathrm{merg,maj,early,50}} > 0.5$), many of these galaxies will also be included when they select for major mergers in the late stages ($p_{\mathrm{merg,maj,late,50}} > 0.5$). 

Quantitatively, we find that 0.18 of galaxies are major mergers in the early stage and that this fraction drops to 0.03 after eliminating galaxies that are more likely to be late and post-coalescence stage major mergers. Similarly, 0.19 of galaxies are major mergers in the late stage; when we consider the clean late stage major merger fraction, this fraction drops to 0.13. The post-coalescence stage has a major merger fraction of 0.35, which drops to 32\% when only considering clean post-coalescence mergers. The implication is that a significant fraction of early stage mergers are likely to be identified as mergers in other stages. This result also holds for the merger stages of the minor mergers. The unclean/clean early stage minor merger fraction is 0.28/0.14. This figure is 0.24/0.16 for the late stage and 0.44/0.32 for the post-coalescence stage.

\begin{table*}
    \centering
    \begin{tabular}{c|ccc|ccc}

     \hline
   Priors & Major & Major   & Major  & Minor & Minor & Minor \\
    & All & Pre-coalescence & Post-coalescence (1.0) & All & Pre-coalescence & Post-coalescence (1.0) \\
    
    \hline
      
        Fiducial$^a$, all clean SDSS &0.18&0.18  &0.30 & 0.37 &0.27
  &  0.39\\
  Flat $[0.05, 0.5]$, all clean SDSS & 0.22$\pm$0.04& $0.22\pm0.03$ &$0.36\pm0.05$ &0.31$\pm$0.08  &$0.29\pm0.09$ & $0.46\pm0.04$  \\
  Fiducial, mass complete& 0.20&0.20  &0.47 &0.41  &0.29   &0.53 \\
   Flat $[0.05, 0.5]$, mass complete & 0.28$\pm$0.07& $0.24\pm0.03$ & $0.53\pm0.06$& 0.35$\pm$0.09 &$0.32\pm0.10$
   &$0.60\pm0.04$ \\
    \end{tabular}
    \caption{Merger fraction for the full sample of SDSS galaxies using different classification thresholds.\\
    $^a$The fiducial model is when $p_{\mathrm{merg}}>0.5$ and the priors are $\pi$ = 0.1 and 0.3 for the major and minor mergers, respectively.}
    \label{tab:nmerg}
\end{table*}

\subsection{Dependence of the major merger fraction on stellar mass and redshift}
\label{results:properties}

In this section, we explore how the measured major merger fraction changes with galaxy stellar mass and redshift. In \S \ref{results:sanity} and \S \ref{ap:sanity}, we further explore if these dependencies reflect biases of the classification or of the galaxy mass selection. 

\begin{figure}
\includegraphics[scale=0.2]{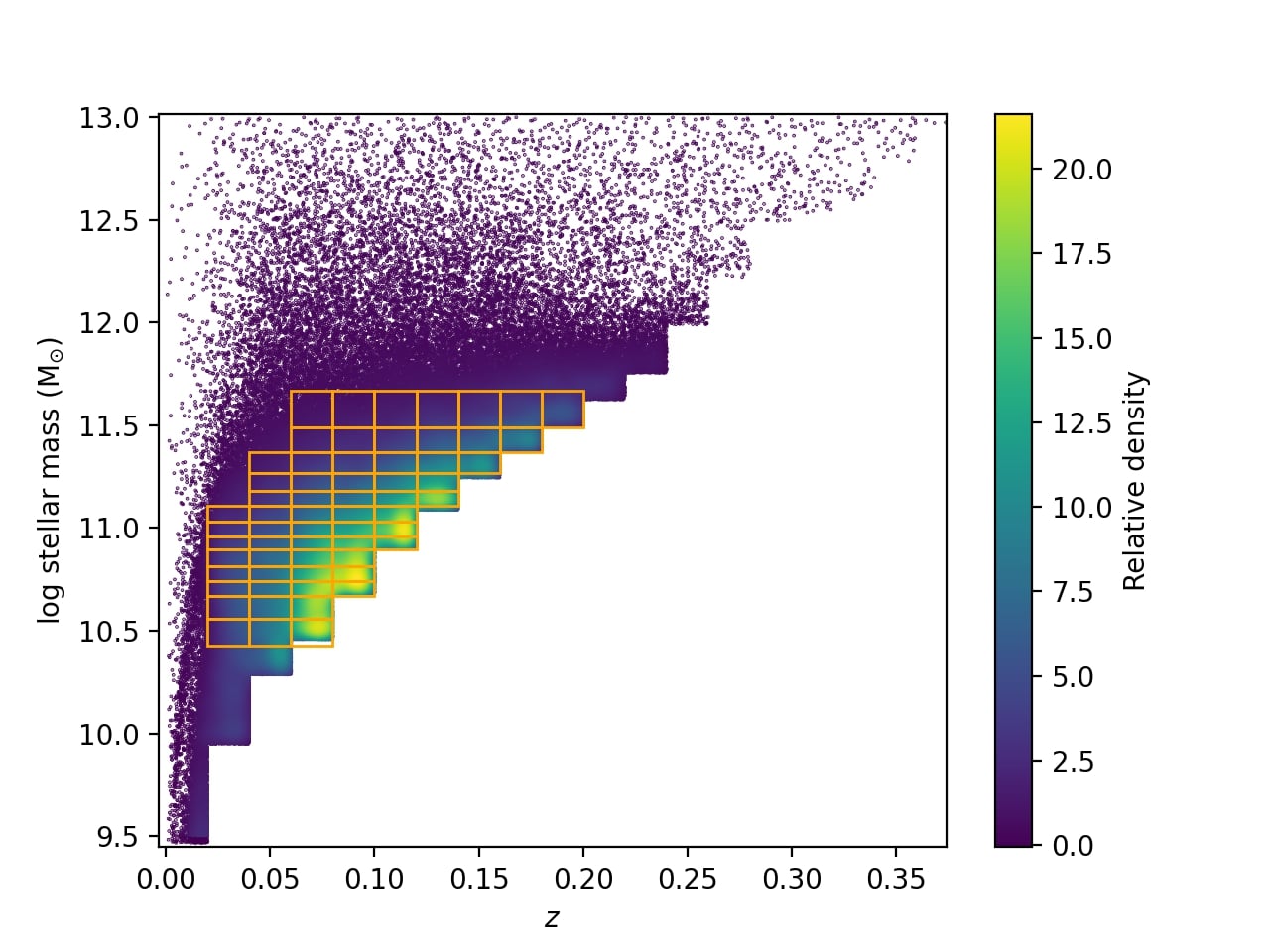}
\caption{Redshift and mass distribution of all galaxies in the mass-complete sample. For the analysis in this section, we select mass and redshift bins that have $>1000$ galaxies and where the mass and redshift distributions are complete (the medians are aligned with the bin center). We outline the final selected bins used for the analysis and annotate the number of galaxies in each bin.}
\label{fig:z_and_mass}
\end{figure}

First, in Figure \ref{fig:z_and_mass}, we separate the mass-complete sample into 15 evenly-sized bins in stellar mass (meaning there are the same number of galaxies in each one-dimensional bin) and bins of $\Delta z = 0.02$ in redshift. After binning the distribution, we eliminate bins where the median values of redshift and mass for the galaxies in that bin are significantly different from the bin centers, which we define as $> 1\sigma$ above or below the bin center, where $\sigma$ is the standard deviation of the values for the galaxies in the bin. This eliminates bins where incompleteness in redshift and/or mass could bias our results. We show the final binning scheme with the number of galaxies in each complete bin in Figure \ref{fig:z_and_mass}. All bins (red) have at least 1000 galaxies. This conservative approach restricts the final sample to 310,012 galaxies.

\begin{figure*}
\centering
\includegraphics[scale=0.45]{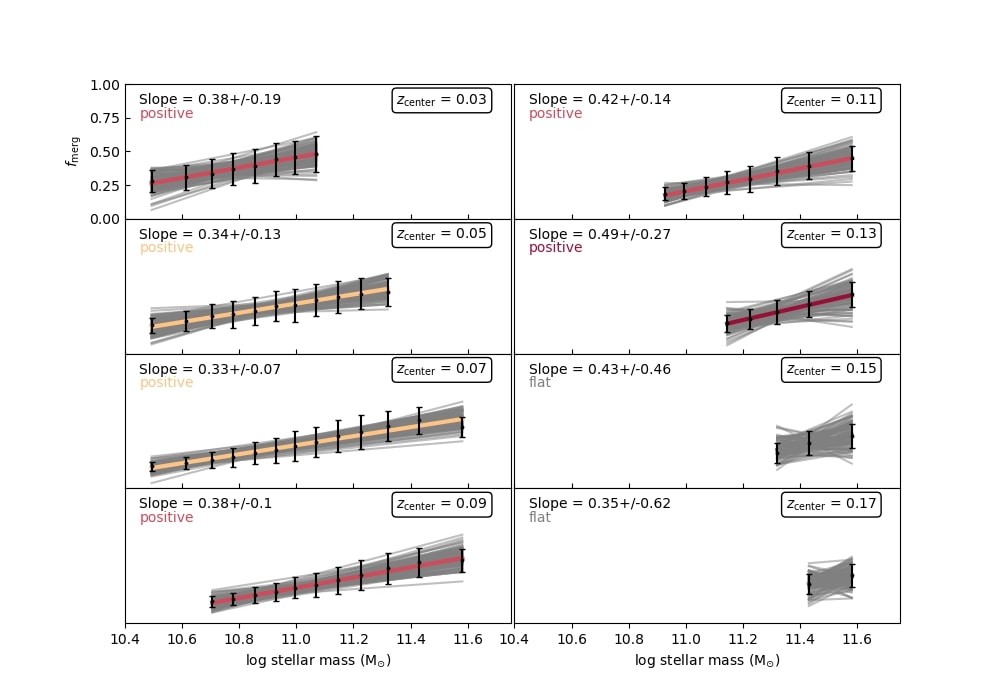}
\caption{Linear fits to the binned $f_{\mathrm{merg}}$ values for the major merger classification as a function of stellar mass for bins at fixed redshift (bin spacing is $\Delta z = 0.02$). We show the average line fit in color and the MCMC iterative fits in grey. We conclude that $f_{\mathrm{merg}}$ has a positive relationship with increasing mass for the majority of the redshift bins. All panels have the same y range.}
\label{fig:redshift_bins}
\end{figure*}

\begin{figure*}
\centering
\includegraphics[scale=0.45]{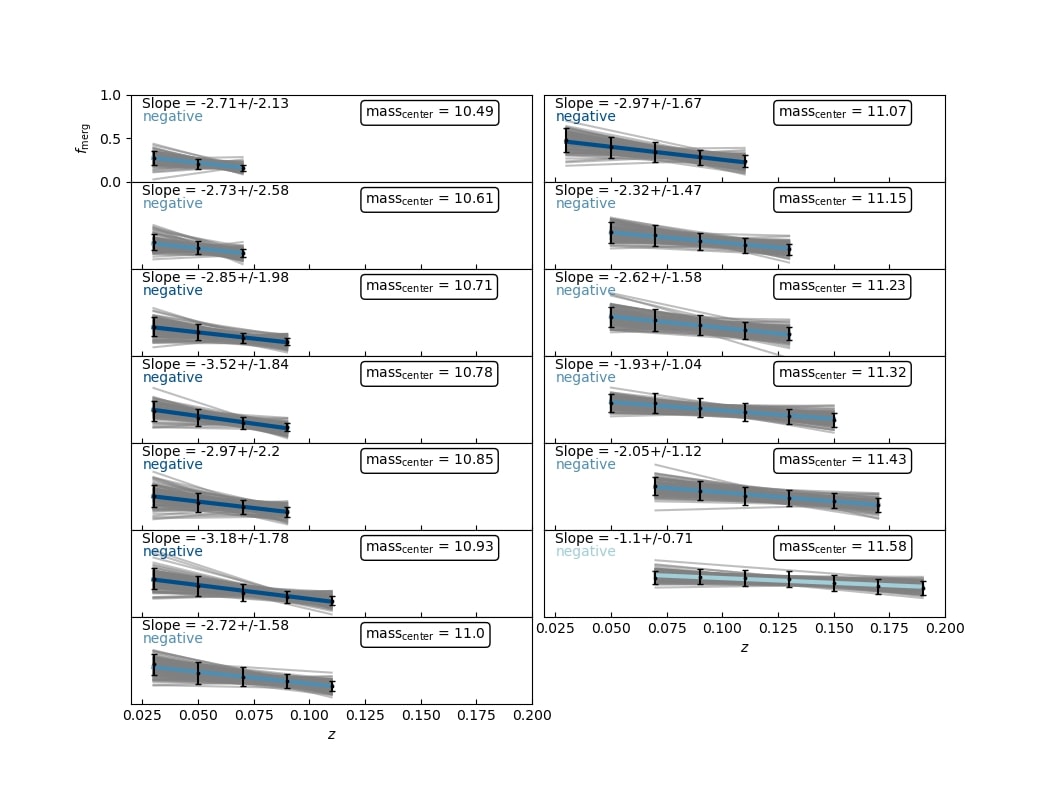}
\caption{Same as Figure \ref{fig:redshift_bins} but analyzing the slope of the major merger fraction with redshift for bins of fixed stellar mass. Here the slope is significantly negative with redshift for all mass bins.}
\label{fig:mass_bins}
\end{figure*}

We determine the median and standard deviation of the $f_{\mathrm{merg}}$ value in each bin by marginalizing across all priors. Next, for each redshift bin, we fit a line to the data points at each stellar mass by running a Markov Chain Monte Carlo (MCMC) analysis; we add the standard deviation (error bar) multiplied by a value drawn from a random normal distribution to each $f_{\mathrm{merg}}$ value and use \texttt{statsmodels} to fit a linear regression. We show the key results for the major merger classification in Figure \ref{fig:redshift_bins} and \ref{fig:mass_bins}, where we find a positive slope of $f_{\mathrm{merg}}$ with stellar mass and a negative slope with redshift, respectively. 

The slope of the major merger fraction with mass (Figure \ref{fig:redshift_bins}) is mostly positive; for 6/8 bins this is a significantly positive slope to 1$\sigma$, where $\sigma$ is the variation in the slope value found via the MCMC iterative analysis. In 2/8 cases, the slope is significantly positive to 2$\sigma$, and in 3/8 cases, it is significantly to 3$\sigma$. The slope of the major merger fraction with redshift (Figure \ref{fig:mass_bins}) is significantly negative in 13/13 bins to 1$\sigma$ confidence.

Considering only the bins that have statistically significant slopes, we find that the value of the slope ranges between $0.31 < \alpha < 0.53$ with mass (for the $z$ bins) and between $-3.35 < \alpha < -1.08$ with redshift (for the mass bins). Generally, the trend is more steeply positive towards higher redshift and more steeply negative towards low and intermediate masses.

\subsection{Is S/N confounding the redshift-dependent major merger fraction?}
\label{results:S_N}
A statistical confound is a variable that distorts the apparent causal relationship between the independent and dependent variables because it is independently associated with both. To investigate if S/N is a confound that is causing the apparent negative slope in the major merger fraction with redshift, stratify, or bin, by S/N. We first restrict the S/N range to $0 < \mathrm{S/N} < 50$ because galaxies with S/N $> 50$ have a sparse distribution in the 3D parameter space. This restricts the sample from 363,644 to 305,321 galaxies.

\begin{figure*}
    \centering
    \includegraphics[scale=0.25]{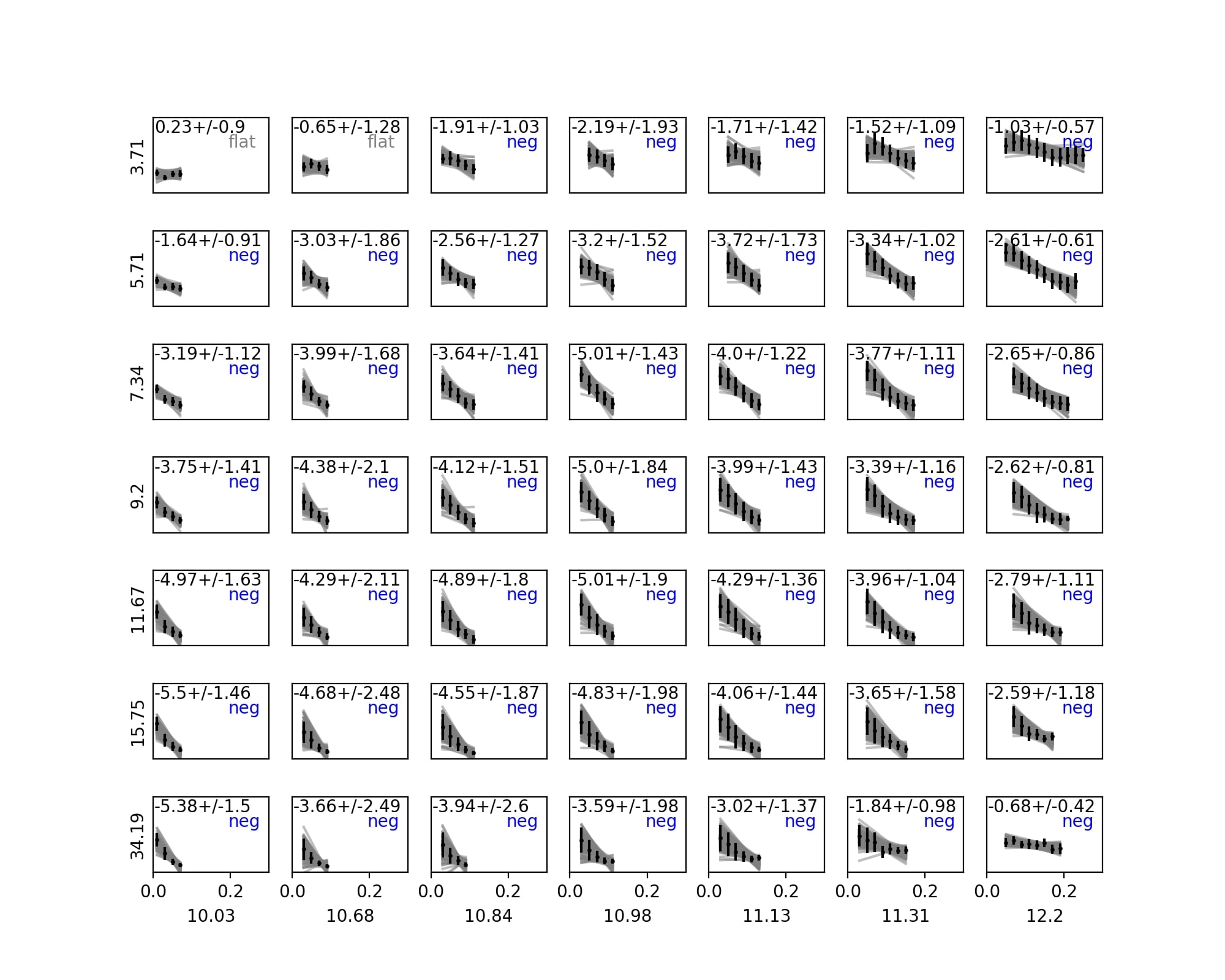}
    \caption{The slope of the major merger fraction with redshift (inset subplot x-axis) for almost all S/N (figure y-axis) and mass (figure x-axis) bins is significantly negative. This indicates that the negative redshift trend for $f_{\mathrm{merg}}$ cannot be attributed to increasing S/N with increasing redshift.}
    \label{fig:S_N_binning_inset}
\end{figure*}

We present our results in Figure \ref{fig:S_N_binning_inset}, where redshift is the target independent variable and S/N and mass bins are the y and x-axis of the figure, respectively. We demonstrate that for almost all 2D bins (in mass and S/N), the slope of $f_{\mathrm{merg}}$ is significantly negative with increasing redshift. In many cases, the slope is slightly more negative than the 2D binning analysis with mass and redshift. We can conclude that a projection of the S/N-dependence of $f_{\mathrm{merg}}$ does not explain the negative slope with redshift; when the sample is stratified by S/N, the slope of the major merger fraction is even more negative with redshift.

We also run this analysis with S/N as the independent variable of interest and find that when we stratify by mass and redshift that the major merger fraction has a mostly negative trend with S/N, meaning that we find higher merger fractions for lower S/N galaxies. This trend is not well fit by a linear relationship; the slope is either flat or negative but very close to flat. This result is distinct from many studies that find a positive trend of merger fraction with S/N, where they are biased to detect brighter galaxies due the merger identification technique's reliance on faint tidal features (e.g. \citealt{Bickley2021}).

\subsection{Are morphology (bulge-to-total mass ratio) or color confounding the redshift-dependent merger fraction?}
\label{results:morphology}

We investigate if the negative slope of the major merger fraction with redshift could be attributed to a sensitivity to galaxy type. For instance, some studies find a different evolution of the merger fraction with redshift for early-type galaxies (ETGs) and late-type galaxies (e.g. \citealt{Lin2008,Lopez-Sanjuan2012}). In some cases, the ETGs have a negative slope with increasing redshift (\citealt{Lin2008,Groenewald2017MNRAS.467.4101G}).

To conduct this analysis, we repeat the analysis of the previous section, this time treating bulge-to-total mass ratio (B/T) and color ($g-r$) as the suspect confounding variables. This 3D binning analysis is identical to the S/N investigation we describe in \S \ref{results:S_N}; here we replace S/N with B/T and $g-r$ color and re-calculate the major merger fraction. By stratifying by these nuisance parameters, we remove their influence from the other parameters of interest (stellar mass and redshift). We find that the slope of the major merger fraction with mass and redshift does not significantly change as a function of galaxy color or B/T mass ratio. This is strong evidence that neither color nor B/T are responsible for the mass and redshift trends. The exception is our reddest bin, where the slope of the major merger fraction with redshift is flat or positive. 

It is important to make the distinction that while B/T and color are not a confounding variables that are responsible for the negative redshift dependence, they can still have independent influence on the merger fraction. For instance, when we stratify by mass, redshift, and B/T, we find that the major merger fraction is mostly flat as a function of B/T but increases with B/T for some bins, peaking around a B/T mass ratio of 0.7. When we stratify by mass, redshift, and color, the major merger fraction is positive with $g-r$, meaning the major merger fraction increases for redder galaxies at high masses and redshifts. At low masses and redshifts, the slope is instead negative or flat. This is consistent with a picture where the major merger fraction increases with B/T and $g-r$ mostly for higher mass galaxies.

\subsection{Dependence of the minor merger fraction on stellar mass and redshift}
\label{results:properties_minor}

\begin{figure*}
\centering
\hbox{\includegraphics[scale=0.25]{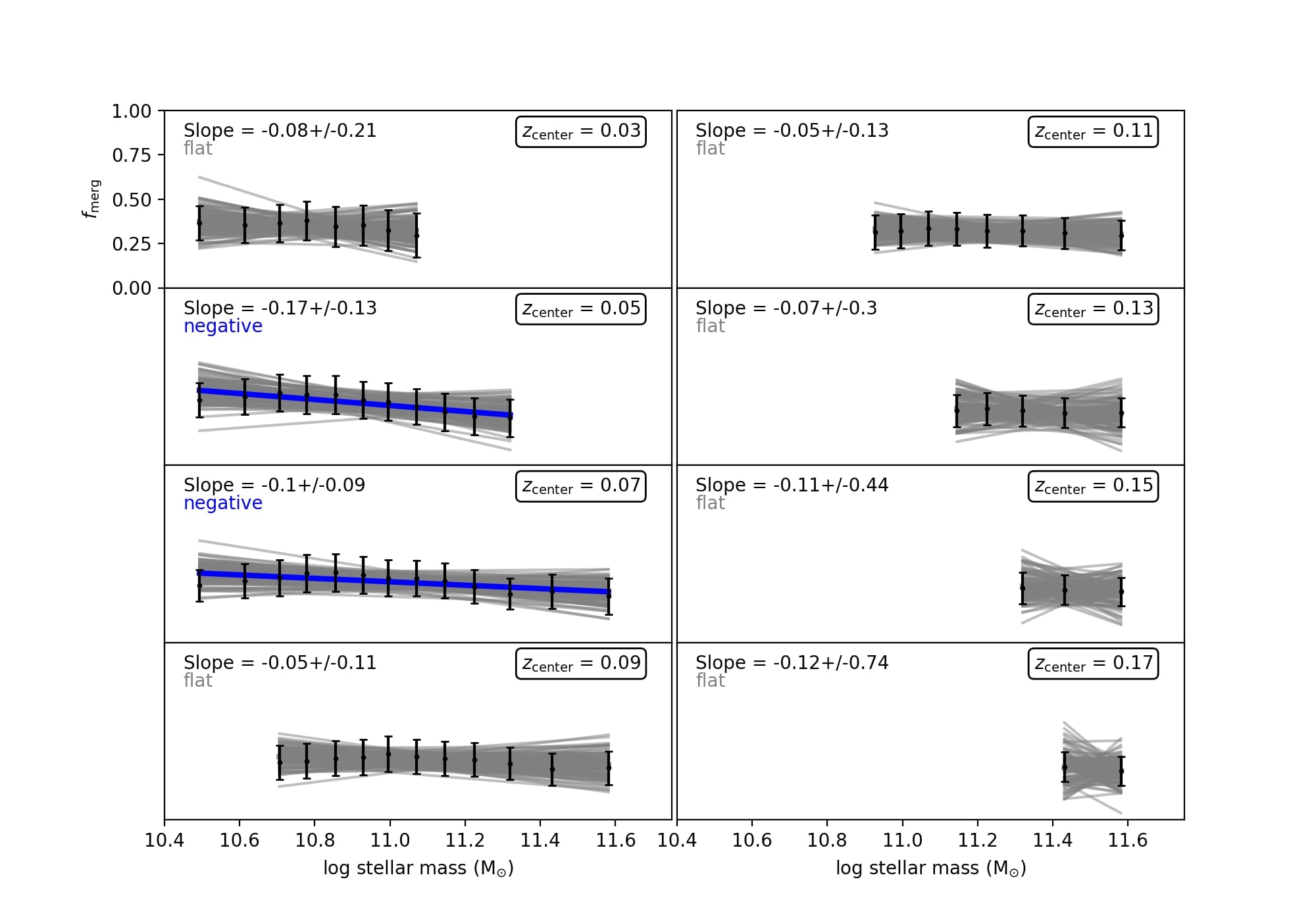}}
\caption{Same as Figure \ref{fig:redshift_bins} but for the minor merger fraction.}
\label{fig:redshift_bins_minor}
\end{figure*}

\begin{figure*}
\centering
\includegraphics[scale=0.25]{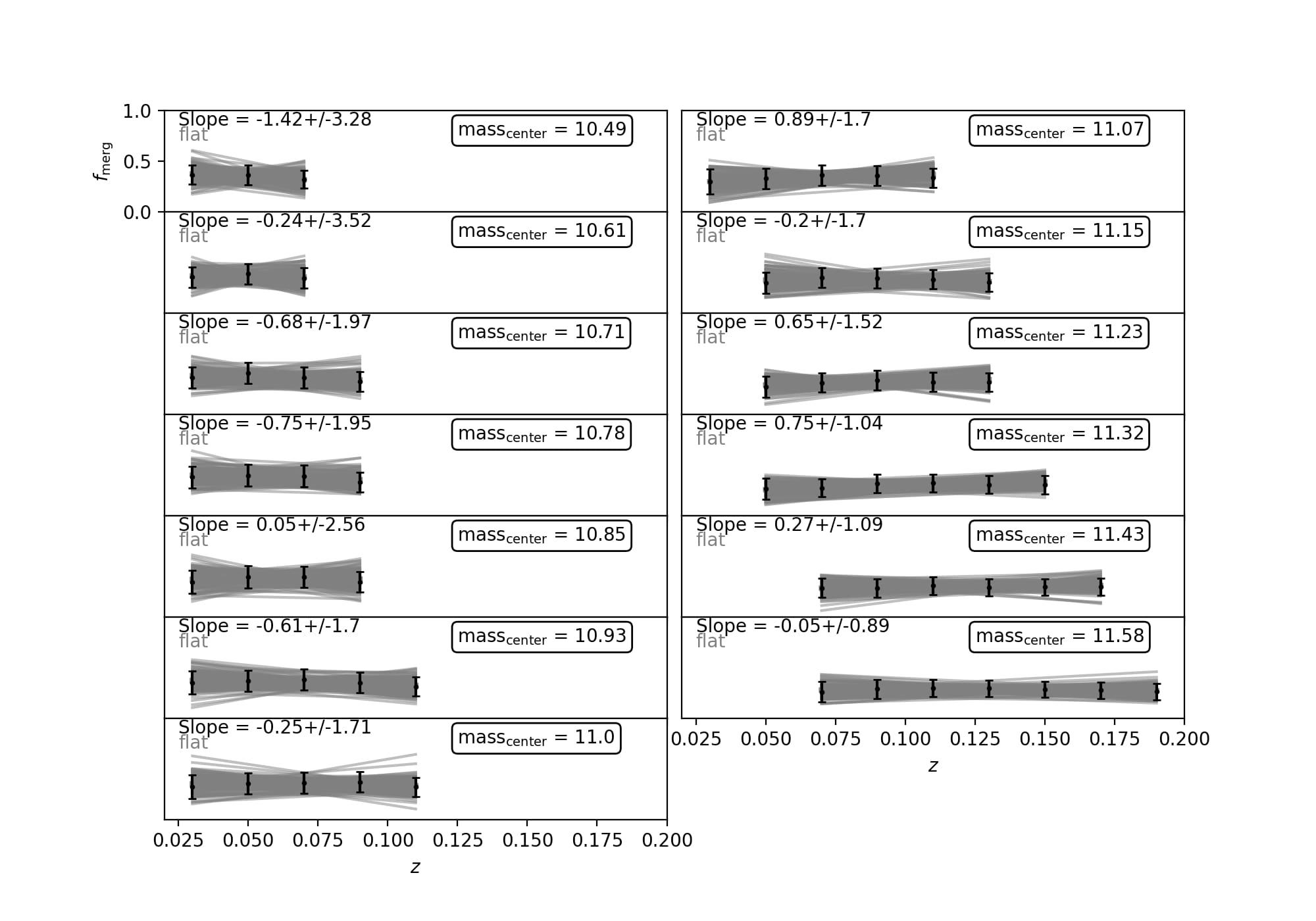}
\caption{Same as Figure \ref{fig:mass_bins} but for the minor merger fraction.}
\label{fig:mass_bins_minor}
\end{figure*}

Here we repeat the analysis, instead using the minor merger classification to identify merging galaxies. We show the results for the binned analysis in Figures \ref{fig:redshift_bins_minor} and \ref{fig:mass_bins_minor} for the slope of the merger fraction with mass and with redshift, respectively. We find that the slope of the merger fraction is mostly flat with stellar mass except for two redshift bins where it is negative. The slope of the merger fraction with redshift is flat for all mass bins. In other words, the minor merger fraction shows little dependence on mass or redshift. We discuss the implications of this in \S \ref{discuss:minor}.

\subsection{Accounting for contamination in the major/minor merger samples by minor/major mergers}
\label{results:double_count}

In \S \ref{results:properties} and \ref{results:properties_minor}, we empirically measure the merger fraction as a function of stellar mass and redshift for the major and minor merger classifications, respectively. These results include overlap between classifications, since we consider all galaxies with median $p_{\mathrm{merg}}$ values greater than 0.5 as mergers. Here we investigate if these results change when we calculate the merger fraction for the sample of major and minor mergers without overlap between classification.

To calculate the clean major and minor merger fraction, we require that $p_{\mathrm{merg,med}} > 0.5$ and $p_{\mathrm{merg,med,maj}} > p_{\mathrm{merg,med,min}} $ for the major mergers and $p_{\mathrm{merg,med}} > 0.5$ and $p_{\mathrm{merg,med,min}} > p_{\mathrm{merg,med,maj}} $ for the minor mergers. The second requirement significantly reduces the sample size of major mergers from 86,843 galaxies to 53,573 galaxies and reduces the sample size of minor mergers from 103,907 to 86,837 galaxies. The major merger sample therefore has a greater contamination contribution from minor mergers, which is to be expected given the larger overall merger fraction for minor mergers.

When we re-calculate the mass- and redshift-dependent merger fraction for the clean samples, we find similar results. The clean major merger fraction has a positive slope with mass and a negative slope with redshift for all bins. Most slopes are slightly flatter than the not clean case; however, this difference is not statistically significant (to $1\sigma$ errors). This slight flattening could be due to a contamination from the minor mergers, where the trend with mass and redshift is flatter. The clean minor merger fraction slopes are consistent to $1\sigma$ with the not clean minor merger fraction slopes. 

In conclusion, while double counting in the major and minor merger samples has a significant effect on the overall number of mergers, double counting does not affect our conclusions about the slope of the major and minor merger fraction with redshift and mass. \textit{The implication is that the slope of the merger fraction is robust to these levels of contamination (38\% and 16\% of the major and minor merger samples, respectively, are contaminated).}

\subsection{Sanity checks}
\label{results:sanity}

As we will address in the discussion section, the result of increasing merger fraction with stellar mass has precedent in the literature. However, the result of decreasing merger fraction with redshift is unprecedented. Given this surprising result, we explore in Appendix \ref{ap:sanity} whether the result of decreasing merger fraction with increasing redshift is physical (real) or whether we can attribute it to sample systematics (i.e. mass incompleteness at higher redshift or errors in the mass calculation or determination of the photometric redshift).

After running our merger sample through multiple sanity checks in Appendix \ref{ap:sanity}, we can conclude that the trend of the major merger fraction increasing with stellar mass (for constant redshift) and decreasing with $z$ for constant stellar mass is robust to changes in how we measure redshift and stellar mass. It is also robust to changes in how we bin the data for this analysis and how we compute the mass completeness. These steps were all taken to rule out the leading culprits of systematic bias in the sample that could lead to our surprising result of the negative evolution of the major merger fraction with redshift.

Finally, we compare our major merger sample to a different merger sample (A18). We find a mostly flat result with redshift for the A18 merger sample. Since we use the same cross-matched sample to rerun the LDA classification and still find a negative trend with redshift for the cross-matched sample, we are able to conclude that this result is not due to peculiarities of the galaxy sample but instead can be attributed to differences due to the merger selection itself.

\subsection{In the absence of mass binning the major merger fraction has an artificial positive trend with redshift}
\label{sanity:no_mass_bins}

We have taken one final step towards understanding the negative trend of the major merger fraction with redshift in the context of other work. Here we run our analysis without mass binning, as other work has done in the past in the absence of enough data to bin and still retrieve a statistically significant result. 

We re-run the analysis without mass bins to determine the confounding role of the positive mass trend in the redshift slope when we do not control for mass. We additionally experiment with eliminating the completeness correction (of Figure \ref{fig:z_and_mass}) and with using spectroscopic redshifts. 

\begin{figure}
\centering
\includegraphics[scale=0.035]{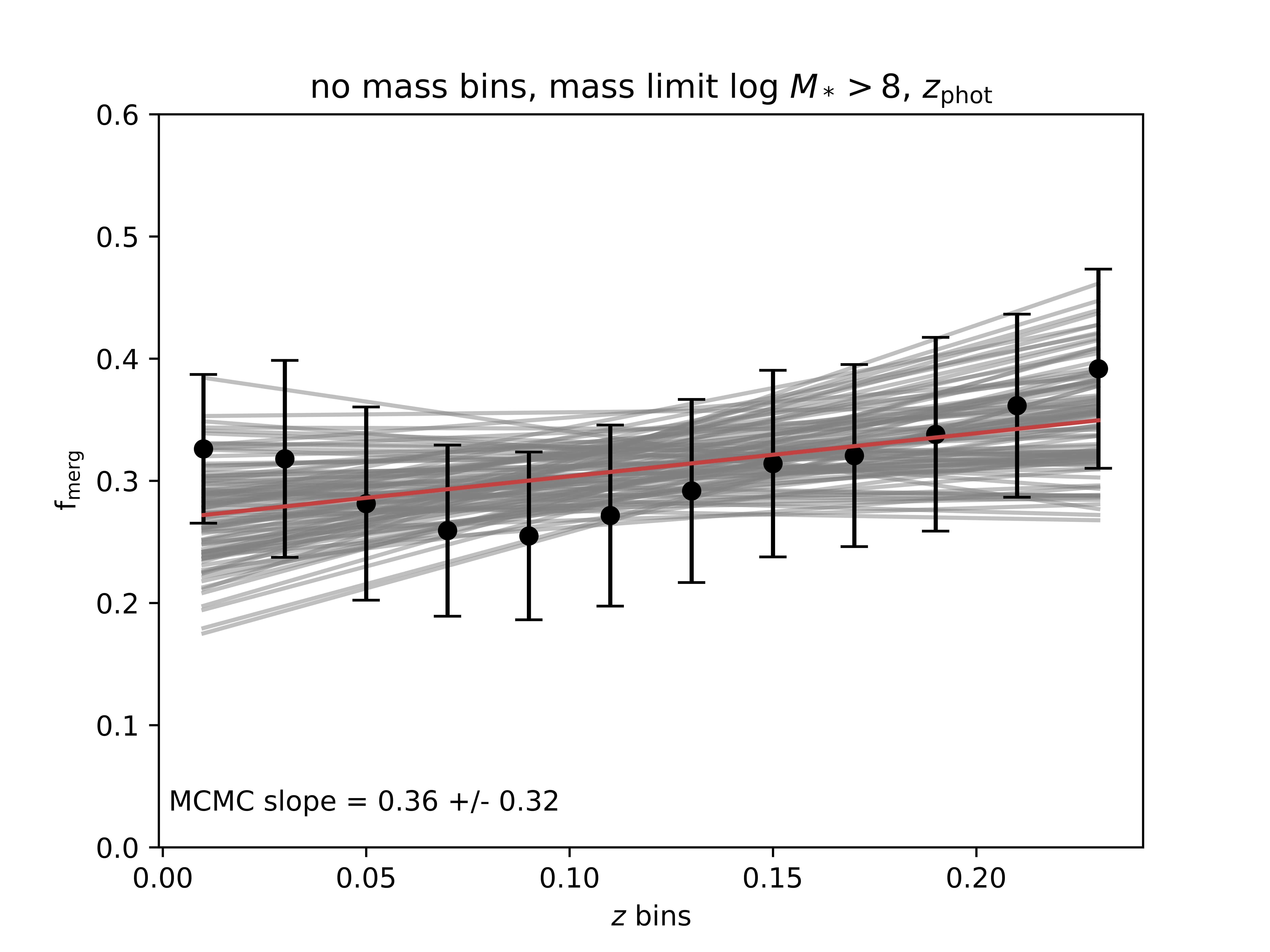}
\includegraphics[scale=0.035]{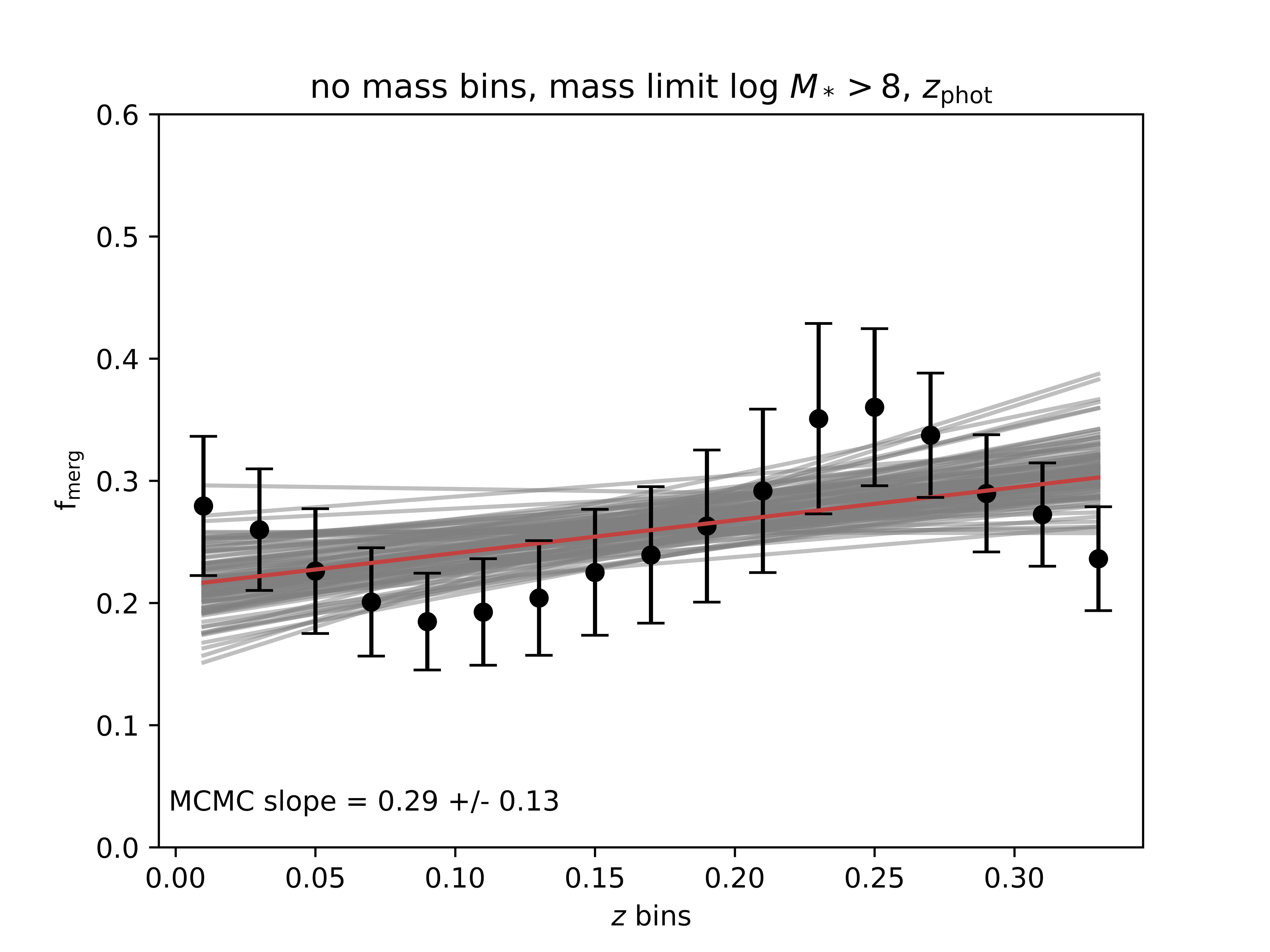}
\includegraphics[scale=0.035]{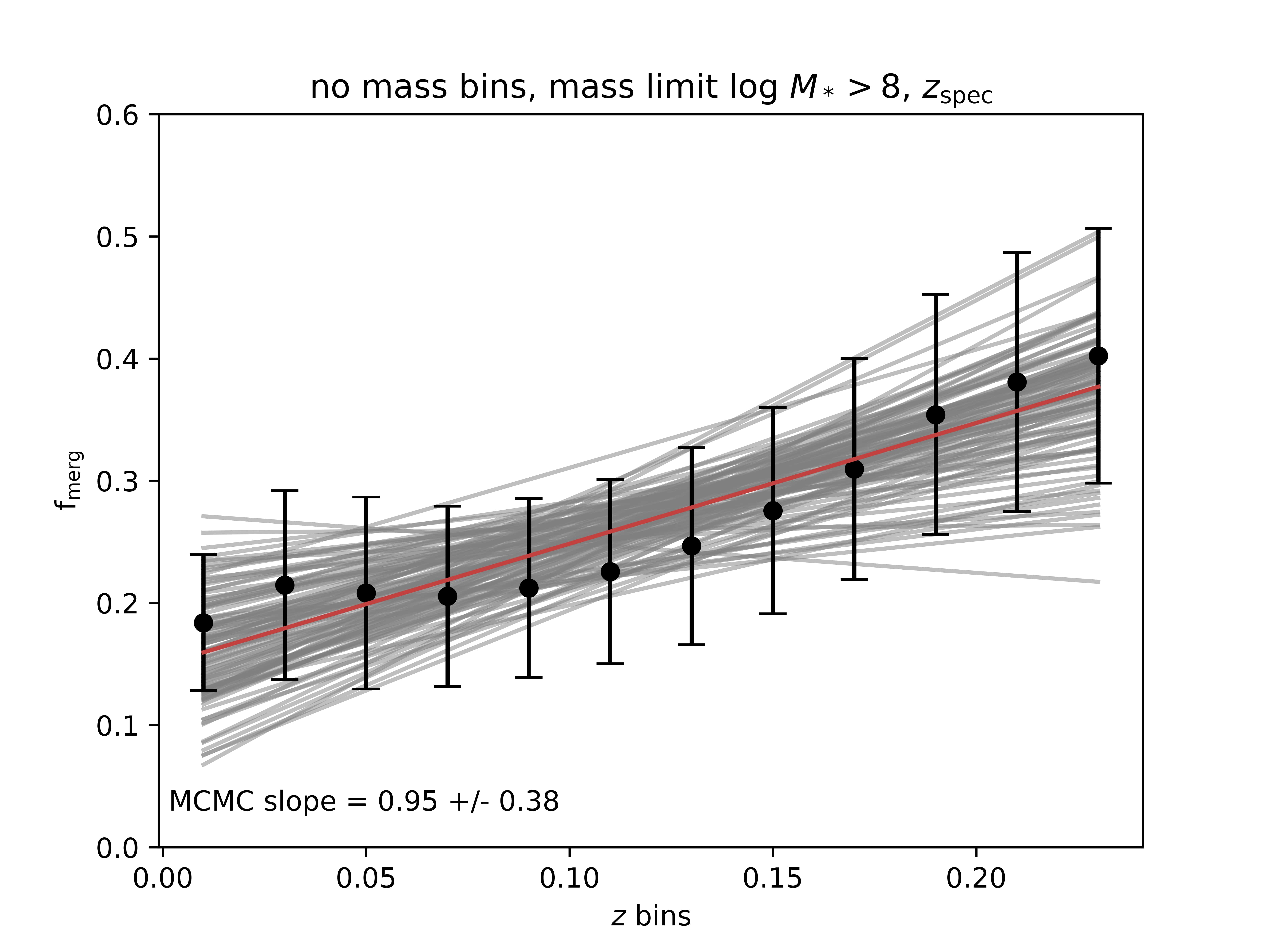}
\caption{Slope of the major merger fraction with redshift when we do not bin for mass. We show the results for the mass complete sample with photometric redshifts (top), for the mass incomplete sample with photometric redshifts (middle), and for the mass complete sample with spectroscopic redshifts (bottom). The black data points and accompanying error bars give the average and standard deviation of the merger fraction for each redshift bin, the red lines show the linear fits under the MCMC realizations (as outlined in \S \ref{results:properties}), and the red line is the average line fit with slope and error given at the bottom of the plot. The average slope is positive for all samples, yet experiences a significant downturn at redshifts $z < 0.1$ for the two top plots, which measure redshift using the photometric redshifts. These plots demonstrates the importance of binning for mass; if this is not considered, the positive trend of the major merger fraction with mass is projected onto the redshift axis resulting in an artificial positive trend of the major merger fraction with redshift.}
\label{fig:no_mass}
\end{figure}

We present our results in Figure \ref{fig:no_mass}. We find a significant positive slope of the major merger fraction with redshift in all cases where we do not bin for mass. This includes the sample that is mass complete with photometric redshifts (top), the sample that is mass incomplete with photometric redshifts (middle), and the sample that is mass complete with spectroscopic redshifts (bottom). All of plots in this figure use color-derived masses, but we find similar results with SPS-derived stellar masses. \textit{Figure \ref{fig:no_mass} is therefore an important reminder that what looks like a positive slope with redshift is actually the projection of a positive slope in mass onto redshift.} 

This figure additionally highlights that while the overall trend is positive, there are different features in each plot produced by the slightly different sample selections. For instance, the peak at low redshift can most likely be attributed to the bias produced by photometric redshift that artificially increases the stellar masses of low mass galaxies. Additionally, the peak at higher redshift in the middle plot (mass incomplete sample) can most likely be attributed to the mass incompleteness of the sample. 

The conclusions from this section can be directly connected to our overall conclusions from this work. While we cannot completely rule out that our negative trend with redshift is not the result of some other systematic bias or a combination of biases (i.e., confounding factors like mass incompleteness and redshift bias), we can at least clearly show the most simple and likely scenario: that mass binning versus no mass binning produce dramatically different trends of the evolution of the major merger fraction with redshift. This demonstrates the importance of running this type of analysis on large samples of galaxies and with a merger classification technique such as the LDA that demonstrates broad reliability across a range of galaxy types. Both of these elements of this paper were essential to be able to bin the sample in both redshift and mass and do a careful completeness correction.

\section{Discussion}
\label{discuss}

Our mass-complete binning analysis of a large sample of galaxies using a carefully calibrated set of classification techniques allows us to make clear conclusions about the evolution of the merger fraction locally ($0.03 < z < 0.19$). The additional novelty of this study is that the large sample size allows us to do this over a finely spaced grid of redshift and mass bins.

Here we discuss our measurements of the mass and redshift-dependent merger fraction in the broader context of previous work. We focus all discussion on predictions and measurements of the merger fraction and reserve all discussion of the merger rate for future work (Simon et al. 2023, in prep). We begin with a brief review of predictions for the local merger fraction from cosmological models (\S \ref{discuss:theory}) and a review of recent empirical estimates of the merger fraction (\S \ref{discuss:empirical}).  We then discuss the implications of our findings of the mass and redshift dependence of the major merger fraction for galaxy evolution in the local Universe (\S \ref{discuss:mass} and \S \ref{discuss:redshift}, respectively). We also discuss the implications of the distinct results we find for the minor merger fraction evolution (\S \ref{discuss:minor}). We summarize our precautions throughout this paper to prevent morphological biases in the results in \S \ref{discuss:robust}. We end with a discussion of the relative strengths of the methodology presented here (\S \ref{discuss:strengths}) as well as the caveats and future work motivated by this study (\S \ref{discuss:caveats}).

\subsection{Predictions of the redshift and mass-dependence of the merger fraction from cosmological models}
\label{discuss:theory}

The $\Lambda$CDM model of structural assembly (e.g. \citealt{White1978}) predicts hierarchical, or bottom-up assembly, meaning that mergers assemble smaller halos first followed by more massive halos at later times (e.g. \citealt{Blumenthal1984Natur.311..517B}). This predicts a merger rate that evolves with the density of galaxies and space in the Universe. The measured fraction of merging galaxies should therefore increase with redshift back to $z \sim$ 2 - 3. Additionally, the merger fraction should have a steep dependence on mass in the local Universe, since the most massive galaxies are predicted to assemble at later times. 

An alternate assembly scenario is cosmic downsizing, where the largest galaxies form early and then stall (e.g. \citealt{Cowie1996AJ....112..839C,Juneau2005ApJ...619L.135J,Treu2005ApJ...622L...5T,Cowie2008ApJ...686...72C}). Mergers have been invoked as a mechanism to drive this compact star formation followed by rapid quenching. While \citet{Bridge2010ApJ...709.1067B} invoke downsizing as a mechanism to drive a negative mass dependence in the merger fraction between redshifts $0.2 < z < 1.2$, \citet{Estrada-Carpenter2020ApJ...898..171E} show that the phenomenon of downsizing has a minimum redshift that ranges between $1.5 < z < 8$. The implication is that this assembly scenario does not apply to the local Universe.

Baryonic evolutionary processes (i.e. feedback) play an important role in galaxy assembly. Simulation work finds that when baryonic feedback is combined with the bottom-up formation model (hierarchical assembly), this can manifest as top-down assembly, i.e. downsizing (\citealt{Stringer2009MNRAS.393.1127S} and references therein). Baryonic feedback suppresses the growth of stellar mass in galaxies above and below $\sim10^{11} M_{\odot}$. This results in a higher number of intermediate mass galaxies. If this effect is strong, it could result in more major mergers between equal-mass galaxies locally. 

In reality, the picture is probably far more complicated than any one of the above formation scenarios. Different processes likely dominate for different mass scales and at various epochs over the age of the Universe. Directly observing the galaxy-galaxy merger fraction as a function of redshift and mass and separating major from minor mergers is therefore critical for constraining the relative contributions of these competing processes.

\subsection{Reviewing past empirical measurements of the mass- and redshift-dependent merger fraction}
\label{discuss:empirical}

Characterizing the mass-dependence of the major merger fraction can help us understand how elliptical galaxies and the bulges of galaxies are being built up over different mass ranges. Accurately measuring the mass-dependent merger fraction locally is especially important for anchoring the redshift-dependent merger fraction and directly testing the hierarchical assembly prediction that the most massive galaxies are assembled locally. 

As with the redshift-dependent merger fraction, previous work in this area relies on either close pair methods (e.g. \citealt{Xu2004ApJ...603L..73X,Patton2008ApJ...685..235P,Domingue2009ApJ...695.1559D,Xu2012ApJ...747...85X}) or morphological studies (e.g. \citealt{Bridge2010ApJ...709.1067B,Casteels2014}) to measure the mass-dependent merger fraction. Most studies find a constant or slightly increasing merger fraction with mass (e.g. \citealt{Xu2004ApJ...603L..73X,Patton2008ApJ...685..235P,Domingue2009ApJ...695.1559D,Xu2012ApJ...747...85X,Robotham2014MNRAS.444.3986R}). Of particular note is the work of \citet{Robotham2014MNRAS.444.3986R}, which focuses on galaxies in the GAMA survey ($0.05 < z < 0.2$). They find that the merger fraction increases with mass by a factor of $\sim$3 between stellar masses of $10^9 < M_* (log\ M_{\odot}) < 10^{11}$. 

Other work finds a decreasing fraction with mass (e.g. the morphological studies of \citealt{Bridge2010ApJ...709.1067B} and \citealt{Casteels2014}). \citet{Bridge2010ApJ...709.1067B} ($0.2 < z < 1.2$) claim that the decreasing fraction with mass is due to cosmic downsizing, while \citet{Casteels2014} ($0.001 <z<0.2$) argue that the decrease is due to an increasing observability timescale for lower mass galaxies.

A number of studies have measured the major merger fraction and how it trends with redshift. It is important to note that most of the past work that has examined the redshift-dependence of the merger fraction has done so for redshift intervals that do not overlap with our study. The consensus among these studies is mostly for a higher merger fraction at higher redshift (\citealt{Lin2008,Conselice2009,Lopez-Sanjuan2012,Robotham2014MNRAS.444.3986R,Mundy2017MNRAS.470.3507M,Mantha2018,Snyder2019,Kim2021MNRAS.507.3113K}), although some studies find a relatively flat merger fraction (\citealt{Bundy2009ApJ...697.1369B,Jogee2009,Keenan2014}).

Only the GAMA-focused studies of \citet{Robotham2014MNRAS.444.3986R} and \citet{Keenan2014} have more than one redshift bin below $z = 0.2$. Additionally, due to small sample sizes, the above work is often unable to bin finely in both stellar mass and redshift, and therefore unable to explore both simultaneously. \textit{Our study is unique in that we have a large enough sample size to create bins in both stellar mass and redshift and this is the first study to do so for fine redshift bins locally ($0.03 < z < 0.19$).}

\subsection{Implications of a positive mass dependence of the major merger fraction}
\label{discuss:mass}

\begin{sloppypar}
We find a positive relationship between the major merger fraction and stellar mass between a range $10.5 < M_* (log\ M_{\odot}) < 11.5$. This is consistent with a hierarchical assembly picture, where more massive halos (hence galaxies) are assembling locally. In this case, since we observe this trend for all redshift bins, we can conclude that this trend holds for the last $\sim$2 Gyrs of galaxy evolution. As mentioned in \S \ref{discuss:theory}, if baryonic processes such as feedback are coupling with hierarchical assembly, this could manifest as top-down assembly (cosmic downsizing). While we cannot rule this scenario out entirely, we can conclude that if this is happening, it is not strong enough to invert or flatten our observed positive trend for the major merger fraction.
\end{sloppypar}

\subsection{Implications of a negative redshift dependence for the major merger fraction}
\label{discuss:redshift}

Our key result is a decreasing major merger fraction with redshift over the range $0.03 < z < 0.19$ (Figure \ref{fig:mass_bins}). The implication is that major mergers become more important in the nearby Universe. This result cannot be explained by either hierarchical assembly or cosmic downsizing. Hierarchical assembly predicts an increase of the merger fraction out to high redshifts, while cosmic downsizing likely does not operate in the local Universe and does not make explicit predictions for the merger fraction. We find it most likely that baryonic feedback is dominating locally, overriding the positive slope predicted by hierarchical assembly.

Here we focus mostly on the implications of our finding in the context of past studies and how these results merit a revision of the current techniques used to measure the evolution of the major merger fraction.

\textit{Most other close-pair studies find a positive trend of major merger fraction with redshift, yet it is important to note that the majority of these studies do not cover the same redshift range as this work ($z < 0.2$) and that none of these studies control for both mass and redshift simultaneously.} As we have shown, the major merger fraction varies as a function of both mass and redshift and the mass dependence can project onto the redshift axis, resulting in an artificial positive relationship with redshift. 

Our recommendation is for the community to: 1) Revisit past analyses of the redshift-dependence of the major merger fraction using a careful mass and redshift binning analysis, and 2) Design future studies that cohesively span the local Universe and the higher redshift Universe. Currently, it is unclear if our findings represent a local inversion in the higher redshift ($z > 0.2$) trend of a positive evolution of the major merger fraction with redshift or if higher redshift studies will be inverted when mass binning and mass completeness are accounted for. 

Additionally, many of the past close-pair studies that find a positive slope with redshift for the major merger fraction are sampling from a severely restricted volume. For a more in-depth analysis of the volume probed by various merger fraction studies, see the discussion of the role of cosmic variance in the calculation of the merger fraction in \cite{LopezSanjuan2014A&A...564A.127L}. Furthermore, \citet{Patton2008ApJ...685..235P} find that the cosmic variance from the SDSS survey is negligible. While cosmic variance is one important consideration, surveys that are limited in volume due to survey size, depth, or additional mass selections, will suffer from the inability to achieve statistically meaningful results from their decreased number statistics because they are unable to finely bin in mass and/or redshift.

\subsection{Implications of distinct trends for the minor merger fraction}
\label{discuss:minor}

It is noteworthy that the minor merger fraction shows remarkably different mass- and redshift-evolution relative to the major merger fraction. The minor merger fraction has a flat dependence on both of these properties. While the slope is flatter, the error bars on the minor merger fraction tend to be larger than those on the major merger fraction for each bin. This could reflect the decreased accuracy of the minor merger classification. 

We explore a few explanations for the flat trends: 1) the minor mergers are subject to different structural assembly processes in the local universe (relative to major mergers), 2) the error bars on the minor merger fraction are obfuscating trends that are positive with stellar mass and negative with redshift, or 3) there are some systematic biases at play in the minor merger classification. 

First, we consider option 1. If this result is not due to a bias but is a physical finding, this demonstrates that minor mergers are about equally important for the assembly of all galaxy masses locally as well as all redshifts within our range. This could further motivate the above explanation explanation that a baryonic process such as feedback is driving the negative redshift trend for the major merger fraction. A process like feedback that increases the fraction of intermediate mass galaxies (hence, increasing the major merger fraction locally) could also lead to a relatively smaller fraction of minor mergers.

We next consider options 2 and 3. We observe underlying structure in the mass-dependent minor merger fraction (a peak at intermediate masses). In our analysis of the properties of the different merger classifications (\S \ref{results:properties}), we find that minor mergers have a tendency towards intermediate masses. This could reflect a bias against identifying low mass and high mass galaxies, which could result in a flatter trend for the minor merger fraction since we should expect to miss both low and high mass galaxies. This makes sense given that the minor merger classification relies upon shape asymmetry to identify faint tidal tails or faint companions. In the case of a bright primary galaxy, this task becomes much more difficult for the classification. This bias is also related to a lower accuracy for the minor merger classification (explanation 2). Our hypothesis is that while this slight bias could exist, we find it unlikely that this slight bias alone is driving the flat evolution. In future work, it would be worth exploring the biases related to the minor merger classification; here we choose instead to focus on the biases related to the major merger classification.

Regardless of whether it is a physical trend or related to classification biases, \textit{the flatness of both of these relations for the minor merger sample further motivates the importance of separating minor mergers from major mergers; if there is significant minor merger contamination in a major merger sample, this would act to flatten out both the mass- and redshift-dependence, resulting in a flat relationship for both.} While we find that this does not have a significant influence on our results (\S \ref{results:double_count}), we recommend that future studies of the redshift dependence of the major merger fraction take this result into consideration.

\subsection{Do the limitations of the simulated training set affect the robustness of these results?}
\label{discuss:robust}

The LDA training set consists of intermediate mass, initially disk-dominated simulated galaxies with initial stellar masses $3.9 \times 10^{10}< M_* (M_{\odot}) < 4.7 \times 10^{10} $ and initial B/T mass ratios $0.0 < B$/$T < 0.2$. \footnote{The stellar masses and B/T mass ratios evolve throughout the time duration of the simulations. For instance, the simulated mergers increase in mass and the major mergers remnants are bulge-dominated (N19).}
Since this training set is limited, we have taken measures to minimize potential biases and find no significant impact on our main result of the mass- and redshift-dependence of the major merger fraction.

To minimize potential morphological biases, the SDSS galaxies used in our merger fraction analysis are restricted to regions familiar to the LDA classifier using the `outlier predictor' flag (Figure \ref{fig:distabc}).
\textit{While the classifier may be morphologically biased for galaxy morphologies outside of the training set, this does not concern the results presented here, which are limited to morphologies that are familiar to the LDA classifier.}

To assess if the galaxies we classify are morphologically biased despite the above precaution, we explore the properties of the SDSS merger sample in \S \ref{results:properties_of_sample} and find no distinction in S/N, $r-$band magnitude, $g-r$ color, stellar mass, and redshift between the merger and parent samples. 
This reflects a major success or our method; that the merger classifications are not biased by galaxy property. In \S \ref{results:compare} when we compare the fraction of ellipticals and mergers in GalaxyZoo classified as LDA major mergers, we find the same fraction (14\%), which is further evidence that the technique does not retain a morphological bias.

In \S \ref{results:morphology}, we confirm that the major merger fraction trends with redshift and stellar mass persist when we control for galaxy morphology ($g-r$ color or B/T ratio). This means that our results hold for all galaxy morphologies in the photometrically clean sample.

Fully investigating the mass and morphological biases of the classification, especially for galaxies with the `outlier predictor' flag are beyond the scope of this work. Future work could investigate the performance of the classifier across different galaxy morphologies.

\subsection{Strengths of this approach}
\label{discuss:strengths}

For many past studies that focus on measuring the major merger fraction, small number statistics are a concern. Cosmic (or sample) variance due to small fields (i.e. see the discussion of Xu et al. 2012) can result in large error bars, leading to a conclusion of flat redshift or mass evolution of the merger fraction. Of additional concern, many of the close pair studies (which constitute the bulk of this literature) suffer from spectroscopic incompleteness at small angular separation, while morphological methods suffer from surface brightness limitations, and as a result are biased towards identifying high mass gas-rich major mergers only. Many morphological methods also suffer from small sample sizes and with a variety of systematics related to different methodologies. 

In this work, we begin from a merger identification technique that is based on a set of well-understood simulations of mergers. This technique has four distinct advantages over past merger identification techniques. 

\begin{enumerate}
    \item We are able to calibrate our methodology, which will become critically important in future work (Simon et al. 2023, in prep), where we plan to constrain the merger rate. In order to determine the merger rate, the merger observability timescale is important, which we are able to measure from the set of simulated mergers. 
    \item Since the technique does not rely on spectroscopic detections, we apply the method to the full SDSS photometric dataset and return the largest-yet sample of merging galaxies. With this large sample, we are able to control for both mass and redshift when we measure the merger fraction as a function of both of these quantities, which we have shown is essential. 
    \item Our technique spans a variety of merger stages, including pre- and post-coalescence stages. It therefore overlaps in stages with both close-pair and morphological studies, which will be crucial for comparing different types of studies when we measure the merger rate. 
    \item Our technique shows significant gains in accuracy and completeness relative to past work, allowing us to build a more complete (and larger) sample of merging galaxies.
\end{enumerate}

\subsection{Caveats and future work}
\label{discuss:caveats}
There are three types of double counting of mergers that can occur under this classification technique: 1) The overlap between major and minor mergers, 2) The overlap between merger stages, and 3) Sometimes in the early stage of the merger, the technique identifies both galaxies as mergers, which is double counting compared to a close-pair technique.

We have already discussed the overlap between merger stages and types in previous sections (\S \ref{results:interpret_stage} and \ref{results:double_count}, respectively). In \S \ref{results:interpret_stage} we find that the early and late stages have significant overlap in classifications but the post-coalescence stage tends to have less overlap; we discuss the implications of this in terms of classification interpretation. In \S \ref{results:double_count} we conclude that the slope of the merger fraction with mass and redshift is unchanged when we account for the double counting of major and minor mergers. 

While splitting mergers by stage was not a primary focus of the merger fraction analysis in this work, in future work (Simon et al. 2023, in prep), we plan to compare our galaxy sample with the close-pair sample from Simon et al. 2022, in prep, in order to constrain the absolute merger fraction. This will be especially important for the early stage mergers, which are most similar to close pair studies. 

In addition, in this work we conduct a brief analysis of the overlap of merger type classifications, in other words, the contamination of the major merger fraction by minor mergers. Our focus is primarily on if this affects our findings related to the merger fraction slope with stellar mass and redshift. In future work we plan to characterize the overlapping merger populations.

In future work it will also be necessary to further address the third type of double counting. For early stage mergers, we find that the LDA method sometimes (but not always) identifies both galaxies in a pair as merging galaxies. This represents a double count relative to close-pair studies where the duo of merging galaxies would be considered to be one `pair’. On the other hand, the LDA method also identifies mergers in the late and post-coalescence stages, which boosts our derived merger fraction relative to that of close-pair methods. Both of these considerations mean that directly comparing our method to close-pair studies is difficult. For this reason, in this work, we have not attempted to directly compare the absolute number of mergers and have instead compared the slope of $f_{\mathrm{merg}}$ with stellar mass and redshift. In Simon et al. 2023, in prep, we plan to compare our galaxy sample with the close-pair sample from Simon et al. 2022, in prep, in order to constrain the absolute merger fraction. In this future work (Simon et al. 2023, in prep), we will also be able to determine the calibration factor, $C_{\mathrm{merg}}$, to convert between the close pair fraction and the fraction of close pairs that will ultimately merge.

It is also important to mention a fundamental difference between morphologically-reliant merger identification techniques like the LDA technique and spectroscopic-based techniques like the close-pair method. Galaxies like those shown in the top left panel of Figure \ref{fig:ack_fps} that are identified as mergers by the LDA technique may in fact be chance projections of unrelated galaxies along the line of sight. Fully characterizing the expected frequency of these chance projections is beyond the scope of this work, although we plan to discuss this in more depth in Simon et al. 2022, in prep, when we compare our merger sample to that of the close pair method.

\section{Conclusions}
\label{conclusions}

In this work we apply the merger classification method from \citet{Nevin2019} to the 1.3 million galaxies in the Sloan Digital Sky Survey DR16 photometric catalog. We additionally expand the merger classifications from N19 to include the different stages of the merger in addition to major versus minor classifications. This results in twelve different merger classifications: major and minor and then we further split by stage: early, late, pre-coalescence (includes early and late), and two different post-coalescence classifications (one extends to 0.5 Gyr post-merger, one extends to 1.0 Gyr). 

We apply all of these classifications to image cutouts from SDSS, calculate the $p_{\mathrm{merg}}$ values and repeat this process for a range of different input priors, marginalizing over these priors to retrieve the posterior distribution of $p_{\mathrm{merg}}$ values for all galaxies for all classifications. We provide these classifications to the reader in the form of online-available tables in addition to an interpretable classification repo known as \texttt{MergerMonger}. In the text we provide examples for how to interpret the results and distinguish between different merger types. 

We next analyze the properties of the merger samples and compare these properties to other merger samples in the literature. We conclude that the properties of the different types of mergers span the full range of properties of the parent SDSS distribution (in S/N, $r-$band magnitude, color, stellar mass, and redshift), which is a major success of the method. We also find that the LDA technique retrieves the majority of the GalaxyZoo and \citet{Ackermann2018MNRAS.479..415A} mergers and further identifies a large sample of galaxies as mergers that were missed by these techniques, demonstrating its success in finding less-obvious mergers than visually identified samples. 

The main goal of this paper is to retrieve the merger fraction ($f_{\mathrm{merg}}$) as a function of galaxy properties, which we do by measuring stellar masses, carefully building a mass-complete sample (our final sample is 310,012 galaxies), and binning by both stellar mass and redshift. For the major merger sample we find a significantly positive trend (1-3$\sigma$ confidence) between $f_{\mathrm{merg}}$ and stellar mass and a significantly negative (to 1$\sigma$ confidence) trend with redshift. We show these key results in Figures \ref{fig:redshift_bins} and \ref{fig:mass_bins}, respectively. This trend is robust between stellar masses of $10.5 < M_* (log\ M_{\odot}) < 11.6$ and redshifts of $0.03 < z < 0.19$. We show that when we do not correct for completeness or bin for mass, the strong dependence of the major merger fraction on mass results in a positive redshift slope, underscoring the importance of a careful binning analysis with a large sample size to recover this result.

Examining these results in the context of past theoretical and observational work, we find that the positive trend of the major merger rate with stellar mass agrees with past results and is consistent with a hierarchical assembly scenario for the Universe. On the other hand, this is the first time a study has focused on the measuring the merger fraction locally ($z < 0.2$) for finely spaced mass and redshift bins, which underscores the uniqueness of the finding of a negative trend for the major merger fraction with redshift.

In future work (Simon et al. 2023, in prep) we plan to use these results in combination with the SDSS-derived close pair fraction from Simon et al. 2022, in prep to calculate a merger rate. From this, we can constrain the gravitational wave background from SMBH mergers.

\section*{Acknowledgements}
To Dr. Aaron Stemo and Dr. Plamen G Krastev for some phenomenal supercomputing support.

This research is supported by NSF AST-1714503 and NSF AST-1847938.

JS is supported by an NSF Astronomy and Astrophysics Postdoctoral Fellowship under award AST-2202388.

The computations in this paper were run on the FASRC Cannon cluster supported by the FAS Division of Science Research Computing Group at Harvard University.

Funding for the Sloan Digital Sky 
Survey IV has been provided by the 
Alfred P. Sloan Foundation, the U.S. 
Department of Energy Office of 
Science, and the Participating 
Institutions. 

SDSS-IV acknowledges support and 
resources from the Center for High 
Performance Computing  at the 
University of Utah. The SDSS 
website is www.sdss.org.

SDSS-IV is managed by the 
Astrophysical Research Consortium 
for the Participating Institutions 
of the SDSS Collaboration including 
the Brazilian Participation Group, 
the Carnegie Institution for Science, 
Carnegie Mellon University, Center for 
Astrophysics | Harvard \& 
Smithsonian, the Chilean Participation 
Group, the French Participation Group, 
Instituto de Astrof\'isica de 
Canarias, The Johns Hopkins 
University, Kavli Institute for the 
Physics and Mathematics of the 
Universe (IPMU) / University of 
Tokyo, the Korean Participation Group, 
Lawrence Berkeley National Laboratory, 
Leibniz Institut f\"ur Astrophysik 
Potsdam (AIP),  Max-Planck-Institut 
f\"ur Astronomie (MPIA Heidelberg), 
Max-Planck-Institut f\"ur 
Astrophysik (MPA Garching), 
Max-Planck-Institut f\"ur 
Extraterrestrische Physik (MPE), 
National Astronomical Observatories of 
China, New Mexico State University, 
New York University, University of 
Notre Dame, Observat\'ario 
Nacional / MCTI, The Ohio State 
University, Pennsylvania State 
University, Shanghai 
Astronomical Observatory, United 
Kingdom Participation Group, 
Universidad Nacional Aut\'onoma 
de M\'exico, University of Arizona, 
University of Colorado Boulder, 
University of Oxford, University of 
Portsmouth, University of Utah, 
University of Virginia, University 
of Washington, University of 
Wisconsin, Vanderbilt University, 
and Yale University.

\section*{Data Availability}
All data products detailed in \S \ref{results:class} are available on \href{https://zenodo.org/record/7438610#.Y5o8MuzMLYI}{Zenodo}\footnote{DOI: 10.5281/zenodo.7438610}. For the \texttt{MergerMonger} code, see the \href{https://github.com/beckynevin/MergerMonger-Public}{Github} repo \github{https://github.com/beckynevin/MergerMonger-Public}. This includes all of the analysis utilities used to generate the results of this paper.

\section*{Software} 
astropy \citep{astropy2013}, pandas \citep{reback2020pandas}, seaborn \citep{seaborn}, scikit-learn \citep{scikit-learn}, and gnu parallel \citep{gnu_parallel}

\bibliographystyle{mnras}
\bibliography{library_overleaf} 

\appendix

\section{Photometric versus spectroscopic redshifts}
\label{ap:mass}

\begin{figure}
\centering
\includegraphics[scale=0.17]{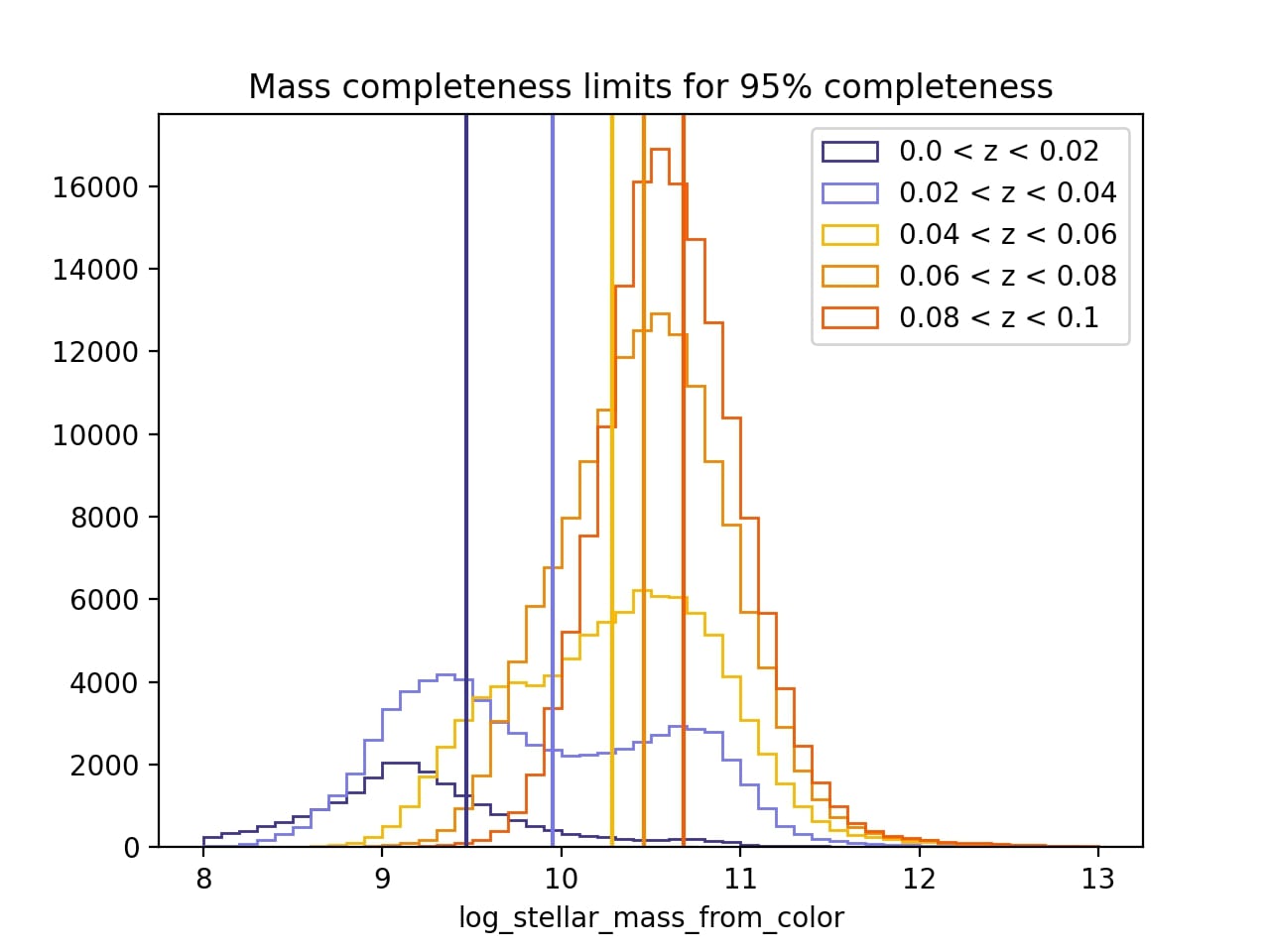}
\includegraphics[scale=0.17]{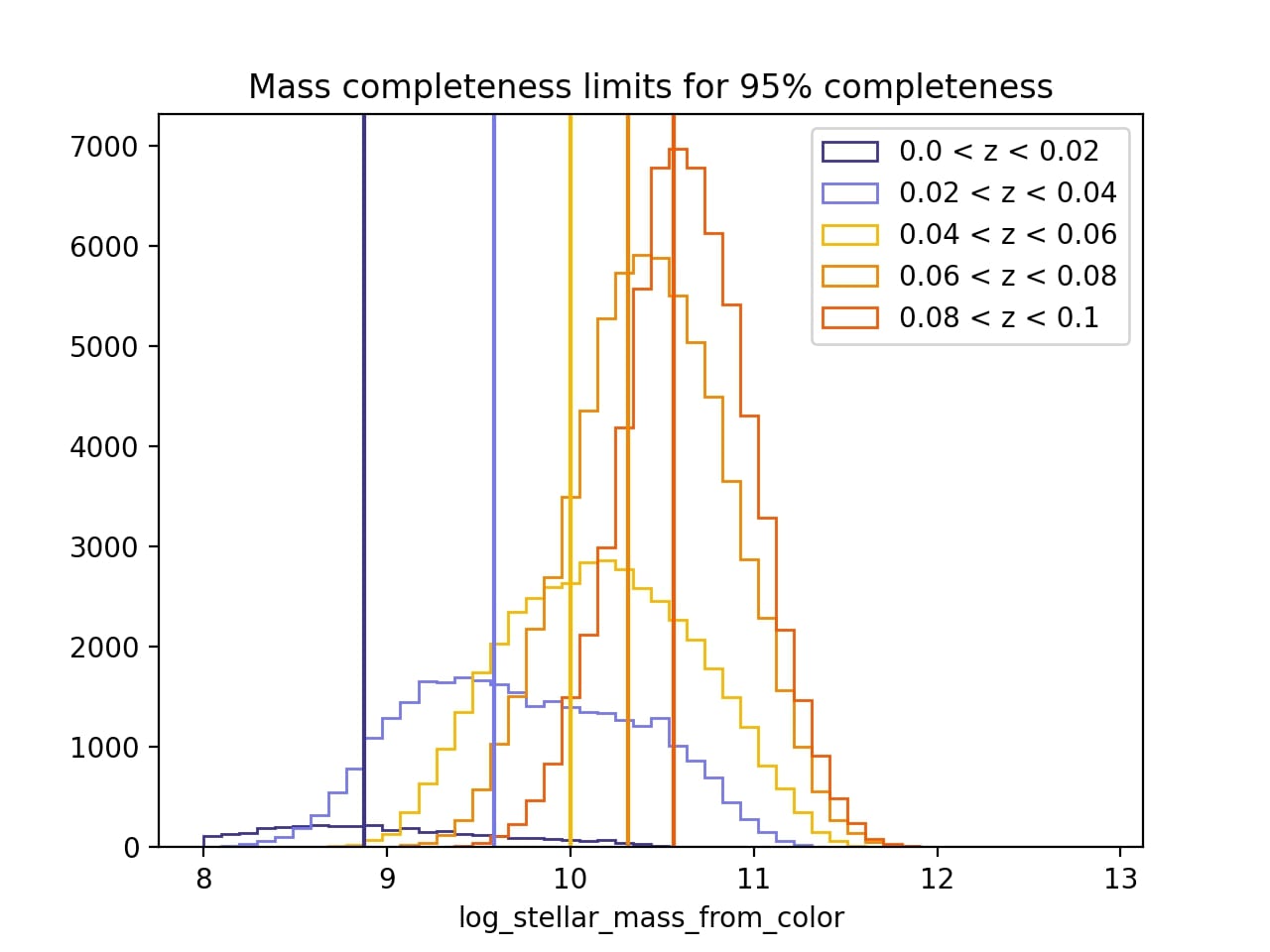}
\caption{Mass distribution by redshift bin for stellar masses derived using photometric redshifts (top) and spectroscopic redshifts (bottom). Our ultimate results are unchanged, yet we note that the mass distributions have very different shapes depending on which redshift prescription we use.}
\label{fig:ap_mass}

\end{figure}

Here we explore the effect of using photometric redshifts on the mass calculation and subsequent mass-completeness cut. While we ultimately find that the merger fraction evolution as a function of redshift is unchanged by using spectroscopic masses, we nevertheless find that the mass distributions as a function of redshift bin (Figure \ref{fig:ap_mass}) are different. 

Figure \ref{fig:ap_mass} shows the mass distribution and 95\% completeness cut for five redshift bins with spacings $\Delta z = 0.02$ for color-based stellar masses measured using photometric (left) and spectroscopic redshifts (right). The distributions on the left are distinctly double-peaked which leads us to conclude that photometric-based redshifts are biasing a population of low-redshift galaxies towards higher masses. When we directly compare photometric-based redshift measurements to spectroscopic-based redshift measurements, we find a bias towards higher redshifts among the photometric measurements at low redshift, which could be resulting in a population of boosted masses, hence the artificial double-peaked profile.

\section{Sanity checks for the result of a negative slope of the major merger fraction with redshift}
\label{ap:sanity}
As we will address in the discussion section, the result of increasing merger fraction with stellar mass has precedent in the literature. However, the result of decreasing merger fraction with redshift over the range $0.03 < z < 0.19$ is relatively unprecedented. Given this surprising result, we use this section to explore whether this result of decreasing merger fraction with increasing redshift is physical (real) or whether we can attribute it to sample systematics (i.e. mass incompleteness at higher redshift or errors in the mass calculation or determination of the photometric redshift).

To test the result of a negative trend with redshift for the major merger fraction, we re-run the major merger fraction measurement in several ways: 1) Using the full, mass-incomplete sample (\S \ref{sanity:incomplete}), 2) Adjusting the redshift binning scheme (\S \ref{sanity:changing_bins}), 3) Using spectroscopic redshifts (\S \ref{sanity:redshift}), 4) Using SPS-derived stellar masses (\S \ref{sanity:mendel_mass}), 5) Running the analysis with the A18 mergers (\S \ref{sanity:A18}), and 6) Re-running the analysis with different merger classifications (\S \ref{sanity:classifications}). We conclude in \S \ref{results:sanity} by arguing that the decreasing merger fraction with redshift is not a result of sample systematics and is instead a physical result.
 
\subsection{Mass incompleteness}
\label{sanity:incomplete}

We run the major merger fraction calculation for the full sample (i.e. mass incomplete) and find that the trends persist. In this case, the slope of the major merger fraction with mass and redshift are less steep in both cases. Additionally, visually, the trend with redshift is very steep at low redshift followed by a flattening out of the redshift trend at redshifts $z > 0.1$. We hypothesize that this flattening could be due to significant mass incompleteness at high redshift. We discuss this trend in \S \ref{sanity:no_mass_bins}.

\subsection{Changing bins}
\label{sanity:changing_bins}

We adjust the redshift bin sizes and use these new binning schemes to re-create the mass complete sample and re-calculate the mass- and redshift-dependence of the merger fraction. We use five different linear spacing schemes for the redshift bins ($\Delta z = 0.01, 0.02, 0.03, 0.04, 0.05$). We also use an adaptive binning scheme where the redshift bin spacing is determined using a k-means approach; this constructs redshift bins that have the same number of galaxies in each bin. We also rerun all of these calculations for the mass incomplete sample. For all binning schemes we still find a positive slope with mass for the majority of the redshift bins and a negative slope with redshift for the majority of the mass bins. This confirms that the redshift bin spacing and/or the associated number of galaxies in each bin are not responsible for the negative trend of the major merger fraction with redshift.

\subsection{Spectroscopic redshifts}
\label{sanity:redshift}

As described in \S \ref{methods:mass}, we cross-match our clean (no photometric flags) merger sample with the \citet{Mendel2014} catalog, which is mass-incomplete. We then use the spectroscopic redshifts from this sample to re-run the color-based mass measurement and redo the mass completeness calculation. Finally, we re-run the major merger fraction analysis. We find that the photometric-based redshifts, which are available for our full sample, exhibit a bias towards higher redshifts (as described in Appendix \ref{ap:mass}), which shifts some galaxies at low redshift out of their redshift bins and results in higher stellar mass estimates. Despite these biases, we still find a significant positive trend for $f_{\mathrm{merg}}$ with mass and a negative trend for $f_{\mathrm{merg}}$ with redshift for the majority of the mass and redshift bins.

\subsection{Mendel masses}
\label{sanity:mendel_mass}

To test the robustness of this result with respect to the mass calculation, we re-run the analysis from the previous section, but instead of color-based stellar masses, we use the stellar masses from \citet{Mendel2014}, which are derived using a S\'ersic decomposition coupled with an SPS-based approach. We also use the spectroscopic redshifts for this analysis. We find that the negative slope of the major merger fraction with respect to redshift is maintained even for the different method of mass measurement, meaning that the color-based approach to mass calculation is not responsible for the negative trend.

\subsection{A18 mergers}
\label{sanity:A18}

To explore whether the negative trend with redshift is a peculiarity of the sample or due to the merger classification, we re-run the analysis with the merger classification of A18. Cross-matching the A18 sample with the SDSS sample reduces the sample size to 97k galaxies. Since the sample size is significantly reduced, we adjust the mass binning to include fewer bins. Due to this reduced sample size, our investigation spans a smaller range in redshift ($0.02 < z < 0.08$) and stellar mass ($10.5 < M_* (log\ M_{\odot}) < 11$). We first verify that the positive slope with stellar mass and the negative slope with redshift persist for the LDA merger classifications. 

When we use the A18 mergers instead (using a threshold of 0.95), we find that the slope is positive with mass for the three redshift bins but that the slope is no longer universally negative with redshift for the mass bins. Instead, it is positive for two of the mass bins, flat for two of the bins, and negative for the two highest mass bins. By using the same cross-matched galaxy subsample with both the LDA classification and the A18 classification, we can confirm that the negative trend we observe for the LDA sample of mergers is due to the merger classification and not a peculiarity of the sample selection.

Given that we have not conducted a full analysis focusing on A18 mergers and are simply comparing them to our merger sample, an explanation of why the merger fraction trends differ for the A18 sample is outside the scope of this paper. However, we can speculate on some differences in merger selection that could lead to these different conclusions. Examining the properties of the two different merger samples, we find some notable differences: The A18 sample of mergers have lower concentrations, higher $A$ values, lower $A_S$ values, tend to be at higher redshifts, and are redder than the LDA sample. As discussed in \S \ref{results:compare}, in most other properties (i.e. S/N and stellar mass), the merger samples are similar. The A18 sample may be slightly biased towards identifying redder galaxies with higher redshifts, which may result in a higher merger fraction at higher redshifts for some of the mass bins.

\subsection{Different classifications}
\label{sanity:classifications}
Here we calculate the dependence of the merger fraction on mass and redshift for the different merger classifications, including the major merger pre- and post-coalescence classifications and the minor merger combined and pre- and post-coalescence classifications. All variants of the major merger classification give similar results; both pre- and post-coalescence have positive slopes for how the merger fraction trends with mass. The pre-coalescence major merger fraction has a negative evolution with redshift. In the case of the post-coalescence major merger classification, the slope of $f_{\mathrm{merg}}$ with redshift is mostly negative and flat for some mass bins.

The minor merger classifications give very different results; as presented in \S \ref{results:properties_minor}, the slope of the combined minor merger fraction with mass is flat (often with a peak at intermediate masses) and it is flat with redshift. The same goes for the pre-coalescence minor mergers and early stage minor mergers. The late stage minor mergers are positive with mass and negative with redshift for most bins, just like the major mergers. The post-coalescence minor mergers are positive with mass and negative or flat with redshift, so very similar to the post-coalescence major mergers. 

There are two important lessons here. First, that minor mergers do not show mass-dependent or redshift-dependent evolution. We will discuss the physical implications of this in \S \ref{discuss:minor}. Second, the dependence of $f_{\mathrm{merg}}$ on stellar mass and redshift is affected by merger stage as well as mass ratio. The implication is that studies that identify mergers should pay careful attention to the biases of the merger sample. However, it is important to note that the different stage classifications of the major merger fraction give similar results. Our finding of the negative trend of the major merger fraction with redshift therefore cannot be attributed to a difference in merger observability timescale of our method relative to close pair techniques. 

\bsp	
\label{lastpage}
\end{document}